\documentclass[aps,prd,showpacs,preprintnumbers,nofootinbib,twocolumn,
superscriptaddress]{revtex4-1}

\usepackage{amsmath}
\usepackage{amssymb}
\usepackage{amsfonts}
\usepackage{dsfont}
\usepackage{slashed}
\usepackage{graphicx}
\usepackage{graphics}
\usepackage{epsfig}
\usepackage{bbm,bm}
\usepackage{psfrag}
\usepackage{epstopdf}
\usepackage[breaklinks]{hyperref}
\usepackage{color}
\graphicspath{{figures/}}

\newcommand{\pat}{\partial_t}
\newcommand{\Eqref}[1]{Eq.~\eqref{#1}}

\newcommand{\trho}{\tilde{\rho}}
\newcommand{\widesim}[2][1.5]{
  \mathrel{\underset{#2}{\scalebox{#1}[1]{$\sim$}}}
}

\newcommand{\dG}{N^2-1}
\newcommand{\Np}{N_{\text{p}}}
\newcommand{\fgFP}{quasi-fixed point}
\newcommand{\fgFPh}{quasi-fixed-point}
\newcommand{\fgFPs}{quasi-fixed points}

\newcommand{\FgFPs}{Quasi-fixed points}

\begin{document}

\title{Nonabelian Higgs models: paving the way for asymptotic freedom}

\author{Holger Gies}
\email{holger.gies@uni-jena.de}
\affiliation{\mbox{\it Theoretisch-Physikalisches Institut, Abbe Center of 
Photonics, Friedrich-Schiller-Universit{\"a}t Jena,}
\mbox{\it D-07743 Jena, Germany}}
\affiliation{Helmholtz-Institut Jena, Fr\"obelstieg 3, D-07743 Jena, Germany}

\author{Luca Zambelli}
\email{luca.zambelli@uni-jena.de}
\affiliation{\mbox{\it Theoretisch-Physikalisches Institut, Abbe Center of 
Photonics, Friedrich-Schiller-Universit{\"a}t Jena,}
\mbox{\it D-07743 Jena, Germany}}


\begin{abstract}
Asymptotically free renormalization group trajectories can be
constructed in nonabelian Higgs models with the aid of generalized
boundary conditions imposed on the renormalized action. We detail this
construction within the languages of simple low-order perturbation
theory, effective field theory, as well as modern functional
renormalization group equations. We construct a family of explicit
scaling solutions using a controlled weak-coupling expansion in the
ultraviolet, and obtain a standard Wilsonian RG relevance
classification of perturbations about scaling solutions. We obtain
global information about the quasi-fixed function for the scalar potential
by means of analytic asymptotic expansions and numerical shooting
methods. Further analytical evidence for such asymptotically free
theories is provided in the large-$N$ limit. We estimate the
long-range properties of these theories, and identify initial/boundary
conditions giving rise to a conventional Higgs phase.
\end{abstract}

\maketitle

\section{Introduction}
\label{sec:intro}

A fascinating aspect of quantum field theories is the fact
that they can represent truly fundamental theories valid on all energy
or length scales as a matter of principle -- or predict their own
failure. An example of the first class are nonabelian gauge theories
that represent \textit{perfect} quantum field theories in the sense of
being potentially valid on all scales. In particular, their
high-energy behavior is governed by asymptotic freedom
\cite{Gross:1973id,Politzer:1973fx}. This property persists also upon
the inclusion of a suitable matter content
\cite{Gross:1973ju,Gross:1974cs,Politzer:1974fr,Cheng:1973nv,Bais:1978fv,
Callaway:1988ya}. 

An example of the second class are pure scalar theories in $d=4$
spacetime dimensions which suffer from triviality
\cite{Wilson:1973jj}, implying that a meaningful continuum theory
exists only for the noninteracting theory. While rigorous proofs for
triviality exist for $d>4$ \cite{Frohlich:1982tw}, strong evidence for
triviality in $d=4$ has been collected by lattice simulations
\cite{Luscher:1987ay,Luscher:1987ek,Luscher:1988uq,Hasenfratz:1987eh,
Heller:1992js,Wolff:2009ke,Buividovich:2011zy}
as well as by functional renormalization group (RG) studies
\cite{Rosten:2008ts}. Similar conclusions appear to hold for QED
\cite{Gockeler:1997dn,Gies:2004hy}.

While triviality rather represents a mathematical consequence that
arises from insisting on sending the maximum validity scale of a
theory (technically corresponding to an ultraviolet (UV) cutoff) to
infinity, the physical viewpoint is slightly different: by keeping the
interaction at low energies finite, triviality translates into a
breakdown of the quantum-field-theory description at a finite UV
scale. For instance, in perturbation theory this is signaled by the
artifact of a Landau pole singularity in the RG evolution of couplings
\cite{Landau:1955,GellMann:1954fq}. Nonperturbatively, this
singularity might be screened such as in QED
\cite{Gockeler:1997dn,Gies:2004hy}, but the conclusion persists that
the physically observed infrared (IR) behavior cannot be connected to
the same theory at an arbitrarily high scale with suitably renormalized
couplings.

Triviality appears to threaten the standard model of particle physics
not only with respect to its U(1) sector. It is well conceivable that
the U(1) factor arises from symmetry breaking of a compact grand
unified gauge theory. It is rather the potential triviality of the
Higgs sector which we consider as the more substantial obstacle of
building a perfect quantum field theory for particle physics. Hence,
we concentrate on nonabelian Higgs systems in the present work. 

A naive expectation would be that nonabelian Higgs systems should be
trivial, as the gauge sector is asymptotically free such that the
scalar triviality problem remains. This is not necessarily true:
already a standard perturbative analysis
\cite{Gross:1973ju,Chang:1974bv,Chang:1978nu,Fradkin:1975yt,Salam:1978dk,%
  Salam:1980ss,Callaway:1988ya,Giudice:2014tma,Holdom:2014hla} reveals
the existence of gauged Yukawa models, where also the seemingly
problematic scalar self-interaction $\sim \lambda\phi^4$ can become
asymptotically free as well. This happens along suitable RG
trajectories depending on the precise matter content of the model.
Such scenarios in principle have the advantage of yielding a
\textit{reduction of couplings}
\cite{Zimmermann:1984sx,Oehme:1984yy,Heinemeyer:2014vxa}, as the
scalar self-interaction may then be induced by the gauge sector. In
essence, this can fix the Higgs-to-gauge-boson mass ratio. So far, a
concrete model building has not been satisfactory, as no unequivocally
convincing model sufficiently similar to the standard model has been
identified, though the search is ongoing, see, e.g.,
\cite{Hetzel:2015bla,Pelaggi:2015kna,Pica:2016krb,Molgaard:2016bqf}.

In the present work, we detail and extend our recent results on the
construction of asymptotically free non-abelian Higgs models which
become visible upon the use of generalized boundary conditions for the
correlation functions of the theory \cite{Gies:2015lia}. The fact that
the RG behavior of a model is sensitive to boundary conditions is well
known from interacting fixed-points in statistical field theories. For
instance, the scaling solution for the potential at the Wilson-Fisher
fixed point yields the correct critical exponents of the Ising
universality class, once it satisfies suitable boundary conditions for
small and large fields as well as self-similarity conditions
\cite{Hasenfratz:1985dm,Morris:1998da,Morris:1996xq,Morris:1997xj}. The
new ingredient in our construction is that we allow for boundary
conditions for the scalar potential which depend on the gauge coupling
$g$.

Such a dependence appears natural in view of the fact that the scalar
potential governing the scalar self-interactions is expected to become
absolutely flat for an asymptotically free theory, vanishing
synchronously with the asymptotically free gauge coupling $g$. In
contrast to the role of boundary conditions at interacting fixed
points, which typically constrain the number of possible scaling
solutions severely (see, e.g., the \textit{singularity count} in
\cite{Dietz:2012ic,Demmel:2015oqa}), we find a larger set of
possibilities of imposing boundary conditions on the scalar potential
in the case of asymptotic freedom, yielding a family of scaling
solutions.

The present work intends to give a detailed account of our
construction. For this, we use several approaches of increasing
sophistication in order to explain our results from various
viewpoints. In Sect.~\ref{sec:PT}, we start from simple one-loop
perturbation theory in order to make contact with the standard
language of describing asymptotic freedom. Sections \ref{sec:EFT1} and
\ref{sec:EFT2} use the language of effective field theory that allows
to study the renormalization properties of the model upon the
inclusion of higher-dimensional operators.  This approach makes the
asymptotically free scaling solutions already visible and facilitates
a first glance at their connection to RG boundary conditions. In
Sect.~\ref{sec:FRG}, the picture unfolds more comprehensively on the
basis of the functional RG flow of nonabelian Higgs models. This
modern tool gives access to the global behavior of the scalar
potential at high energies and allows for a classification of the
scaling solutions as well as a RG relevance count of perturbations
about the noninteracting Gau\ss{}ian fixed point. Further insights are
obtained on the basis of a large-$N$ approximation in
Sect.~\ref{sec:largeN}, which also allows for a first but rough
estimate of the full flow from the UV to the IR. The latter is also
discussed more phenomenologically in Sect.~\ref{sec:UVIR} in order to
estimate the long-range properties of our models. We conclude in
Sect.~\ref{sec:conc}.

\section{Perturbative one-loop analysis}
\label{sec:PT}

We consider nonabelian Higgs models with an SU($N$) gauge sector
coupled to a charged scalar $\phi^a$ in the fundamental
representation. The classical action reads
\begin{equation}
S_\mathrm{cl}=\int\!\! d^4x\Big[ \frac{1}{4}F_{\mu \nu}^iF^{i\mu \nu}
+(D^{\mu}\phi)^{\dagger}(D_{\mu}\phi)+\bar{m}^2\rho+\frac{\bar{\lambda}}{2}\rho^2\Big],
\label{eq:classicalaction}
\end{equation}
where $\rho:=\phi^{a\dagger}\phi^a$. Classically, the model has three
(bare) parameters, the scalar mass $\bar{m}$, self-coupling
$\bar{\lambda}$, and gauge coupling $\bar{g}$. From a perturbative
viewpoint, the two couplings correspond to deformations of the free
theory. The noninteracting limit $\bar\lambda=0=\bar g$ denotes the
Gau\ss{}ian fixed point. The covariant derivative reads
\begin{equation}
D_\nu^{ab}=\partial_\nu\delta^{ab}-i\bar{g} W_\nu^i (T^i)^{ab},
\label{eq:covder}
\end{equation}
where $[T^i,T^j]=if^{ijk}T^k$ are the generators of the fundamental
representation and $W_\nu^i$ denotes the Yang-Mills vector potential
with field strength $F_{\mu \nu}^i=\partial_\mu W_\nu^i-\partial_\mu
W_\nu^i+\bar{g} f^{ijl}W^j_\mu W^l_\nu$.

For the present purpose, perturbative quantization of the model is
most-conveniently performed with the Faddeev-Popov method applied to a
background-field $R_\alpha$ gauge, as detailed below in
Sect.~\ref{sec:FRG}.  For a first perturbative glance at the system,
it suffices to consider the conventional one-loop $\beta$ functions
for the renormalized couplings $g$ and $\lambda$, as presented for
example by Gross and Wilczek~\cite{Gross:1973ju},
\begin{eqnarray}
\pat g^2=\beta_{g^2}&=&-b_0 g^4\label{eq:1lbetag}\\
\pat\lambda =\beta_{\lambda}&=&A\lambda^2 +B^\prime\lambda g^2 + C g^4,
\label{eq:1lbetal}
\end{eqnarray}
where $\pat\equiv k \frac{d}{dk}$ denotes the derivative with respect
to an RG scale $k$. The constants are given by
\begin{eqnarray}
b_0&=&\frac{1}{8\pi^2} \left(\frac{11}{3}N-\frac{1}{6}\right), \quad A=\frac{N+4}{8\pi^2}\ , \label{eq:constSUN}\\
B^\prime&=&-\frac{3}{8\pi^2} \frac{N^2-1}{N}\ , \quad C=\frac{3}{8\pi^2} \frac{(N-1)(N^2+2N-2)}{4N^2}. \nonumber
\end{eqnarray}
We mostly specialize to the simplest case of SU(2), where
\begin{equation}
b_0=\frac{43}{48\pi^2} , \,\,\, A=\frac{3}{4\pi^2} , \,\,\,
B^\prime=-\frac{9}{16\pi^2} , \,\,\, C=\frac{9}{64\pi^2}. \label{eq:constSU2}
\end{equation}
We emphasize that these $\beta$ functions approximate perturbatively
the RG flow of the couplings in the deep Euclidean region, where all
mass scales are neglected compared to energy, momentum, or RG
scales. In this regime, the flow of the mass parameter decouples from
Eqs.~\eqref{eq:1lbetag},\eqref{eq:1lbetal}, and is thus ignored at this
point. The $\beta$ functions define a vector field on the coupling space. 
The zeros of this vector field read 
\begin{eqnarray}\label{eq:roots_oneloop}
g^2&=&0\nonumber\\
\lambda_{1,2}&=&\left[\frac{ -B^\prime \pm \sqrt{B^{\prime\, 2}-4AC}}{2}\right] g^2=\xi_{1,2} g^2=0\ .
\end{eqnarray}
The notation suggests that the Gau\ss{}ian fixed point $g=\lambda=0$
can be read as being governed by the vanishing of the gauge coupling
$g\to0$. At finite $g$, the constant combinations $\xi_{1,2}$ define the
zeros of the scalar sector, $\beta_\lambda=0$. For SU(2), these roots are complex, 
\begin{equation}
\xi_{1,2}=\frac{3\pm i\sqrt{3}}{8}, \label{eq:xipm}
\end{equation}
(they turn real for $N\geq(\sqrt{21}+1)/2\simeq2.79$). In the present
simple case, the RG flow can be integrated straightforwardly. The flow
of $\lambda$ can be written as
\begin{equation}\label{eq:integratedflow_oneloop}
\lambda(g^2)=-\frac{g^2}{2A}\left[B+\sqrt{\Delta}\tanh\!\left(\frac{\sqrt{\Delta}}{2 b_0}\ln\frac{g_\Lambda^2}{g^2}\right)\right],
\end{equation}
where
\begin{equation}
B=B^\prime+b_0\quad , \quad \Delta=B^2-4AC, \label{eq:defBDelta}
\end{equation}
and $g_\Lambda^2$ is an integration constant. The flow of the gauge coupling
obeys the standard asymptotically free log-like running involving a
separate integration constant.

\begin{figure}[!t]
\begin{center}
 \includegraphics[width=0.23\textwidth]{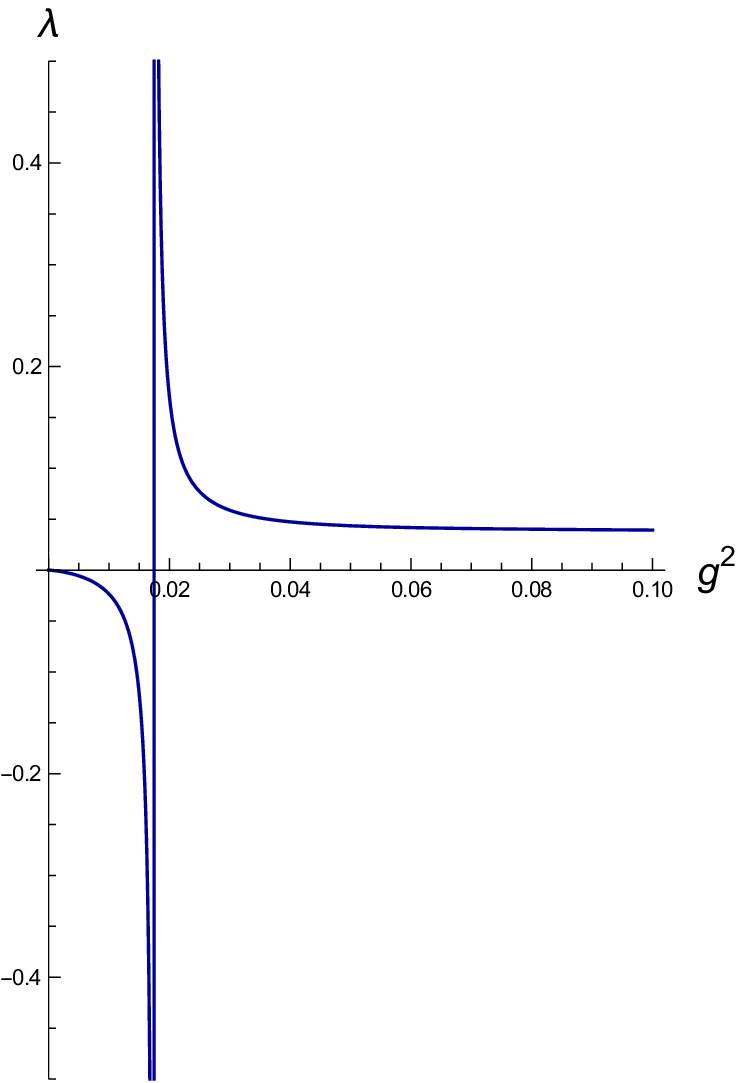}\ \ \
 \includegraphics[width=0.23\textwidth]{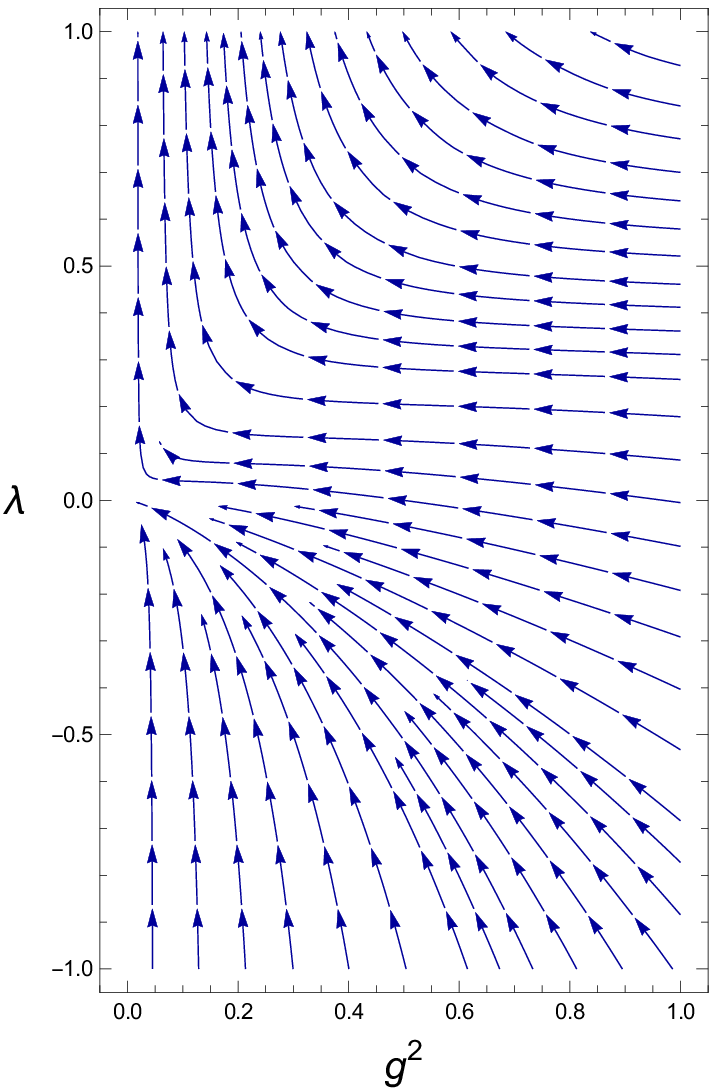}
 \caption{One-loop flow of the SU(2) model. Left panel: integrated
   flow with initial condition $\ln g_\Lambda^2=1$, illustrating the
   triviality problem signaled by the perturbative Landau pole at
   finite $g^2$.  Right panel: phase
   diagram with the Landau pole being visible as a UV run-away of trajectories towards
   $\lambda\to+\infty$ for physical low-energy boundary conditions
   $\lambda,g^2>0$ at some IR scale.}
\label{fig:integratedflow_trivial_oneloop}
\end{center}
\end{figure}

If $\Delta$ were positive, one could follow a smooth trajectory in the
positive $(g^2,\lambda)$ plane down to the origin corresponding to the
Gau\ss{}ian fixed point, and the theory would be asymptotically
free. However, since $\Delta$ is negative for all SU($N$) with
$N\geq2$, the $\tanh$ turns into a tangent. This induces branch cuts
at positions depending on the initial conditions. One cut inevitably
occurs in between a positive $(g^2,\lambda)$ initial point and the
origin. This is the nonabelian Higgs version of the Landau pole
occurring in the pure scalar theory that prevents the theory to be
perturbatively meaningful at all scales.  This is plotted for the
present model in the left panel of Fig.~\ref{fig:integratedflow_trivial_oneloop}.

Incidentally, $\Delta$ could be made positive by suitably tuning $b_0$
to small values upon introducing a balanced number of
fermions. This line of model building has been pursued  since the early
days of asymptotic freedom
\cite{Gross:1973ju,Chang:1974bv,Chang:1978nu,Fradkin:1975yt,Salam:1978dk,%
  Salam:1980ss,Callaway:1988ya,Giudice:2014tma,Holdom:2014hla}. However
in this work, we stay within the class of nonabelian Higgs models, and
intend to construct asymptotically free trajectories without further
degrees of freedom.

The perturbative Landau-pole problem can also be illustrated directly
with the RG flow vector field provided by the $\beta$ functions on the
$g^2,\lambda$ plane. Such a phase diagram is shown in the right panel
of Fig.~\ref{fig:integratedflow_trivial_oneloop}. The branch cut
manifests itself in the form of a separatrix making its way from the
lower right corner of the plane to the origin.  Trajectories starting
at $\lambda,g^2>0$ at an IR scale are repelled from this separatrix
during their flow towards the UV (direction of arrows) and run away
towards $\lambda\to+\infty$ in a finite RG time. Trajectories on the
left of this line at positive $g^2$ but negative $\lambda$ are on the
left of their own Landau pole and approach the Gau\ss ian fixed point
in the UV. The latter trajectories are considered unphysical as the
scalar potential appears unstable.

\begin{figure}[!t]
\begin{center}
\includegraphics[width=0.23\textwidth]{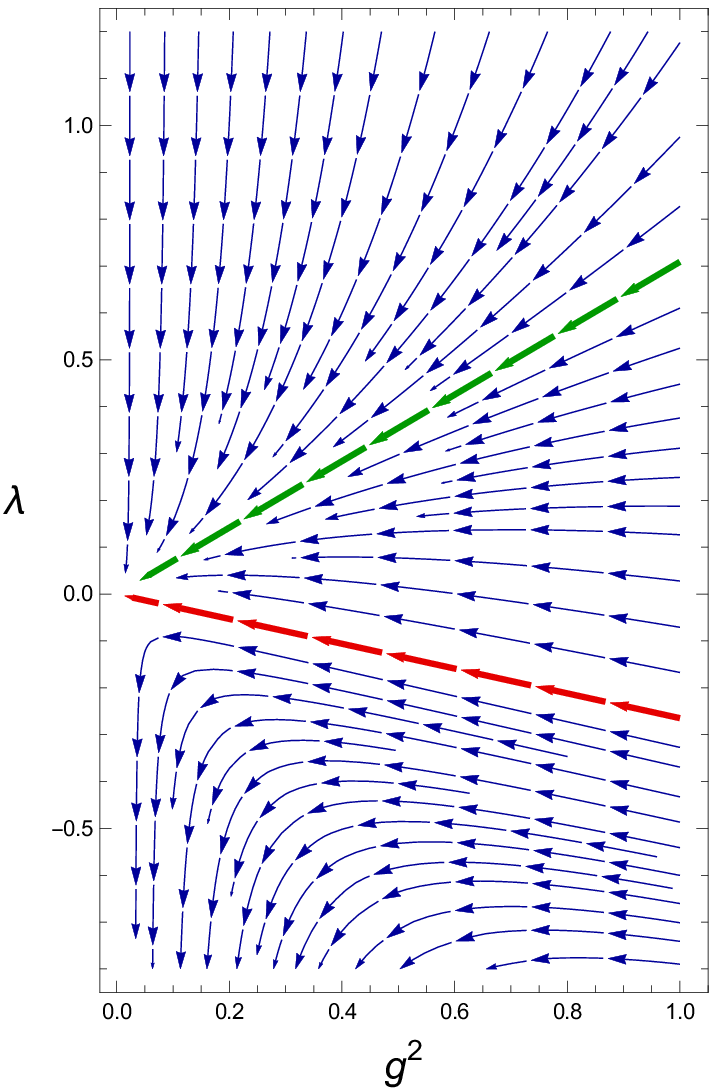}\ \ \ 
\includegraphics[width=0.23\textwidth]{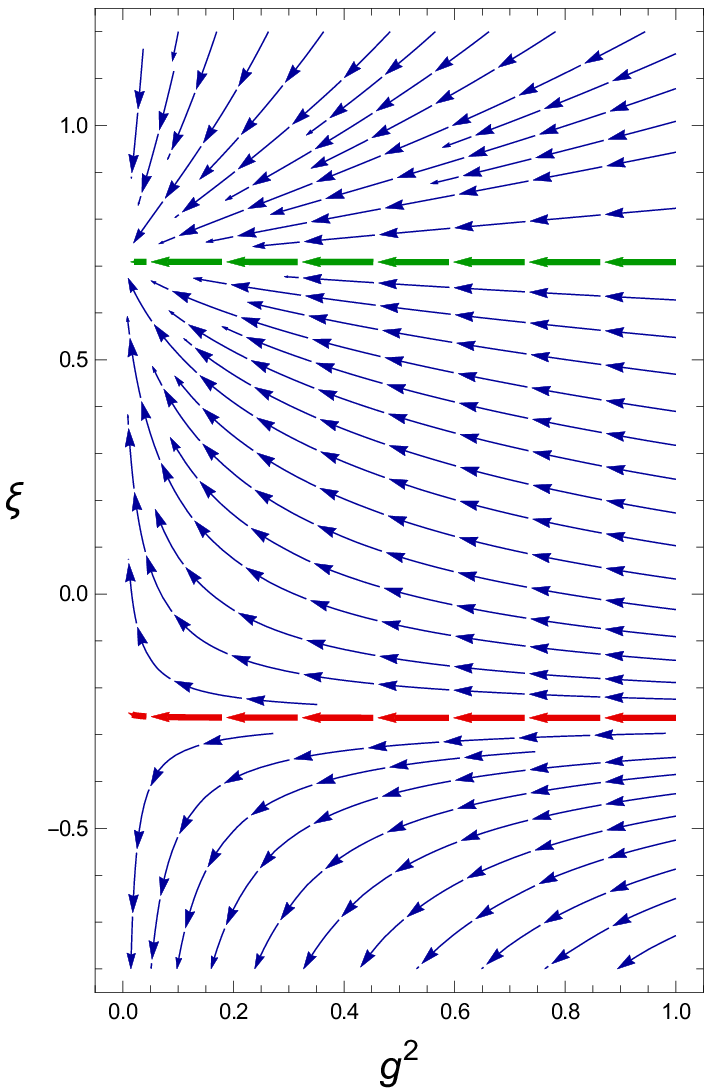}
 \caption{One-loop phase diagram of a fake model with negative $A$
   coefficient and $\Delta>0$, possessing real roots of the equation
   $\beta_\xi=0$, cf. \Eqref{eq:betaxi}, and featuring asymptotic
   freedom: trajectories satisfying physical low-energy boundary
   conditions $\lambda,g^2>0$ at some IR scale asymptotically approach
   the free Gau\ss{}ian fixed point towards the UV. 
   This is shown in terms of both the ordinary quartic coupling $\lambda$
   (left panel), and the rescaled version $\xi=\lambda/g^2$ (right panel).
   The two colored trajectories correspond to the two roots of
   \Eqref{eq:lambdapm}. They exemplify the auxiliary concept of
   \fgFPs, cf. main text.}
\label{fig:streamplot_nontrivial_oneloop}
\end{center}
\end{figure}

As emphasized above, the existence of this Landau-pole behavior of
perturbation theory is tied to the fact that $\Delta<0$ for the
present model. In order to illustrate the expected UV behavior for
asymptotically free theories, let us discuss a fake model with
positive $\Delta$ for the remainder of the section. For this, we
simply change the sign of $A$ by hand. As mentioned above, there is no
Landau pole in this case which also becomes obvious in the streamplot
in the left panel of Fig.~\ref{fig:streamplot_nontrivial_oneloop}.  As a consequence of
the negative sign of $A$ implying $\Delta>0$, the trajectories that
satisfy physical boundary conditions with positive $\lambda,g^2>0$ in
the IR are now asymptotically attracted by the Gau\ss{}ian fixed
point towards the UV, and no run-away to $+\infty$ occurs.
In the left panel of Fig.~\ref{fig:streamplot_nontrivial_oneloop}, we highlight two lines
that correspond to the trajectories with
\begin{equation}
\lambda_\pm=\frac{-B\pm\sqrt{\Delta}}{-2A}g^2, \label{eq:lambdapm}
\end{equation}
which can be obtained in the limit $g_\Lambda^2\to\infty$ or
$g_\Lambda^2\to 0$, respectively. Along these trajectories, the system
shows a peculiar behavior since $\lambda$ is proportional to $g^2$ at
all scales.  Eq.~\ref{eq:integratedflow_oneloop} shows that all the asymptotically free
trajectories of this model close to the Gau\ss{}ian fixed point have a
leading linear $g^2$-dependence with a $g_\Lambda^2$-independent
proportionality constant $(-B+\sqrt{\Delta})/({-2A})$; i.e., all
trajectories exhibit the same UV asymptotics for $g_\Lambda^2>0$. This
is in agreement with the observation that zeros of $\beta_\lambda$
need to be of order $g^2$, but it might be surprising that the
constant of proportionality does not agree with these zeros,
cf. \Eqref{eq:roots_oneloop}.  The mismatch has a deep meaning and is
of key relevance for what follows.

In a one-loop set-up, asymptotic freedom occurs with an asymptotic
scaling of $\lambda$ proportional to $g^2$.  The latter piece of
information can be encoded in the condition that the ratio
$\xi=\lambda/g^2$ is frozen in the far UV, namely that
$\beta_\xi=0$~\cite{Gross:1973ju}.  Since
\begin{equation}
\beta_{\xi}=g^2\left[A\xi^2 +B\xi + C \right], \label{eq:betaxi}
\end{equation}
we can understand the above-mentioned mismatch by realizing that
$\beta_\xi=0$ is the relevant condition that characterizes all the asymptotically free
trajectories, rather than simply $\beta_\lambda=0$ (which is
nevertheless satisfied at $g^2=0$).  Thus, one can translate the flow diagram of
the model in terms of $\xi$, as in the right panel of
Fig.~\ref{fig:streamplot_nontrivial_oneloop}. This makes the
properties of the asymptotically free trajectories more transparent.
We observe that the flow of $\xi$ has two fixed points corresponding
to $\xi_\pm=\lambda_\pm/g^2$ at finite $g^2$ with different properties.
For $\xi_+=\lambda_+/g^2$, the $\xi$-direction is UV attractive (RG
relevant), implying that we can build a one-parameter family of
trajectories hitting this fixed point.  For $\xi_-=\lambda_-/g^2$, the
$\xi$-direction is UV repulsive (RG irrelevant), such that only one
trajectory hits this fixed point. This trajectory is obtained by moving off
this fixed point along the marginally relevant direction $g^2$, without
switching on any component along the $\xi$-direction. This trajectory
is an example of a reduced number of physical parameters for the IR
physics (in the present fake model, however, it corresponds to
unphysical negative values of the coupling $\lambda$).

In conclusion, the roots of the finite-$g^2$ fixed-point equation
$\beta_\xi=0$ for $\xi$ classify the asymptotic (small $g^2$) behavior
of the possible asymptotically free trajectories. The latter
trajectories exist in the real space of couplings if and only if there
are real roots of $\beta_\xi=0$.  These roots are a first example for
the auxiliary concept of \textit{\fgFPs}, which is heavily used below: such
\fgFPs\ denote zeroes of the $\beta$ functions of the appropriately
$g^2$-rescaled scalar sector
even for finite values of the gauge coupling $g^2$. 
Incidentally, this fixed point condition for ratios of couplings,
has been used in the literature for a long time, starting with~\cite{Gross:1973ju},
under different names, such as \textit{eigenvalue conditions}~\cite{Chang:1974bv}~\cite{Callaway:1988ya}
or \textit{fixed-flows}~\cite{Giudice:2014tma}.
As a seemingly
trivial consequence of the shift from $\lambda$ to $\xi$, we have
$\beta_{\xi}=0$ for any $\xi$ for vanishing gauge coupling $g^2=0$. In
other words, the Gau\ss{}ian fixed point $g^2=0$, $\lambda=0$ in the
old parametrization, becomes the whole $g^2=0$ axis in the new
parametrization. This trivial statement has an immediate consequence
for the search for nonvanishing-$g^2$ fixed-points. Whereas the
standard parametrization demands for a $g^2$ trajectory hitting a
single point in the far UV, the new parametrization requires to
search for $g^2$ trajectories hitting a whole axis. In the fake model
and within the one-loop flows discussed above, trajectories exist that
asymptotically reach one of two possible points on the $\xi$-axis.

\section{Effective field-theory analysis in the deep Euclidean region}
\label{sec:EFT1}

The previous standard one-loop analysis of the nonabelian Higgs model
(with the correct positive sign for $A$) reveals the absence of
perturbative asymptotic freedom. This result clearly holds under the
standard assumptions for a perturbative analysis, being typically
extendible to any finite order in an expansion in terms of
$g^2$. Further implicit assumptions include, for instance, the
anticipated appropriateness of an analysis in the deep Euclidean
region, such that the running of mass terms and threshold effects can
be ignored.

Despite the seemingly negative answer from the perturbative analysis
of the real model, we keep the lesson from the fake model in mind that
asymptotically free trajectories -- if they exist -- can be
characterized by a peculiar kind of scaling, for which the scalar
self-interactions are governed not only by the renormalization scale
but also by the running gauge coupling. For the correspondingly
rescaled coupling $\xi=\lambda/g^2$, the Gau\ss{}ian fixed point
unfolds to a whole line which may or may not be reachable by
legitimate RG trajectories.

Let us now slightly change the viewpoint: considering the interaction
term $\lambda \phi^4$, we can also reinterpret the coupling rescaling
as a field rescaling
\begin{equation}
\lambda\phi^4 = \xi \left({\sqrt{g}}\ \phi \right)^4.
\label{eq:simplefieldrescaling}
\end{equation}
Now, if $\xi$ happens to approach the line $\xi=\text{const.}$ for
$g^2\to 0$ on an asymptotically free trajectory (as in the fake
model), fluctuations of the rescaled field variable $(\sqrt{g}\phi)$
will correspond to large amplitude fluctuations in $\phi$. This
motivates to go beyond the lowest-order interaction term $\sim\phi^4$
and more generally study the full running potential. In the spirit of
effective field-theory, we can span the full potential in terms of
operators of increasing mass dimension,
\begin{equation}
U(\phi)=\sum_{n=1}^{\Np} \frac{\lambda_n}{n! k^{2(n-2)}} \rho^n, \quad \rho=\phi^\dagger \phi,
\label{eq:efffieldpot}
\end{equation}
where the couplings $\lambda_n$ are dimensionless because of an appropriate
scaling with the RG scale $k$, and $\Np$ labels the highest order
included in this operator expansion. For simplicity, we ignore here
the necessary wave function renormalizations for the proper definition
of the renormalized couplings. They are included in the full
calculation and will be more carefully introduced in the next section.

In standard effective field theory, a potential as in
\Eqref{eq:efffieldpot} or further higher-dimensional operators are
used to parametrize the physics at a fixed (high-energy) scale
$k=\Lambda_{\text{eff}}$, and then fluctuations are integrated out to
describe the long-range physics at momenta $p\ll
\Lambda_{\text{eff}}$. In the present work, we instead use
\Eqref{eq:efffieldpot} with the running scale $k$ and study the flow
of the couplings towards higher and higher energies. 

In this spirit, each coupling $\lambda_n$ has its own RG flow given by
the corresponding $\beta_{\lambda_n}$ function, which can be derived
by standard effective-field theory techniques (see also next
section). In the massive scheme and within the approximations considered in this work,
the function $\beta_{\lambda_n}=\pat\lambda_n$ generically
depends on the couplings up to order $\lambda_{n+1}$; (precise definitions will be given in Sec.~\ref{sec:FRG}). Truncating the 
expansion in \Eqref{eq:efffieldpot} as well as the set of
$\beta_{\lambda_n}$ functions at a fixed polynomial order $n\leq\Np$,
leaves the next higher coupling $\lambda_{\Np+1}$ undetermined, even
though it enters the flow equation for $\lambda_{\Np}$. To close the
system of equations, one may approximate $\lambda_{\Np+1}=0$. This is
well justified in a perturbative region where the higher-order
operators are generated by the fluctuations involving the leading
operators, and hence $\lambda_{\Np+1}$ parametrizes subleading
higher-loop corrections.

While \Eqref{eq:efffieldpot} is a suitable expansion in the symmetric
regime, the potential can also be expanded about the vacuum
expectation value $v$ in the broken regime, 
\begin{equation}
U(\phi)=\sum_{n=2}^{\Np} \frac{\lambda_n}{n! k^{2(n-2)}} \left(\rho-\frac{v^2}{2}\right)^n, 
\label{eq:efffieldpotSSB}
\end{equation}
The mass-like parameters $\lambda_1$ of \Eqref{eq:efffieldpot}, or
$\lambda_2 v^2$ in \Eqref{eq:efffieldpotSSB} are assumed to be
negligible for a UV analysis in the deep Euclidean region.  In fact,
the validity of this assumption has been challenged in a series of
works \cite{Gies:2009hq,Gies:2009sv,Gies:2013pma}, where nontrivial RG
flows towards the UV have been constructed on the basis of
scale-dependent threshold phenomena. The present work is partly related
with these constructions, but more generally relates the new UV
trajectories to boundary conditions at large fields, allowing for
asymptotically free scaling solutions.

\subsection{First glance at scaling solutions}
\label{subsec:fg}

In order to understand how the higher-dimensional perturbatively
non-renormalizable operators in \Eqref{eq:efffieldpot} can support the
construction of asymptotically free scaling solutions, let us
generalize the concept of rescaling the scalar coupling or fields by
the gauge coupling, cf. \Eqref{eq:simplefieldrescaling}, to the full
potential.  By consistency, this entails consequent rescalings of
higher polynomial couplings, $\lambda_n=g^n\xi_n$.  Yet, $\xi_n$ might
still attain vanishing or diverging values towards the UV.

The presence of fixed-points (in the fake model) at $g^2=0$ for finite
values of $\xi\equiv\xi_2$ now serves as a motivation for a new search
strategy for asymptotically free trajectories.  As a hypothesis, let
us for the moment assume that the model beyond the perturbative realm
admits trajectories hitting any chosen point along the asymptotically
free $\xi_2$ axis. If this were the case, neither perturbation theory
nor truncations of the effective-field theory expansion at a fixed
$\Np$ setting $\lambda_{\Np+1}=0$ would be able to reveal this
feature: there simply is no free parameter in the remaining set of
$\beta$ functions that would allow us to choose the $\xi_2$ value at
which the flow arrives in the UV.

In order to test this hypothesis, we need an approximation that is
able to describe lines of fixed points, i.e. allowing for the presence
of a free parameter.  This can be achieved in the effective-field
theory context by treating the next coupling $\lambda_{\Np+1}$ outside
a given $\Np$ truncation as a free parameter instead of assuming that
it can be ignored.  From the point of view of a full functional
approach discussed in the next section, where we deal with the full
potential $U(\phi)$ and not with simple polynomials, this free
parameter can be interpreted as emerging from the boundary conditions
for the potential. 

Let us now demonstrate in a somewhat oversimplified setting, that such
a parameter freedom can be sufficient to find trajectories that reach
any desired value for $\xi_2$ in the far UV. For this, we consider the
one-loop flow within the effective-theory approach and minimally
include the coupling $\lambda_3=g^3 \xi_3$ as a free parameter,
corresponding to truncating at $\Np=2$. Staying within the deep
Euclidean region ($\lambda_1=0$), the $\beta$ function for
$\lambda\equiv\lambda_2$, or alternatively $\xi=\xi_2$ reads,
\begin{eqnarray}
\beta_{\lambda_2}&=&A\lambda_2^2 +B^\prime\lambda_2 g^2 + C g^4-D\lambda_3,
\label{eq:quasioneloop_lambda}\\
\text{or}\,\,\beta_{\xi_2}&=&g^2(A\xi_2^2 +B\xi_2  + C)-g D \xi_3.
\label{eq:quasioneloop_xi}
\end{eqnarray}
As $\lambda_3$ does not represent a power-counting renormalizable
coupling, the coefficient $D$ is non-universal, i.e., scheme and
regulator dependent. In our scheme detailed below and for SU(2), we
have $D=1/(4\pi^2)$. All other constants are the same as in
Sect.~\ref{sec:PT}.

\begin{figure}[!t]
\begin{center}
 \includegraphics[width=0.23\textwidth]{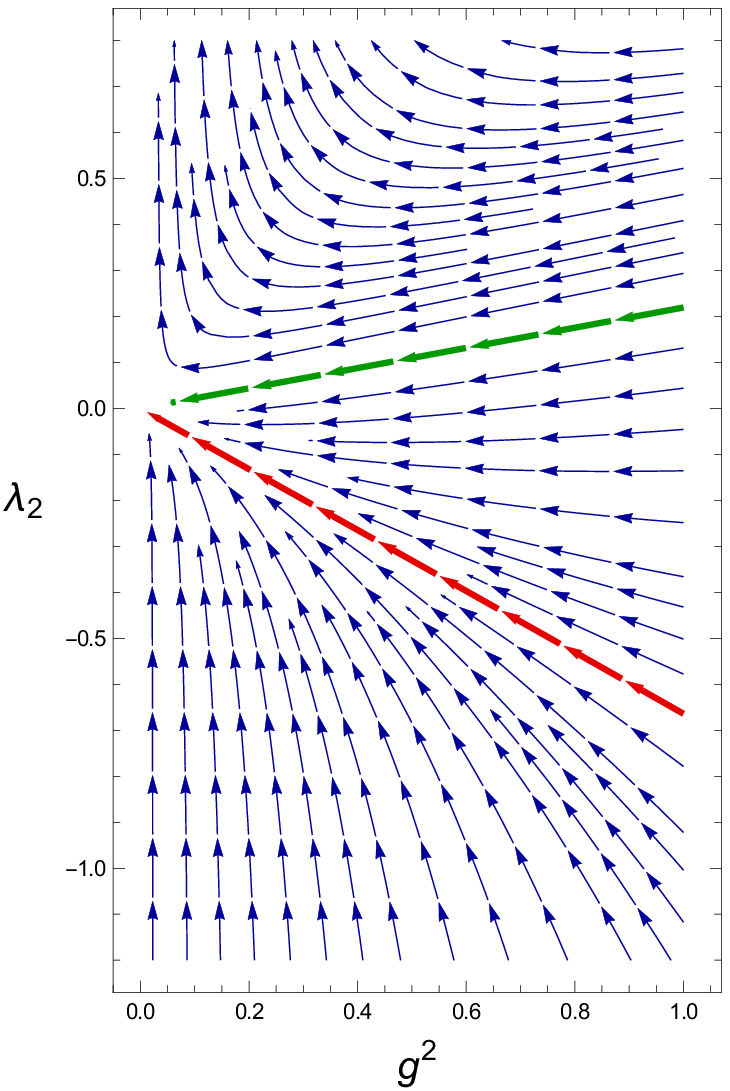}\ \ \
 \includegraphics[width=0.23\textwidth]{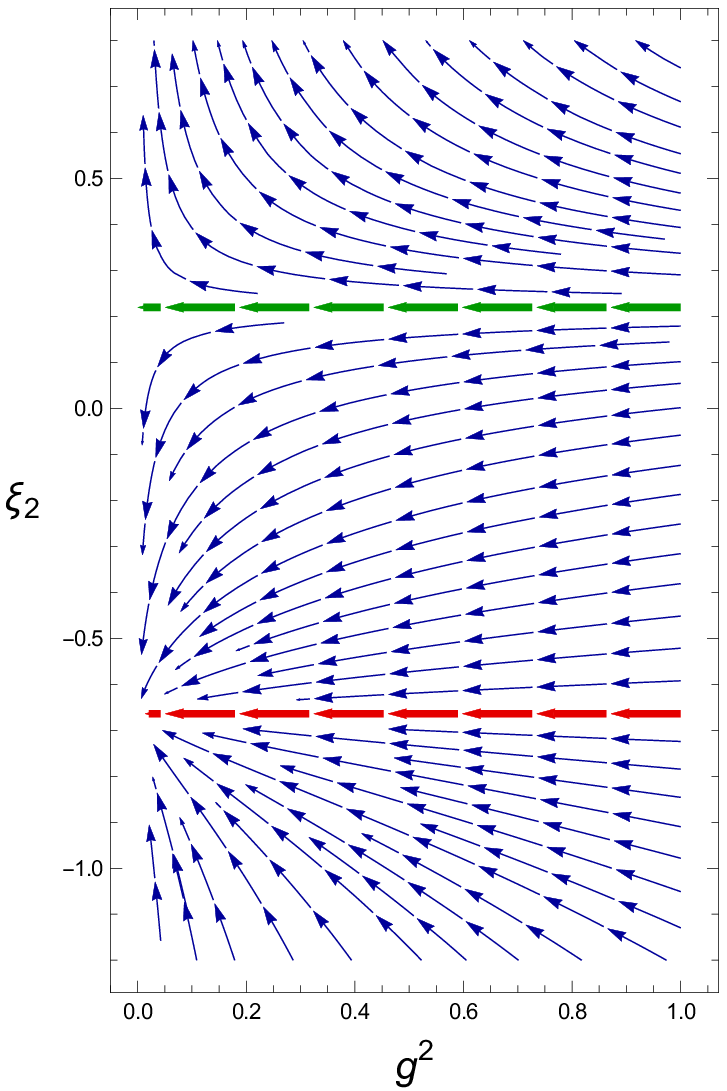}
 \caption{Phase diagram in the $(\lambda_2,g^2)$ plane (left panel)
   or $(\xi_2,g^2)$ plane (right panel) in an effective-field theory
   approximation with $N_\text{p}=2$ as in \Eqref{eq:efffieldpot} in
   the deep Euclidean limit, closing the system of coupled flows with a free
   parameter. We have chosen $\chi=\lambda_3/g^4=1$; the colors
   highlight the asymptotically free trajectories corresponding to the
   two roots of the \fgFPh\ equation $\beta_{\xi_2}=0$.}
\label{fig:streamplot_xi_quasioneloop}
\end{center}
\end{figure}

Having $\xi_3$ as a free parameter in \Eqref{eq:quasioneloop_xi} is
equivalent to a free coefficient $C$ in the pure one-loop case. As a
consequence, we can choose $\xi_3$ such that the roots are real.
Specifically, the inclusion of $\xi_3$ in \Eqref{eq:quasioneloop_xi}
amounts to replacing $C$ in \Eqref{eq:betaxi} with
$C^\prime=C-D\xi_3/g$.  This suggests to keep the ratio $\chi=\xi_3/g$ fixed,
such that $\xi_2$ approaches finite real roots in the $g^2\to0$ limit,
\begin{equation}\label{eq:rootsxi_quasioneloop}
\xi_{2\pm}=\frac{\lambda_{2\pm}}{g^2}=\frac{-B\pm\sqrt{\Delta^\prime}}{2A},
\end{equation}
representing \fgFPs, where
\begin{equation}
\Delta^\prime=B^2-4AC^\prime \ , \quad C^\prime=C-D\chi
\ , \quad \chi=\frac{\xi_3}{g}=\frac{\lambda_3}{g^4}.
\label{eq:Deltaprime}
\end{equation}
The integrated flow is then again
\begin{equation}\label{eq:integratedflow_quasioneloop}
\lambda_2(g^2)=-\frac{g^2}{2A}\left[B+\sqrt{\Delta^\prime}\tanh\!\left(\frac{\sqrt{\Delta^\prime}}{2 b_0}\ln\frac{g_\Lambda^2}{g^2}\right)\right],
\end{equation}
where $g_\Lambda$ is an integration constant. For $\xi_3$ chosen such
that $\Delta^\prime>0$, there is no Landau pole in $\lambda$. Hence,
we obtain asymptotically free trajectories, as can be seen from
Fig.~\ref{fig:streamplot_xi_quasioneloop}.
We observe that the fixed point for a given ratio $\chi$ with a positive
value of $\xi_2$ for $g^2\to 0$ has a marginal-irrelevant direction,
implying that the long-range physics depends on one physical parameter
less then predicted by perturbative power counting. This dynamical
reduction of parameters is, however, compensated by the necessary
choice of a value for $\chi$. Nevertheless, the meaning of these
parameters has slightly changed: whereas in the perturbative framework we fix
parameters \textit{within a theory}, different values of $\chi$ correspond to
different boundary conditions and thus rather to \textit{different theories}.

At this simple stage, the choice of fixing $\chi$ to a constant looks
rather arbitrary. In particular, it implies that we are fixing
specific $g^2$ dependencies for the couplings, e.g., $\lambda_3\sim
g^4$ or $\xi_3\sim g$. A gauge coupling dependence is only natural as
the gauge sector will inevitably drive the running of higher-order
operators. Still, at this simple level of approximation, it seems that
we would have to guess the correct scaling of couplings in the full
system or we are left with an ambiguity of possible different
choices. In the next section, we demonstrate that this ambiguity is
removed in the full system leading to a remaining dependence on the
boundary conditions for the theory.

\begin{figure}[!t]
\begin{center}
 \includegraphics[width=0.35\textwidth]{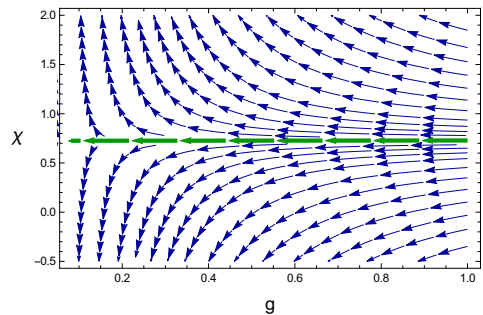}
 \caption{Phase diagram in the $(\chi,g^2)$ plane in an
   effective-field theory approximation with $N_\text{p}=2$ as in
   \Eqref{eq:efffieldpot} in the deep Euclidean limit, closing the
   system of coupled flows with a free parameter. Here, the parameter
   is chosen such that $\xi_2$ stays fixed at
   $\xi_2=\lambda_2/g^2=0.1$ The green asymptotically free trajectory
   corresponds to the root of the \fgFPh\ equation
   $\beta_{\chi}=0$ also at finite $g^2$.}
\label{fig:streamplot_chi_quasioneloop}
\end{center}
\end{figure}

To illustrate the effect of different choices in the present simple
setting, let us insist on a flow that keeps $\xi_2$ constant at the
expense of choosing $\chi$ (or $\xi_3$) accordingly. Insisting on
$0=\beta_{\xi_2}$ as a function of $g^2$ and $\chi$, leads us to a
flow of $\chi$ given by
\begin{equation}
\beta_\chi=-\frac{\partial \beta_{\xi_2}}{\partial g^2}
\left(\frac{\partial \beta_{\xi_2}}{\partial \chi}\right)^{-1} \beta_{g^2}\ .
\label{eq:betachi}
\end{equation}
To lowest order in the gauge coupling, we obtain
\begin{equation}
\beta_\chi=-g^2\frac{b_0}{D}\left(A\xi_2^2+B\xi_2+C-D\chi\right)\ .
\label{eq:betachi2}
\end{equation}
Again, we observe the existence of \fgFPs\ $\beta_\chi=0$ at
finite $g^2$, satisfying the same relations
\eqref{eq:rootsxi_quasioneloop} and \eqref{eq:Deltaprime}, including
the implicit mapping between a choice for $\xi_2$ and the
corresponding value of $\chi$. The corresponding flow at a fixed $\xi_2>0$
is shown in Fig.~\ref{fig:streamplot_chi_quasioneloop}. We conclude
that the asymptotic properties of these asymptotically free
trajectories are independent of the precise choice of the free
parameter. For both choices of finite asymptotic ratios
$\chi=\lambda_3/g^4$ or $\xi_2=\lambda_2/g^2$ we observe asymptotic
freedom and a UV fixed-point behavior with the same type of
marginal-irrelevant perturbation in the mutual dynamical coupling
$\xi_2$ or $\chi$, respectively.

\subsection{Generalized scaling solutions}
\label{subsec:gss}

The preceding simple analysis revealed that an appropriate $g^2$
scaling of the higher-order operators is required in order to build
asymptotically free trajectories. At a first glance, it seems that the
scaling $\xi_3=\lambda_3/g^3 \sim g$ for small $g$ is essential for
the desired result. For instance, if we keep $\xi_3$ constant, the
last term in \Eqref{eq:quasioneloop_xi} makes $\xi_2$ diverge towards
the UV.  Still, for an answer about the (non-)existence of
asymptotically free trajectories in that case, we have to inspect the
UV flow of $\lambda_2$.

In order to understand the set of possible consistent scalings that
feature asymptotically free trajectories, let us try to generalize the
previous analysis. For this, we start from the $\beta_{\lambda_2}$
function \eqref{eq:quasioneloop_lambda} in the effective-theory
setting with $\Np=2$ as before, keeping $\lambda_3$ as a possibly
scale-dependent free parameter. As before, we assume $g^2$ to be
finite, and look for \fgFPs\ defined by
$\beta_{\lambda_2}=0$. For $\lambda_3=\chi g^4$, we precisely discover
the one-parameter family of the preceding subsection parametrized by a
fixed value of $\chi$.

Let us now be more general and set 
\begin{equation}
\lambda_3=\chi g^{2\gamma},
\label{eq:introduceP}
\end{equation}
(in this language, the value $\gamma=2$ corresponds to the previous
analysis). The \fgFPs\ of $\beta_{\lambda_2}$ are
then given by the analogue of \Eqref{eq:rootsxi_quasioneloop} with the
replacement 
\begin{equation}
C'=C-D\chi g^{2(\gamma-2)}.
\label{eq:CprimeP}
\end{equation}
For $\gamma<2$, the $D$ term dominates in the small-coupling limit,
yielding real roots,
\begin{equation}
\lambda_2=\pm g^{\gamma}\sqrt{\frac{D\chi}{A}}+O(g^2)\ .
\label{eq:lambda2P}
\end{equation}
For fixed $\chi>0$, we hence observe a whole set of further
asymptotically free trajectories parametrized by the power
$\gamma$. At this point, it seems that $\gamma$ has to satisfy
$\gamma\leq2$, since the $D$ term vanishes for small gauge coupling
for $\gamma>2$ and the corresponding roots in $\lambda_2$ remain
complex. In the following, we use the terminology \textit{$P$-scaling
  solutions} for the asymptotically free trajectories where
$\lambda_2$ vanishes proportional to $g^{4P}$ in the UV (in the
present case, we have $P=\gamma/4$ for $\gamma\leq2$; as shown below,
the $\gamma>2$ counterparts, in fact, do exist, but only become
visible, once we drop the artificial restriction to the deep Euclidean
regime).
That $\lambda_2$ or $\lambda_3$ could
scale like some arbitrary
non-integer power of $g^2$ might look 
suspicious, and might point to
possible pathological properties of these trajectories. Nevertheless,
our analysis, as discussed more in detail in what follows,
is not
able to rule out this possibility. Moreover,
such kind of asymptotically free trajectories have already been discovered at one loop
in supersymmetric models~\cite{Browne:1975js}\cite{Kaplunovsky:1982mf}, 
see also~\cite{Oehme:1984yy} for a lucid
explanation close to the present treatment.

\begin{figure}[!t]
\begin{center}
 \includegraphics[width=0.23\textwidth]{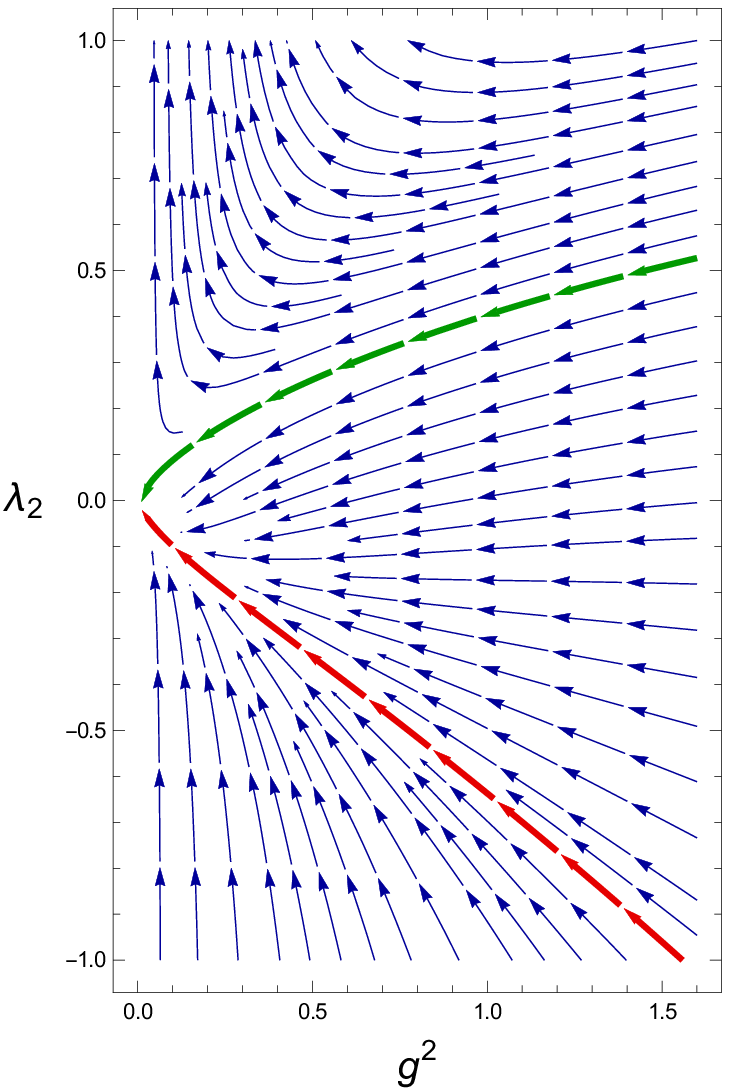}\ \ \
 \includegraphics[width=0.23\textwidth]{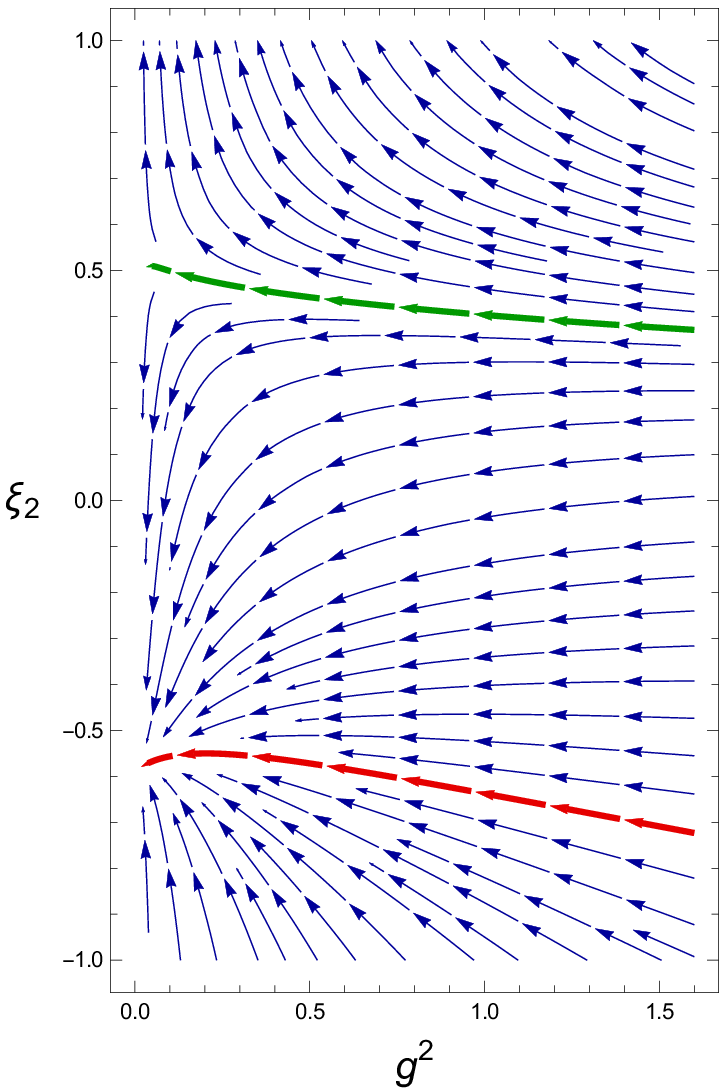}
 \caption{Phase diagram in the $(\lambda_2,g^2)$ plane (left panel)
   or $(\xi_2,g^2)$ plane (right panel) in an effective-field theory
   approximation with $N_\text{p}=2$ exhibiting $P=3/8$ scaling
   solutions for the choice $\chi=\lambda_3/g^3=1$ in the deep
   Euclidean limit. 
   The asymptotically free trajectories, which are
   highlighted in colors, again correspond to the
   two roots of the \fgFPh\ equation $\beta_{\xi_2}=0$, base on
   the generalized rescaling \eqref{eq:xinP}, i.e., $\xi_2=\lambda_2/g^{3/2}$.
   }
\label{fig:P3o8_streamplot_nontrivial_quasioneloop}
\end{center}
\end{figure}

\begin{figure}[!t]
\begin{center}
 \includegraphics[width=0.4\textwidth]{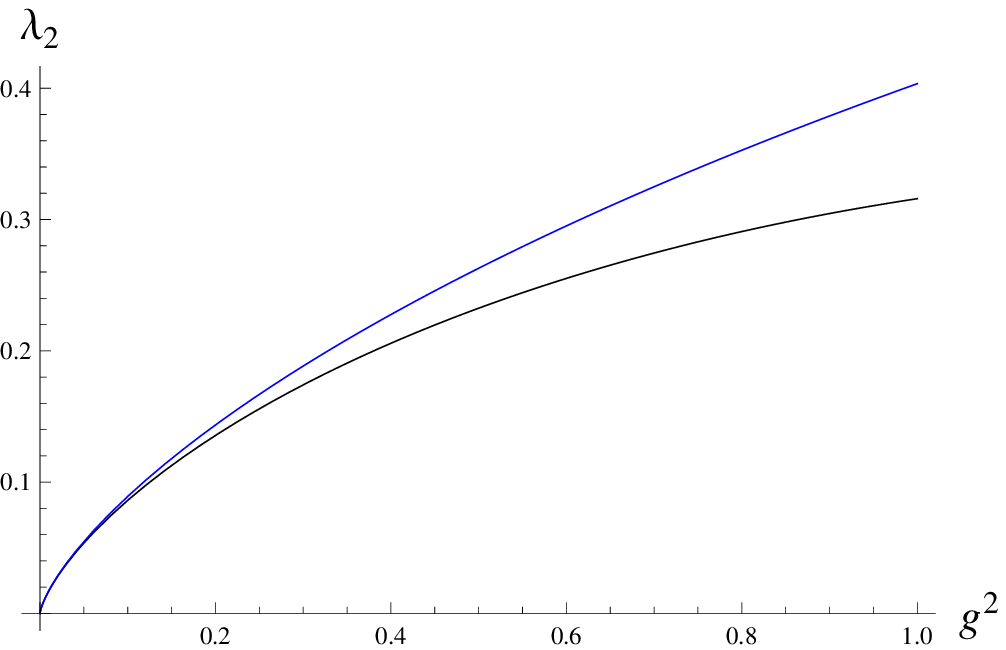}
 \includegraphics[width=0.4\textwidth]{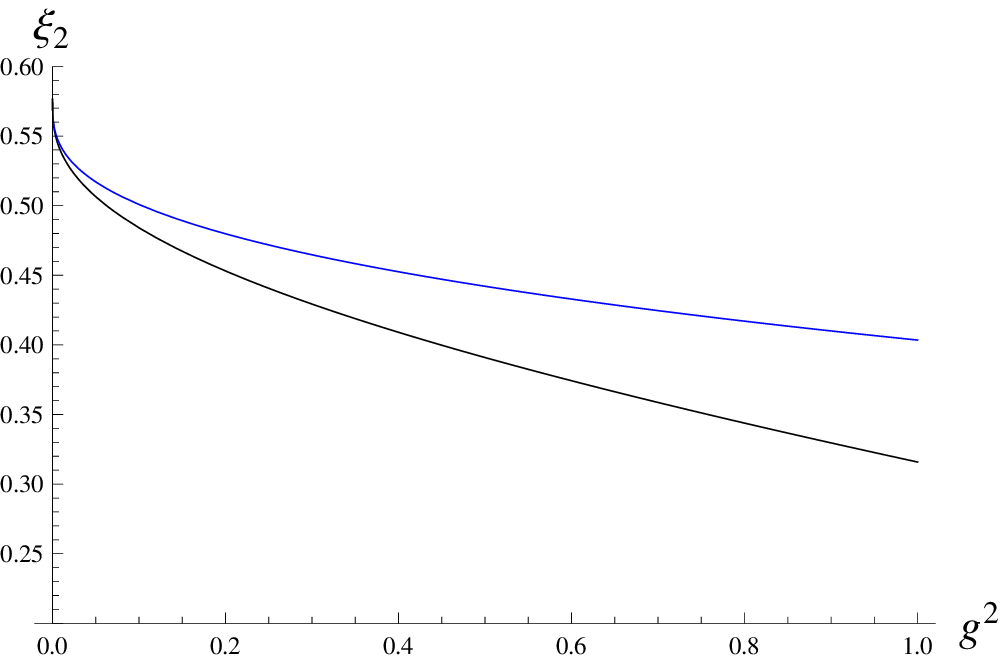}
 \caption{Asymptotically free trajectory in the effective-field theory
   approximation quasi-one-loop truncation with $N_\text{p}=2$ as in
   Eqs.~\eqref{eq:quasioneloop_lambda} for $P=3/8$ at fixed
   $\chi=\lambda_3/g^3=1$ in terms of $\lambda_2$ (upper panel) and
   then in terms of the rescaled coupling $\xi_2=\lambda_2/g^{3/2}$
   (lower panel) as a function of $g^2$.  The blue (upper) curve is
   the numerical result, while the black (lower) curve is the
   approximation given by the \fgFP, i.e.~the positive root of the fixed-point equation
   $\beta_{\xi_2}=0$.  }
\label{fig:P3o8_AF_trajectory}
\end{center}
\end{figure}

Let us look at the case $P<1/2$ more specifically using the example
mentioned above where $\lambda_3 / g^3$ approaches a constant. In the
new notation of Eqs.~\eqref{eq:introduceP} and \eqref{eq:lambda2P},
this corresponds to the case $P=3/8$ and $\chi=\lambda_3/g^3$. From
\Eqref{eq:lambda2P}, we expect the occurrence of asymptotically free
trajectories where $\lambda_2$ vanishes like $\lambda_2\sim g^{3/2}$
in the UV. These trajectories can indeed be localized in the
$(\lambda_2,g^2)$ phase diagram in
Fig.~\ref{fig:P3o8_streamplot_nontrivial_quasioneloop} (upper
panel). From the perspective that the rescaling with the gauge
coupling corresponds to a rescaling of the fields analogously to
\Eqref{eq:simplefieldrescaling}, $\phi\to g^P\phi$, it is useful to
also generalize the definition of the the gauge-rescaled couplings:
\begin{equation}
\xi_n=g^{-2Pn} \lambda_n.
\label{eq:xinP}
\end{equation}
With this generalized definition, the asymptotically free trajectories
become again manifest by ending on the $\xi_2$ axis in the
$(\xi_2,g^2)$ plane, with $\xi_2=\lambda_2/g^{3/2}$ for $P=3/8$, see
Fig.~\ref{fig:P3o8_streamplot_nontrivial_quasioneloop} (lower panel).
Again, there is a trajectory reaching a positive $\xi_2$ fixed point
with a marginally-irrelevant direction. The new feature for the
present case is that the critical trajectory emanating from this fixed
point is no longer $g^2$ independent. In order to verify whether a
given point in the coupling plane in the IR is on the asymptotically
safe trajectory, we have to integrate the flow towards the UV (with
sufficient numerical precision). The explicit trajectories (blue
curves) in the $(\lambda_2,g^2)$ plane (upper panel) or $(\xi_2,g^2)$
plane (lower panel) are shown in Fig.~\ref{fig:P3o8_AF_trajectory}. We
also plot the positive root of the equation $\beta_{\xi_2}=0$,
i.e.~the \fgFP, for any given value of $g^2$ (black curves) which
represents a reasonable approximation for small values of the gauge
coupling.  Though the \fgFPs\ (determined as a function of $g^2$) do
not give the correct physical RG flow, the results of
Fig.~\ref{fig:P3o8_AF_trajectory} demonstrate that the \fgFPs\ can be
used to track or approximate scaling solutions of the full system in
the asymptotically free regime $g^2\to0$.

\begin{figure}[!t]
\begin{center}
 \includegraphics[width=0.4\textwidth]{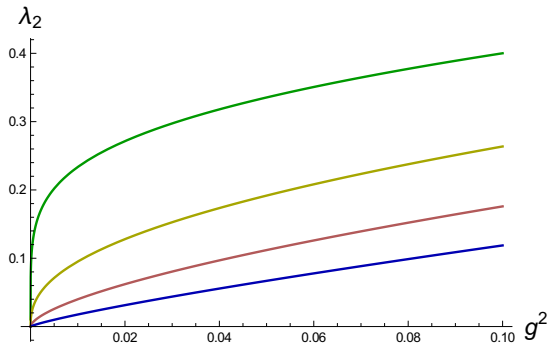}
 \caption{\FgFPs\ approximating asymptotically free trajectories as
 	given by~\Eqref{eq:root_xi_quasioneloop}
 	in the  effective-field theory approximation with $\Np=2$
	at fixed $\chi=\lambda_3/g^{8P}=1$, 
	for several values of $P\in \{0.1, 0.2, 0.3, 0.4\}$ 
from green (upper curve) to blue (lower curve).
}
\label{fig:smallP_AF_trajectory}
\end{center}
\end{figure}

The results are similar for any other value of the scaling power $P\in
(0,1/2)$. If we consider the effective-field-theory flow equations at
fixed $\chi=\lambda_3/g^{8P}$, we can build asymptotically free
trajectories such that the rescaled coupling $\xi_2=\lambda_2/g^{4P}$
attains a finite positive value in the $g^2\to 0$ limit. Any limiting
value of $\xi_2$ can be reached by a suitable choice of $\chi$.  For
small $g^2$ the corresponding trajectory can be approximated by the
\fgFP\ given by the positive root of the equation $\beta_{\xi_2}=0$.
Explicitly,
\begin{equation}
\beta_{\xi_2} = A g^{4P} \xi_2^2 +B g^2 \xi_2  + C g^{4-4P} - D\chi g^{4P},
\label{eq:mussnochschnellrein}
\end{equation}
with the generalized definition,
\begin{equation*}
B=B'+2Pb_0.
\end{equation*}
The resulting \fgFPh\ representing an approximation to the flow trajectory
for small $g^2$ is then
\begin{equation}
\xi_2(g)=-\frac{B}{2A}g^{2(1-2P)}+\frac{1}{2A}\sqrt{\Delta g^{4(1-2P)}+4AD\chi}.
\label{eq:root_xi_quasioneloop}
\end{equation}
Some of these are plotted in terms of $\lambda_2=\xi_2 g^{4P}$ for
different values of $P$ in Fig.~\ref{fig:smallP_AF_trajectory}.

A natural question is, of course, whether the present simplest
approximation of the effective-field theory RG flows truncating at
$\Np=2$ is legitimate and capable of describing the full system
appropriately. In fact, it is straightforward to generalize the
argument to any higher $\Np>2$ by self-consistently solving the
coupled system of flows for $\lambda_{2, \dots, \Np}$ and choosing an
appropriate $g^2$-dependent scaling for the highest order
$\lambda_{\Np+1}$ occurring in the flows. The result is a polynomial
approximation to a full scalar interaction potential which approaches
asymptotic flatness on a suitable trajectory in the limit $g^2\to0$.

Since all couplings $\lambda_n$ scale with certain powers of $g$ to
zero, one may actually worry about the influence of $\lambda_1$, which
according to \Eqref{eq:xinP} scales with the least power of $g$. So
far, we have ignored possible contributions by assuming that it
suffices to stay within the deep Euclidean region. A more careful
discussion is the subject of the following section.

\section{Effective field-theory analysis including thresholds}
\label{sec:EFT2}

Let us now give up the artificial restriction to stay within the deep
Euclidean region. In standard analyses, this region where all momenta
and RG scales are assumed to be larger than any mass scale is used to
define RG functions such as the $\beta$ functions. Together with the
use of a mass-independent RG scheme, this removes any mass-scale
dependence from the RG functions. As long as these mass-scales do not
run fast or grow large in the UV, the analyses in the deep Euclidean
region suffices completely to study the UV properties of a theory also
in the broken phase \cite{Lee:1974gua}. In fact, some of our scaling
solutions turn out to violate the implicit assumptions underlying the
deep Euclidean analysis. Hence, we now include mass scales that can
induce threshold behavior explicitly in our simplified effective
field-theory analysis in the following. 

The deviations from the deep Euclidean behavior show up in the
behavior of the scalar expectation value $v$. Therefore, we
concentrate on the expansion \eqref{eq:efffieldpotSSB} which to lowest
order as required for the $\Np=2$ approximation reads
\begin{equation}
U(\phi)=\frac{\lambda_2}{2} \left(\rho-\frac{v^2}{2}\right)^2+ 
\frac{\lambda_3}{6 k^2} \left(\rho-\frac{v^2}{2}\right)^3+\dots
\label{eq:Uofphiexp}
\end{equation}
In addition to the RG flow of $\lambda_2$, we also consider the flow
of $v^2$ or a suitably gauge-rescaled version thereof. Apart from
possible wave function renormalizations, the gauge/field rescaling suggests
to consider the dimensionless variable
\begin{equation}
x_0=g^{2P} \frac{v^2}{2k^2}\equiv g^{2P} \kappa,
\label{eq:defx0}
\end{equation}
with $\kappa$ denoting the dimensionless expectation value without
gauge rescaling.  If $v^2$ or $\kappa$ is nonzero, the gauge and
scalar propagators acquire mass terms which are also accounted for in
the following. Since $v^2$ does not correspond to a marginal operator,
its flow is not universal, but scheme and regularization
dependent. Throughout this work, we use a natural functional RG scheme,
the details of which are given below in Sect.~\ref{sec:FRG}. Here we
continue with the simplified effective field-theory-type analysis in
the spirit of the preceding section.

\subsection{$(P\!=\!1/2)$-scaling solutions}
\label{subsec:P=1/2}

Let us start by analyzing the role of the flow of the scalar
expectation value for the example $P=1/2$, where the analysis in the
deep Euclidean region revealed a scaling solution with $\lambda_2\sim
g^2$ for a fixed $\chi=\xi_3/g$. We are specifically interested in the
role of the rescaled expectation value $x_0= g \kappa$. To one-loop
order, we obtain the following flow equations to lowest order in the
gauge coupling
\begin{eqnarray}
\beta_{x_0}&=&-2x_0+ g \left(\frac{3}{16\pi^2} +\frac{9}{64\pi^2\xi_2}\right)\!\!+O(g^2)\nonumber\\
&=&g \left(\frac{3}{16\pi^2}-2 \kappa +\frac{9}{64\pi^2\xi_2}\right)\!\!+O(g^2)\label{eq:x0P05}\\
\beta_{\xi_2}\!&=&g^2\left(\frac{9}{64\pi^2}+\frac{\xi_2}{3\pi^2}+\frac{3\xi_2^2}{4\pi^2}-\frac{\chi}{16\pi^2}+\frac{9\chi}{64\pi^2\xi_2}\right)\!\!+\!O(g^4)\nonumber\\
&&\label{eq:xi2P05}
\end{eqnarray}
The $\beta_{\xi_2}$ equation exactly corresponds to
\Eqref{eq:betachi2}, except for the last term, which arises from the
fact that the potential is expanded about the running expectation value
$\kappa$. (Here and in the following, the $\beta$ functions can be
obtained by straightforward expansion of the full RG flow of the
potential given below in \Eqref{floweq:potential}, taking the gauge
rescalings appropriately into account.)

\begin{figure}[!t]
\begin{center}
 \includegraphics[width=0.4\textwidth]{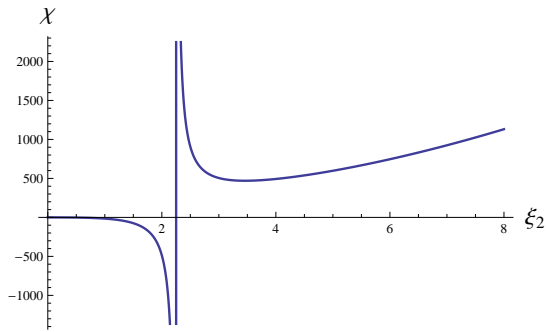}
 \caption{ The zeroes of $\beta_{\xi_2}$ in terms of $\chi$ as a
   function of $\xi_2$ near $g^2\to 0$ for the $P=1/2$ scaling
   solution in the effective-field-theory analysis with $N_p=2$.}
\label{fig:P1o2_FPchi_in_xi2}
\end{center}
\end{figure}

We observe again that Eqs.~\eqref{eq:x0P05} and \eqref{eq:xi2P05}
exhibit \fgFPs\ at finite values of $g^2$. For instance those of
$\xi_2$ can be read off by plotting the required value of $\chi$ for a
given value of $\xi_2$ in order to obtain $\beta_{\xi_2}=0$, see
Fig.~\ref{fig:P1o2_FPchi_in_xi2} .
As a new feature, we observe that also negative values of $\chi$ can
lead to admissible values of $\xi_2$. Naively, a negative
$\chi=\lambda_3/g^4$ seems to indicate that the potential may become
unstable towards large field amplitudes. However, the sign of the
highest term in a truncated expansion does not necessarily capture the
global stability properties. All global solutions given below are
fully stable. The important observation at this point is that an
asymptotically free trajectory appears
to exist for any value of $\xi_2$ which
can be constructed with a suitable choice of fixed $\chi$.

\begin{figure}[!t]
\begin{center}
 \includegraphics[width=0.4\textwidth]{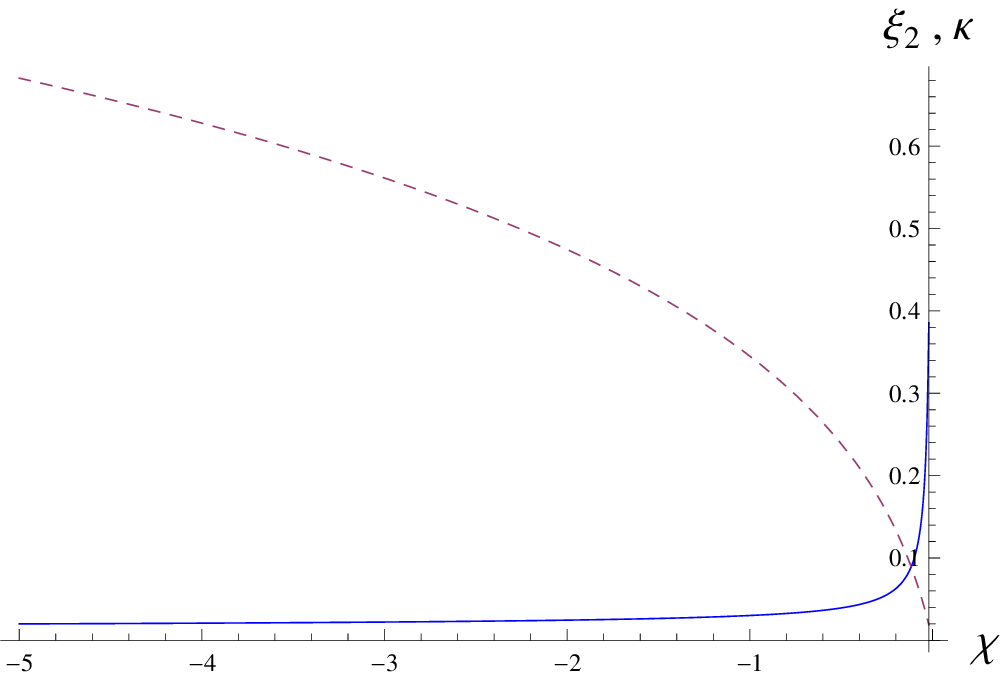}
 \includegraphics[width=0.4\textwidth]{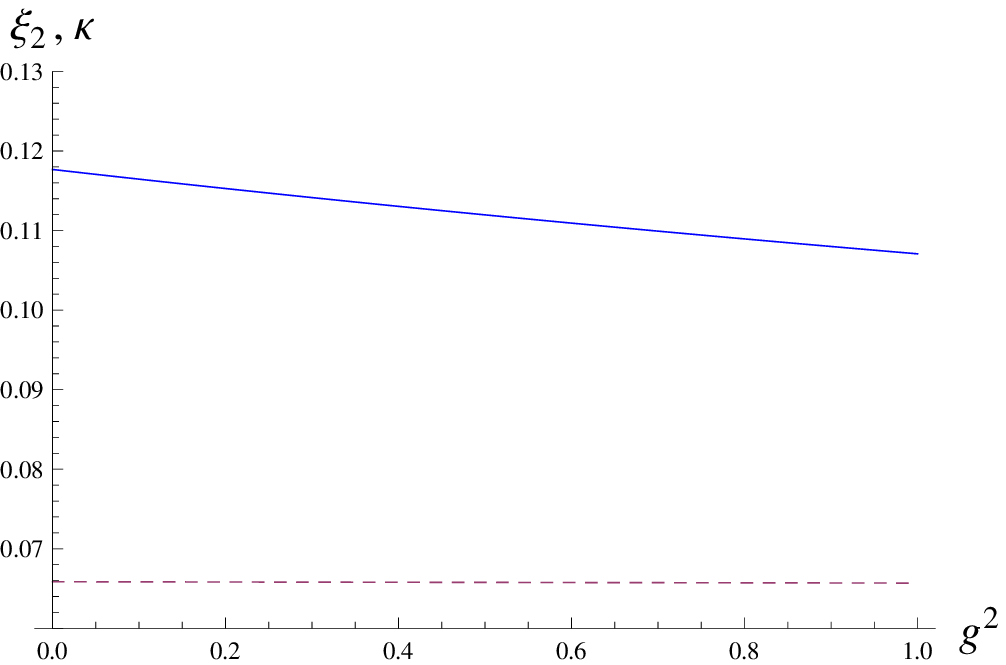}
 \caption{ Roots of the fixed-point equations $\beta_{x_0}=0$ and
   $\beta_{\xi_2}=0$ in terms of the dimensionless field expectation
   value $\kappa$ (solid) and the gauge-rescaled coupling $\xi_2$
   (dashed) as functions of $\chi$ and $g^2$ for the $P=1/2$ scaling
   solution in the effective-theory approximation with $N_p=2$,
   with full $g^2$-dependence.  As 
   example values, the upper panel uses $g^2=10^{-6}$, and the lower
   panel $\chi=-0.08$.}
\label{fig:P1o2_FP}
\end{center}
\end{figure}

Once, the \fgFPh\ value for $\xi_2$ has been identified,
\Eqref{eq:x0P05} yields the corresponding \fgFPh\ value for
$\kappa$, which is approached in the UV in the limit $g^2\to 0$. These
values are shown in Fig.~\ref{fig:P1o2_FP} as a function of
$\chi$ for fixed $g^2=10^{-6}$ (upper panel), or as a function of
$g^2$ for $\chi=-0.08$ (lower panel). In particular, this lower panel
shows that $\kappa$ does not vanish in the deep UV, but approaches a
constant. This implies that the dimensionful expectation value of the
field increases with the scale towards the UV, $v^2=2\kappa k^2\sim
k^2$ for $g^2\to 0$. In other words, the RG flow is never in the deep
Euclidean region along this asymptotically free trajectory.

\begin{figure}[!t]
\begin{center}
 \includegraphics[width=0.23\textwidth]{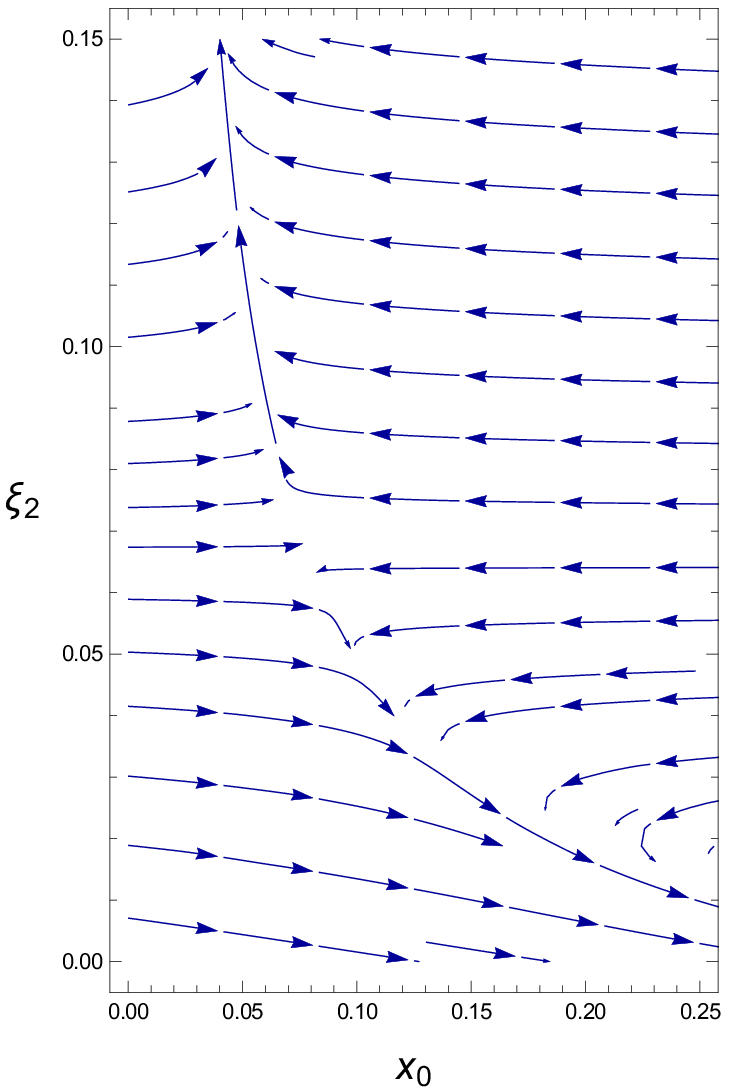}\ \ \
 \includegraphics[width=0.23\textwidth]{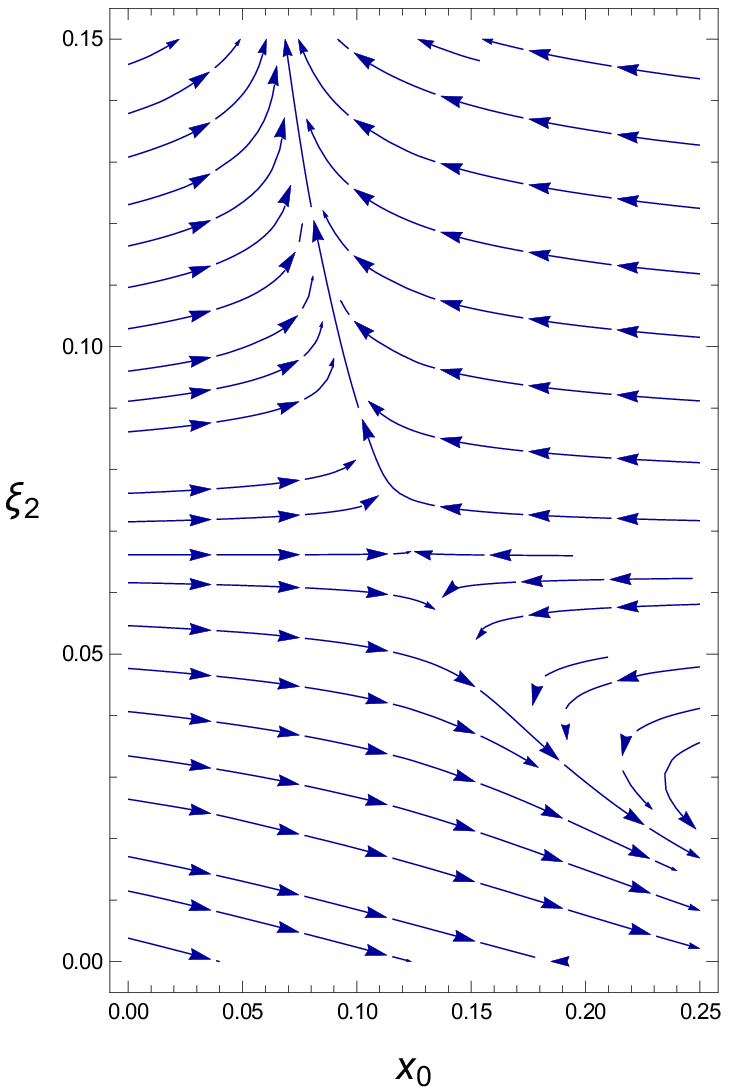}
 \caption{
RG flow in the $N_p=2$ effective-field-theory approximation for $P=1/2$, projected on planes at
fixed value of $g^2$ and $\chi=\xi_3/g$.
Right panel: $\chi=-0.08$ and $g^2=1.5$. Left panel: $\chi=-0.08$ and $g^2=0.5$.
}
\label{fig:streamplot_xi2_x0}
\end{center}
\end{figure}

As the inclusion of $\kappa$ goes along with an operator $\sim
\phi^2$, we have now included a power-counting relevant direction in
the flow. It is straightforward to verify that this also holds for our
approach to the asymptotically free fixed point. In
Fig.~\ref{fig:streamplot_xi2_x0}, we show the flow in the $(x_0,
\xi_2)$ plane at fixed values of $g^2$. The single green (critical)
line in Fig.~\ref{fig:streamplot_xi_quasioneloop} has become a
two-dimensional surface (a line in each $g^2$ slice of
Fig.~\ref{fig:streamplot_xi2_x0}), i.e. a one-parameter family of
trajectories.  In other words, the UV critical surface of each UV fixed point
becomes two-dimensional, and any point on this trajectory lies on top
of an asymptotically free trajectory.

This is consistent with power-counting Gau\ss ian critical exponents,
which suggest that the UV critical surface can be parametrized by
$g^2$ and $x_0$.  This can be verified by computing the critical
exponents (scaling dimensions) within the present approximation of the
beta functions from a linearization near the Gau\ss{}ian fixed
point. Collecting the couplings in a vector with components $g_i$, we consider the stability matrix
\begin{equation}
B_{ij} = \frac{\partial \beta_{g_i}}{\partial g_j}\Big|_{g_i=g_i^\ast},
\label{eq:dedstabmat}
\end{equation}
at the fixed point $g_i^\ast$. The critical exponents $\theta_I$
correspond to minus the eigenvalues of this matrix.  This
linearization has to be done with care by using the coupling $g$ instead
of $g^2$, otherwise we would run into an artificial branch-cut
singularity at vanishing $g^2$. Furthermore, we use $x_0$ and $\xi_2$
as couplings, and parametrize the set of trajectories with the UV
fixed-point value for $\xi_2$ for $g\to0$. The standard diagonalization of the
stability matrix then leads to the expected Gau\ss ian set of
exponents and perturbations, that is, to the following eigenvalues and
eigenvectors of the linearized flow
\begin{eqnarray}
\theta_{0}&=&2\ ,\quad (\delta x_0, \delta\xi_2, \delta g)=(1,0,0),\nonumber\\
\theta_{1}&=&0\ ,\quad (\delta x_0, \delta\xi_2, \delta g)=(0,1,0),\nonumber\\
\theta_{2}&=&0\ ,\quad (\delta x_0, \delta\xi_2, \delta g)=\left(1,0,\frac{128\pi^2\xi_2}{9+12\xi_2}\right).
\label{eq:evP05}
\end{eqnarray}
The first line characterizes the relevant direction corresponding to a
scalar mass term with $\theta_2=2$ in agreement with the
power-counting dimension.  The second line refers to the quartic
coupling, which is marginal at the linear level as usual;
as discussed below,
nonlinearities of the flow make it, in fact, marginally irrelevant.
The second vanishing exponent in the third line corresponds to the marginally relevant
direction due to the $\beta_{g^2}$ function of the gauge coupling.  Notice that
in presence of scalar degrees of freedom this direction necessarily
mixes the gauge and the scalar sector, since gauge loops induce scalar
self-interactions. The UV value for $\xi_2$ parametrizes this set of marginally relevant directions corresponding 
to asymptotically free trajectories ending at $\xi_2$ for $g\to0$.  The fact that the UV critical surface is spanned
by a relevant coupling, which for $ k \to+\infty$ decreases as $\delta x_0\propto k^{-\theta_0}$, and a marginally
relevant coupling, which shows a log decrease $g^2\propto 1/\ln k$,
entails that all the asymptotically free trajectories merge into a single trajectory
for very large RG time.  This single trajectory is characterized by a
vanishing relevant component and, as a consequence, it approaches the
Gau\ss{}ian fixed point with a vanishing critical exponent $\theta_{2}$.

\begin{figure}[!t]
\begin{center}
 \includegraphics[width=0.35\textwidth]{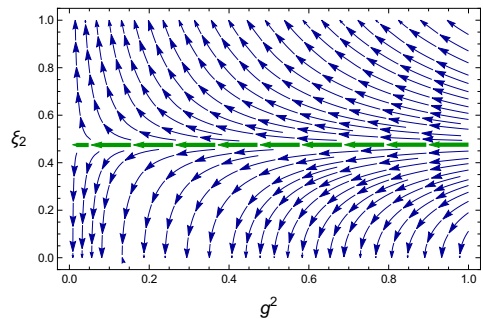}
 \caption{ Projection of the RG flow in the $N_p=2$
   effective-field-theory approximation for $P=1/2$ onto the surface
   with vanishing relevant component for fixed $\chi=\xi_3/g=-2$.
   In green we highlight the purely marginally-relevant asymptotically free
   trajectory. The flow is singular at $\xi_2=0$ because of the
   nontrivial running of the field expectation value: $x_0\neq 0$.  }
\label{fig:P1o2_streamplot_xi_reduced}
\end{center}
\end{figure}

The trajectory asymptotically approaching a given value of $\xi_2$ can
be constructed in the same way as before this time including the
parameter $\sim x_0$. Concentrating on the case, where $x_0$ does not
flow rapidly, we consider $x_0$ values, where $\beta_{x_0}\simeq0$,
i.e., near the glitches in Fig.~\ref{fig:streamplot_xi2_x0}. From the
first line of \Eqref{eq:x0P05}, we can solve for $x_0$ as a function
of $\xi_2$ and $g^2$, thus reducing the problem to a two-dimensional
theory space. The leading order of the beta-function of $\xi_2$ as
displayed in \Eqref{eq:xi2P05} does not depend on $x_0$, such that this
reduction leaves the far UV running of the quartic coupling
unaltered. We plot the corresponding projection of the RG vector field
in Fig.~\ref{fig:P1o2_streamplot_xi_reduced}.
As we had before for the case without a relevant direction,
cf. Fig.~\ref{fig:streamplot_xi_quasioneloop} (right panel), the
asymptotically free trajectory is simply parametrized by a constant
value of $\xi_2$. This is exactly the one corresponding to the
\fgFPs\ of \Eqref{eq:xi2P05} at nonvanishing $g^2$.

We anticipated that the quartic coupling is marginally irrelevant,
but this issue needs a careful analysis, since it reveals some
subtleties.
To this end, one can inspect the eigenvalues of the stability matrix up to order $g^2$,
\begin{eqnarray*}
\theta_{1}&=&-g^2\frac{32\xi_2^2(2+9\xi_2)-27\chi}{192\pi^2\xi_2^2}\\
\theta_{2}&=&g^2\frac{43}{32\pi^2}.
\end{eqnarray*}
Here we set $\chi$ and $\xi_2$ to their relative \fgFPh\ values, but keep $g^2$ nonvanishing.
While $\theta_0$ and $\theta_2$ are always positive, the sign of $\theta_1$
can change depending on $\xi_2$.
Inserting the determination of $\chi$ in terms of $\xi_2$ at the \fgFP\,
one finds
\begin{equation}
 \theta_{1}=-g^2\frac{(9-4\xi_2)}{64\pi^2\xi_2}\frac{\mathrm{d}\chi}{\mathrm{d}\xi_2}.
\end{equation}
Hence, $\theta_1$ is negative everywhere apart for the region $\xi_2\in[9/4,3.4595]$ to 
the right of the pole in  Fig.~\ref{fig:P1o2_FPchi_in_xi2}, where the curve has negative slope.
In other words, the quartic coupling is always irrelevant apart for the latter case.
This region of $\xi_2$ values can be achieved by fixing $\chi$ to a sufficiently positive number,
cf.~Fig.~\ref{fig:P1o2_FPchi_in_xi2}. If
this is the case, there is always 
one additional \fgFP\ at which the quartic coupling is irrelevant.

\begin{figure}[!t]
\begin{center}
 \includegraphics[width=0.23\textwidth]{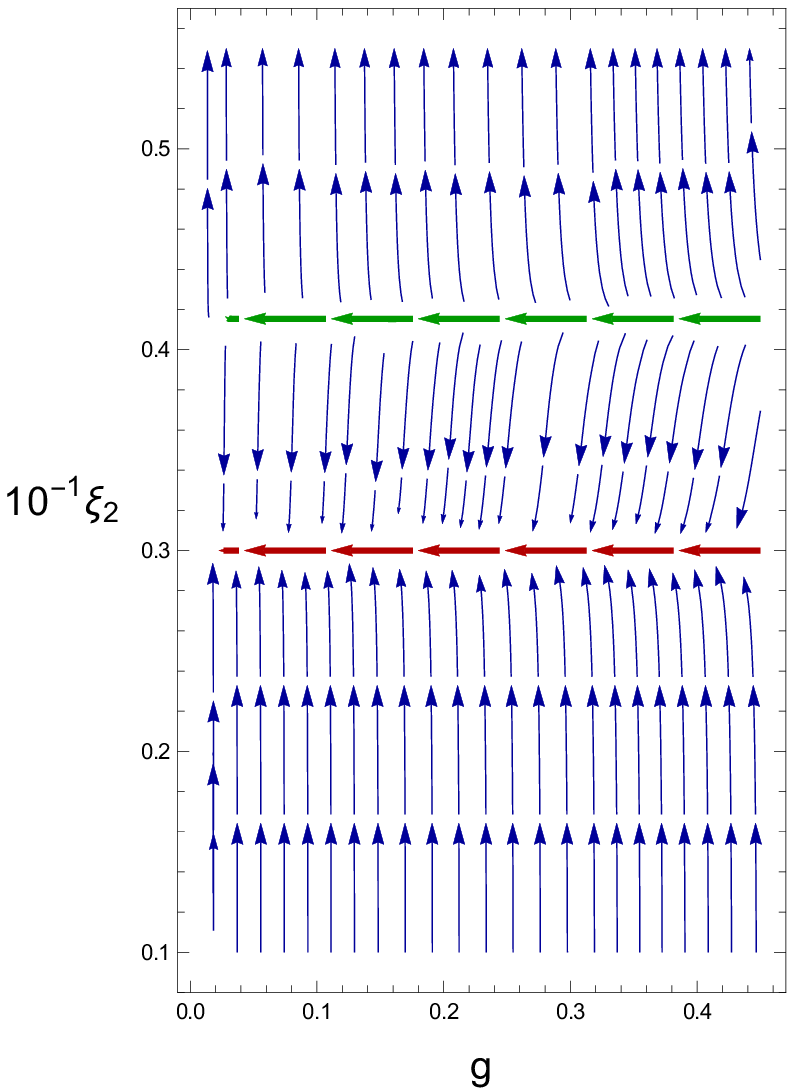}\ \ \
 \includegraphics[width=0.23\textwidth]{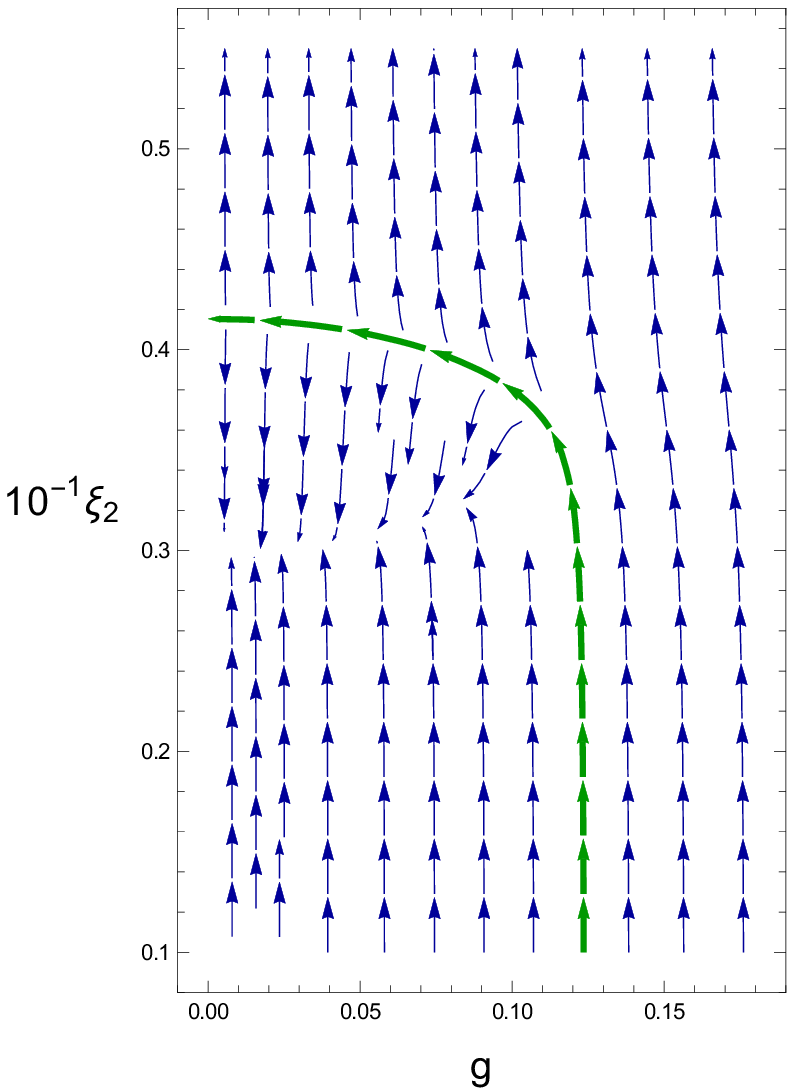}
 \caption{
    Stream-plot of the RG flow in the $N_p=2$ unconventional polynomial truncation for $P=1/2$, 
    projected by replacing $\kappa$ with its \fgFPh value
    and fixing $\chi=\xi_3/g=505$.
    In the left panel is plotted the leading-order expansion in $g^2$ \Eqref{eq:xi2P05},
    and the colors highlight two roots of the \fgFPh\ equation $\beta_{\xi_2}=0$.
    In the right panel is plotted the NLO expansion, where one of the roots clearly 
    hits $\xi_2=0$ at a finite $g^2$; the other root must behave accordingly,
    though this is hard to visualize in a streamplot.
}
\label{fig:P=1/2_twoAFtrajectories}
\end{center}
\end{figure}

These \fgFPs\ should manifest themselves as two asymptotically free trajectories,
a UV stable one, at a smaller value of $\xi_2$, and a UV unstable one, at a bigger value of $\xi_2$.
At the leading order of \Eqref{eq:xi2P05}, these trajectories are straight lines.
Both at next and next-to-next-to-leading order in $g^2$ (NLO and NNLO in the following)
 these trajectories become curves which 
move towards negative values of $\xi_2$, possibly 
merging at some finite value of $g^2$. The corresponding stream plots look 
essentially identical at NLO and NNLO, therefore we show the NLO one
in Fig.~\ref{fig:P=1/2_twoAFtrajectories}. Also the value of $g$ at which $\xi_2$ moves towards zero
is essentially the same. We conclude that this remains true also at higher orders.
Let us recall that we need to replace $\kappa$ by its \fgFPh\ value to produce these plots;
this value 
can be computed analytically not only at leading order \Eqref{eq:xi2P05},
but also at NLO and at NNLO.
By sampling other large positive values of $\chi$, we always observe the same phenomenon,
 which makes us believe that the branch of \fgFPh\ solutions having $\chi>0$ are generically affected by 
an instability driving the quartic coupling towards negative values.
The remaining branch of \fgFPh\ solutions, at which the quartic coupling is marginally irrelevant,
does not show such a behavior, and the potential remains stable at every $g^2$. 
For this reason, we will address only such solutions in what follows.

It is instructive to compare the present trajectories to those
considered by Coleman and Weinberg in the context of scalar
QED~\cite{Coleman:1973jx}. The Coleman-Weinberg (CW) trajectories are
characterized by renormalization conditions that ensure the absence of
a bare and renormalized scalar mass parameter. In the present setting,
this is equivalent to demanding for the absence of a relevant
component at any scale, which can here be arranged for without
spoiling asymptotic freedom. Whereas a nontrivial vacuum expectation
value is generated along CW trajectories towards the IR, the present
marginally relevant asymptotically free trajectory exhibits a finite
value of the scale dependent potential minimum $\kappa$ at and close
to the UV fixed point. For increasing $g^2$ towards the IR we
observe the same property, at least approximately by recalling that the
line of fixed points at nonvanishing $g^2$, plotted in
Fig.~\ref{fig:P1o2_FP}, is a satisfactory approximation of the
asymptotically free trajectories for sufficiently small $g^2$.  The
main difference to the CW trajectories is that our trajectories are
asymptotically free and do not suffer from UV Landau poles.  In this
sense the present trajectories are similar to those considered by
Salam and collaborators~\cite{Salam:1978dk} where the UV behavior is
also characterized by a vanishing quartic coupling $\lambda_2$.
Unlike in the CW mechanism, the absence of a relevant component at an
initial renormalization scale $k=\Lambda$ does not entail that the
bare Lagrangian is classically scale invariant.  Indeed it is not,
since $\kappa$ is always nonvanishing.  Only asymptotically at
$\Lambda\to\infty$ full (not classical) scale invariance is recovered.

Another important difference is that the present case actually offers
a one-parameter family of marginally-relevant asymptotically free
trajectories labeled, e.g., by the value of $\xi_2$ at the fixed
point.  This free parameter does not occur in the original CW
mechanism since the free quartic coupling is unambiguously tied
to the gauge coupling in order to reach the symmetry-broken phase.  As
a consequence, the ratio between the Higgs mass and the gauge boson
mass can be predicted along CW trajectories, being a fixed
number. While the CW trajectories are conceptually attractive as they
do not suffer from a naturalness problem, they seem not relevant for
standard-model phenomenology, as this mass ratio unfortunately comes
out too small for accommodating the Higgs and $W,Z$ boson masses. For
our marginally-relevant asymptotically free trajectories, the existence of a
one-parameter family of trajectories implies that the mass ratio of
scalar and gauge boson masses becomes a function of the free parameter
$\xi_2$. This means, our scenario features marginally-relevant
asymptotically free trajectories with zero relevant direction
(``CW-like'') apparently
free from a naturalness problem,
as the relevant direction
corresponding to the field expectation value $\kappa$ does not run
quadratically with the RG scale $k$. Hence,it is interesting to estimate the IR
mass spectrum emerging from these trajectories. For this, we study the
flow of the dimensionful mass parameters 
\begin{equation}
m_{\mathrm  H}^2=(2\lambda_2\kappa) k^2, \quad m_{\mathrm W}^2=(g^2\kappa/2)
k^2.
\label{eq:massdefs}
\end{equation}
This ratio is at any scale equal to $4\xi_2$ evaluated at that scale,
and the latter strongly depends on its fixed-point value at $g^2=0$,
which is an arbitrary positive number.  Plots of this ratio are given
below within a full FRG analysis. However, these are restricted to the
case of standard trajectories featuring a IR Higgs phase thanks to the
presence of a nonvanishing relevant component.  As far as CW
trajectories in their original sense are concerned, the task of
following their flow over many orders of magnitude while consistently
eliminating
the relevant component and possibly capturing strongly-coupled
dynamics, is beyond the reach of the present work.

To summarize, an effective-field theory treatment including a
parametrization of the influence of higher-dimensional operators by
one free parameter allows to construct a two dimensional UV critical
surface for each fixed point with positive $\xi_2=\lambda_2/g^2$.
This translates into a $2$-dimensional family of asymptotically free trajectories, that
can be labeled by the free parameter $\chi$, which in turn can be
expressed in terms of $\xi_2$ in the far UV, or by the relevant
component $\kappa$ at a finite value of $g^2$.  It is tempting to view
$\chi$ as an exactly marginal coupling, parametrizing a line of fixed
points. However, $\chi$ parametrizes different boundary conditions
for the RG flow. Thus, different values of $\chi$ can rather be
considered as parametrizing different theories.

\subsection{$(P\!<\!1/2)$-scaling solutions}
\label{subsec:P<1/2_with_relevant}
%

For the discovery of asymptotically free trajectories in the
effective-field-theory setting discussed above, it is crucial
to introduce a parametrization of the influence of unknown
higher-dimensional operators. Still, the UV behavior has remained
accessible by standard perturbative expansion techniques with the
gauge coupling $g^2$ as the governing small parameter. For the details
of the expansion, it is important to keep track of the $g^2$
dependence of the boundary condition, which is parametrized by the
rescaling power $P$ in our setting. 

We are therefore looking for the simplest description that preserves
the presence of real finite-$g^2$ \fgFPs, at values of the coupling that
come arbitrarily close to the full description in the $g^2\to 0$
limit. In the previous subsection, this simplest description for
$P=1/2$ is provided by the next-to-leading $g^2$ dependence of
$\beta_{x_0}$ and $\beta_{\xi_2}$. The corresponding roots of the
\fgFPh\ equations -- if expressed in terms of $\kappa$ and
$\xi_2$ -- were $g^2$ independent and therefore equal to the full \fgFPh\
values.  We now want to repeat this kind of analysis for $P<1/2$.  Let
us recall the general rescalings:
\begin{equation}
x_0=g^{2P}\kappa\ , \quad
\xi_2=g^{-4P}\lambda_2\ , \quad
\xi_3=g^{-6P}\lambda_3=g^{2P}\chi\ .
\label{eq:rescale}
\end{equation}
After these rescalings, $g$ appears in the beta functions through
positive integer powers of three elementary powers: $g^2$, $g^{2P}$,
and $g^{2(1-P)}$.  It is useful to think of the weak-coupling
expansion in terms of a Taylor expansion of the beta functions in
powers of these three variables.  Concentrating on $P<1/2$, the
leading power is $g^{2P}$. A coherent picture already arises at second
order,
\begin{eqnarray}
\beta_{x_0}&=&-2x_0+ g^{2P} \left(\frac{3}{16\pi^2}\right)
=
g^{2P} \left(\frac{3}{16\pi^2}-2\kappa\right) \label{eq:x0Pless05}\\
\beta_{\xi_2}&=&g^{4P}\left(\frac{3\xi_2^2}{4\pi^2}-\frac{\chi}{16\pi^2}\right)\label{eq:xi2Pless05}
\end{eqnarray}
Compared with the $P=1/2$ case, the $B$ and $C$ terms of the one-loop
formula, are now sub-leading, because they are of order $g^2$ and
$g^{4(1-P)}$.  As in the $P=1/2$ case, this lowest order description
of $\beta_{\xi_2}$ does not depend on $x_0$. Also the fact that the
fixed-point value of $\kappa$ does not depend on $g^2$ extends to the
whole range $0<P\leq1/2$ in this simple approximation.  It is
interesting to note that $\kappa$ is independent of $P$ for
$0<P\leq1/2$. This implies that each of the $P$-scaling solutions
yields the same nonvanishing dimensionless field expectation value in
the far UV, even though different $P$ values exhibit a different
$g^2$-power-like approach to an asymptotically free flat interaction
potential.

\begin{figure}[!t]
\begin{center}
 \includegraphics[width=0.4\textwidth]{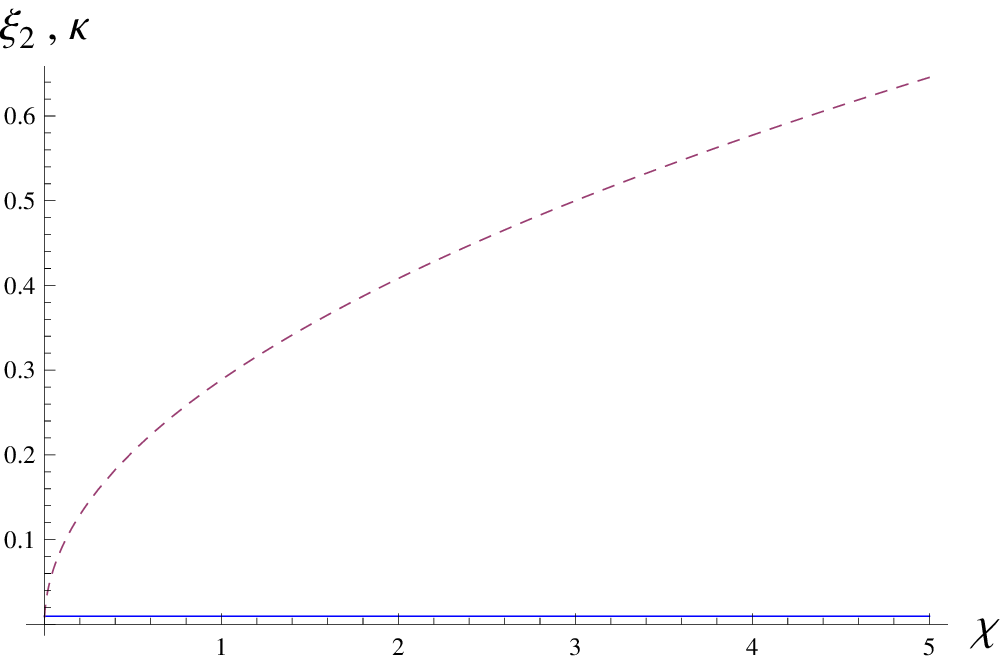}
 \includegraphics[width=0.4\textwidth]{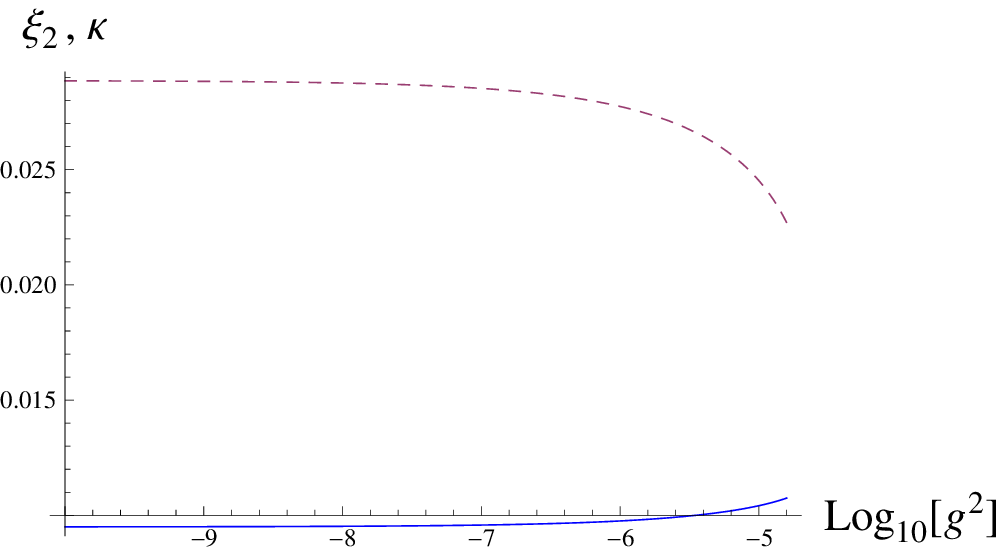}
 \caption{ Roots of the fixed-point equations $\beta_{x_0}=0$ and
   $\beta_{\xi_2}=0$ (\fgFPs) in terms of the dimensionless field expectation
   value $\kappa$ (solid) and the gauge-rescaled coupling $\xi_2$
   (dashed) as functions of $\chi$ and $g^2$ for the $P=1/4$ scaling
   solution in the effective-theory approximation with $N_p=2$.  As
   example values, the upper panel uses $g^2=10^{-12}$, and the lower
   panel $\chi=0.01$.  }
\label{fig:P1o4_FP}
\end{center}
\end{figure}

Going beyond this lowest order expansion, the \fgFPh\ values of
$\kappa$ and $\xi_2$ can straightforwardly be computed from the full
$\beta$ functions in this effective field theory setting for a given
value of $\chi>0$ and $g^2>0$, see e.g., Fig.~\ref{fig:P1o4_FP} (upper
panel) for $P=1/4$. For small coupling, the full numerical result
shown in Fig.~\ref{fig:P1o4_FP} agrees with the values predicted by
Eqs.~\ref{eq:x0Pless05} and \ref{eq:xi2Pless05} to a high
accuracy. Deviations become visible for increasing values of $g^2$
(lower panel).
%

\subsection{$(1/2\!<\!P\!<\!1)$-scaling solutions}
\label{subsec:P>1/2_with_relevant}

Let us now study whether the inclusion of the relevant direction gives
us also access to $P$-scaling solutions for $P>1/2$. For simplicity,
we stay within the simplest effective-field theory approximation
$\Np=2$ and consider the $\beta$ functions of $x_0$ and $\xi_2$,
obtained from the running of the original couplings $\kappa$ and
$\lambda_2$ by an appropriate $P$ rescaling,
cf. \Eqref{eq:rescale}. As before, we Taylor-expand the $\beta$ functions
in powers of $g^2$, $g^{2P}$, and $g^{2(1-P)}$. For $1/2<P<1$, these
powers are ordered as $g^{2P}>g^2>g^{2(1-P)}$ for $g^2\ll 1$.

We follow the same line of argument as before, expanding the $\beta$
functions for weak coupling treating $x_0$, $\xi_2$ and $\xi_3$ as
independent. To leading order in  $g^{2(1-P)}$, we obtain for $\beta_{x_0}$,
\begin{equation}
\beta_{x_0}=-2x_0+ g^{2(1-P)} \frac{9}{64\pi^2\xi_2}\ , \label{eq:x0Pless1}
\end{equation}
which already defines a meaningful nontrivial \fgFP\ for finite
values of $g$.  For $\beta_{\xi_2}$, care has to be taken of the
fact that any integer power of $g^{2(1-P)}$ can be smaller than
$g^{2P}$ itself, depending on $P\in(1/2,1)$. We thus retain both
$g^{2P}$ and the full dependence on $g^{2(1-P)}$, obtaining
\begin{eqnarray}
\beta_{\xi_2}&=&g^{2(1-P)}\frac{9\xi_3}{8\pi^2(2+g^{2(1-P)}x_0)^3\xi_2}\nonumber\\
&&+g^{4(1-P)}\frac{9}{16\pi^2(2+g^{2(1-P)}x_0)^3}
\left(2+\frac{x_0 \xi_3}{\xi_2}\right)\nonumber\\
&&+g^{2P}\frac{\xi_3}{16\pi^2}\left(-1
+\frac{x_0\xi_3}{\xi_2}\right)\ . \label{eq:xi2Pless1}
\end{eqnarray}
This simplified set of equations supports a simple
\fgFPh\ solution for finite $g^2$, at which $\xi_2$ stays nonvanishing
provided $\xi_3=g^{2(1-P)}\chi$, and
\begin{equation}
x_0=g^{2(1-P)}\frac{9}{128\pi^2\xi_2}\quad ,\quad \xi_2=-\chi\ .\label{eq:FPP1}
\end{equation}
This solution is exactly reproduced also with the non-Taylor-expanded
$\beta$ functions.  Knowing the scaling of $\xi_3$ at the fixed point, we can
further simplify $\beta_{\xi_2}$ to
\begin{equation}
\beta_{\xi_2}=g^{4(1-P)}\frac{9}{64\pi^2\xi_2}(\xi_2+\chi)\ .\label{eq:xi2Pless1b}
\end{equation}
Compared to the previous cases, we observe a qualitative change in the
resulting $g^2$ dependence of some couplings in order to obtain
asymptotically free trajectories. For instance, the power dependence
of $\xi_3$ on $g^2$ decreases now for $P\in(1/2,1)$ and goes to zero
for $P\to 1$, whereas it was increasing for $P\in(0,1/2)$. A similar
qualitative change applies to the scaling of $x_0$. We find that
$\chi$ has to be negative as is also legitimate for $P=1/2$.

An important novelty of these $P>1/2$ scaling solutions is the
behavior of the dimensionless field expectation value $\kappa$.  Since
$\kappa=g^{-2P}x_0\sim g^{2-4P}$, it diverges in the limit $g^2\to 0$.
This does not lead to any problem, since the expectation value
increases as the potential gets flatter and flatter, such that only
when the potential is completely flat the expectation value has
reached infinity. This behavior of $\kappa$ is essential for these
scaling solutions, which is why we were not able to see them without
the inclusion of the relevant direction. Despite the fact that
$\kappa$ grows unboundedly towards the UV, no nontrivial threshold
effects appear, since the expectation value enters the denominators
with sufficient powers of $g^2$ vanishing in the UV.

\subsection{$(P\!=\!1)$-scaling solutions}
\label{subsec:P=1_with_relevant}
%
%
\begin{figure}[!t]
\begin{center}
 \includegraphics[width=0.4\textwidth]{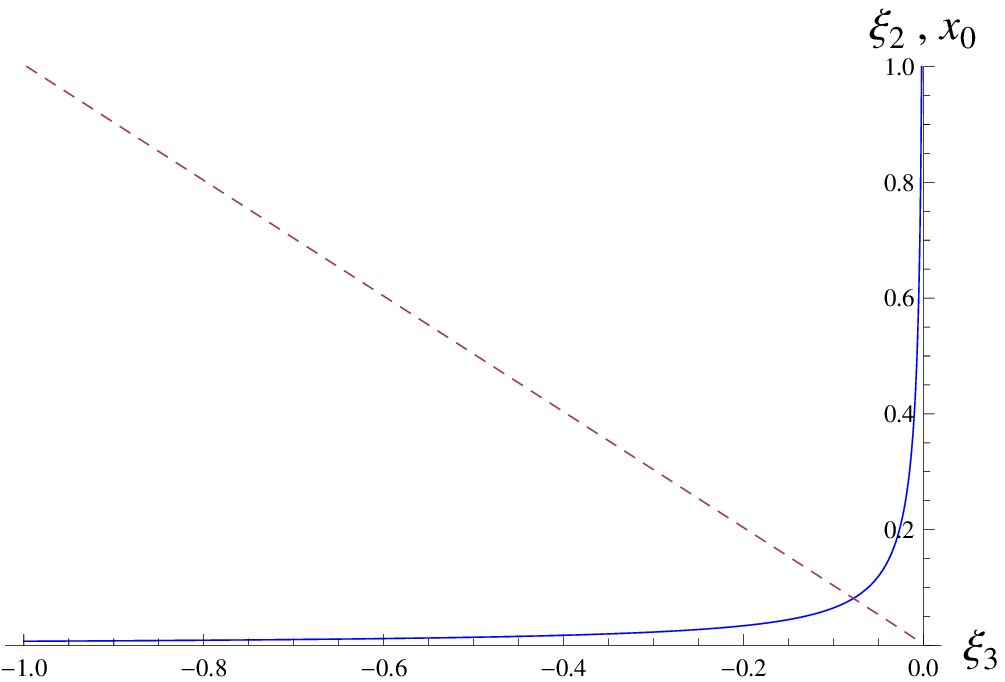}
 \includegraphics[width=0.4\textwidth]{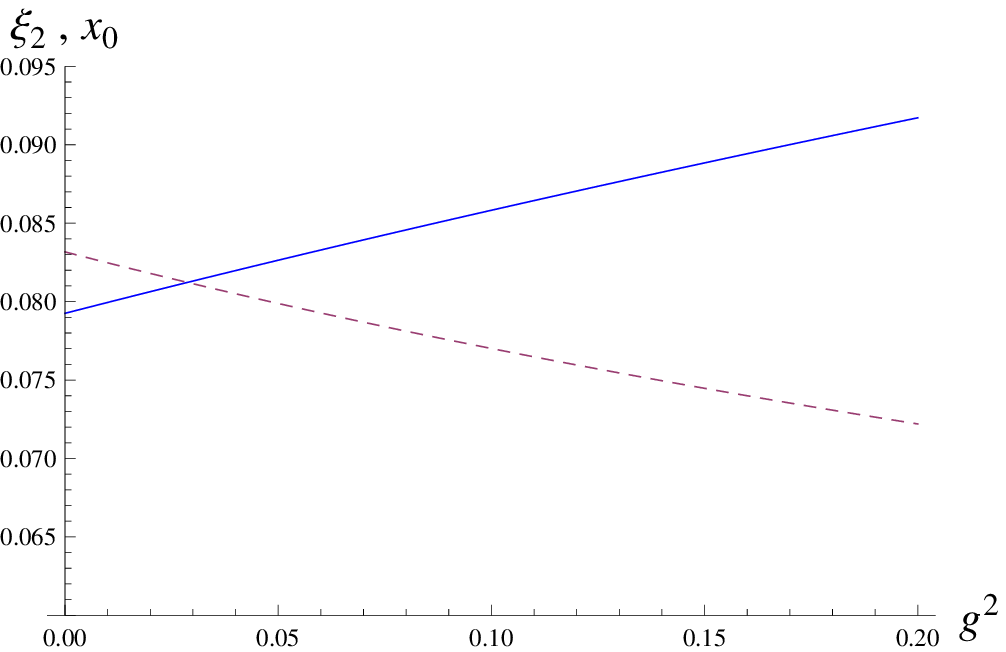}
 \caption{ Roots of the fixed-point equations $\beta_{x_0}=0$ and
   $\beta_{\xi_2}=0$ (\fgFPs) in terms of the dimensionless gauge-rescaled
   field expectation value $x_0$ (solid) and the gauge-rescaled
   coupling $\xi_2$ (dashed) as functions of $\xi_3$ and $g^2$ for the
   $P=1$ scaling solution in the effective-theory approximation with
   $N_p=2$.  As example values, the upper panel uses $g^2=10^{-10}$,
   and the lower panel $\xi_3=-0.08$.}
\label{fig:P1_FP}
\end{center}
\end{figure}

\begin{figure}[!t]
\begin{center}
 \includegraphics[width=0.35\textwidth]{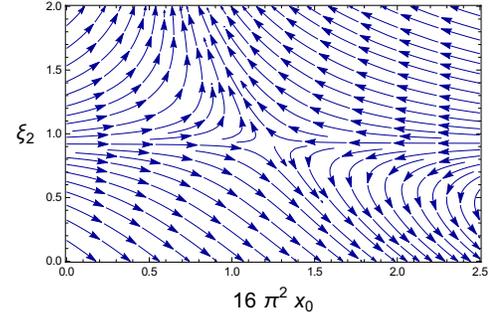}
 \caption{ RG flow in the $N_p=2$ effective-field-theory approximation
   for $P=1$, projected on a plane at fixed value of $g^2=0.01$ and
   $\xi_3=-1$.  }
\label{fig:P1_streamplot_xi2_x0}
\end{center}
\end{figure}

\begin{figure}[!t]
\begin{center}
 \includegraphics[width=0.35\textwidth]{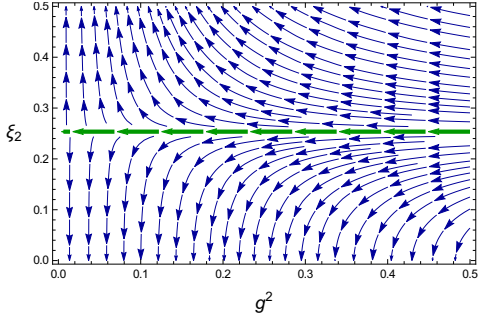}
 \caption{ Projection of the RG flow in the $N_p=2$
   effective-field-theory approximation for $P=1$ onto the surface
   with vanishing relevant component for fixed $\xi_3=-1/4$.  In green
   we highlight the purely marginally-relevant asymptotically free trajectory. The flow
   is singular at $\xi_2=0$ because of the nontrivial running of the
   field expectation value: $x_0\neq 0$.  }
\label{fig:P1_streamplot_xi_reduced}
\end{center}
\end{figure}

The $P=1$ case is particularly interesting as the powers counting in
$g^2$ changes completely, since $g^{2(1-P)}$ is no longer a good
expansion variable, and $g^2=g^{2P}$ becomes the only available small
parameter. Already the leading order collects a lot of terms and is
rather extensive.  On the other hand, the fact that $g^{2(1-P)}=1$
provides already sufficient structure to the zeroth-order $\beta$
functions in order to exhibit nontrivial \fgFPs. As a consequence, the
far UV behavior of the asymptotically free trajectories will be described by finite
nonvanishing values of all the three parameters $x_0$, $\xi_2$, and
$\xi_3$ in the $P=1$ case. The zeroth-order expansion of the beta
functions reads
\begin{eqnarray}
\beta_{x_0}&=&-2x_0+ \frac{9}{16\pi^2(2+x_0)^2\xi_2}+O(g^2)\label{eq:x0P1}\\
\beta_{\xi_2}&=&\frac{9\xi_3}{16\pi^2(2+x_0)^2\xi_2}+\frac{9}{8\pi^2(2+x_0)^3}+O(g^2) \label{eq:xi2P1}
\end{eqnarray}
and provides us with three fixed-point solutions, one of which is real.  For
small $x_0$ the latter is
\begin{eqnarray*}
x_0&=& \frac{18}{27-256\pi^2\xi_3}+O(x_0^2)\\
\xi_2&=&\frac{9}{256\pi^2}-\xi_3+O(x_0^2)\ .
\end{eqnarray*}
which favors negative values of $\xi_3$.  That $\xi_3$ needs to be
negative is confirmed by the analysis of the non-Taylor-expanded full version of the
$\beta$ functions, for generic $x_0$, that provides the \fgFPh\ values 
shown in Fig.~\ref{fig:P1_FP}.  From the simplified $\beta$ functions
of Eqs.~\eqref{eq:x0P1} and \eqref{eq:xi2P1}, these are most simply
described by expressing $\xi_2$ and $\xi_3$ as function of $x_0$,
\begin{eqnarray*}
\xi_2&=& \frac{9}{32\pi^2x_0(2+x_0)^2}\\
\xi_3&=&-\frac{9}{16\pi^2x_0(2+x_0)^3}\ .
\end{eqnarray*}
For the $P=1$ scaling solution, it is useful to take a closer look at
the corresponding RG flow and at the trajectory itself, similar to the
analysis for the $P=1/2$ case. As Fig.~\ref{fig:P1_streamplot_xi2_x0}
shows, the \fgFPs\ have a relevant direction that dies off
exponentially in the UV. Therefore, the far UV behavior of any asymptotically free
trajectory is one and the same at fixed $\xi_3$. The latter can be
described also in the reduced theory space with vanishing relevant
direction (CW-like trajectories). This is obtained by solving
$\beta_{x_0}=0$ for $x_0$ within the zeroth-order approximation
presented above. This process is tantamount to restricting ourselves
to the repulsive line in Fig.~\ref{fig:P1_streamplot_xi2_x0}, which is
a two-dimensional space with coordinates $\xi_2$ and $g^2$. This
system of coordinates is singular at $\xi_2=0$, since we can solve for
$x_0$ as a function of $g^2$ and $\xi_2$ only if $\xi_2$ is
nonvanishing.  The projection of the RG flow onto the two-dimensional
theory space with vanishing relevant component is shown in
Fig.~\ref{fig:P1_streamplot_xi_reduced}.  As a trivial consequence of
the fact that the simplified beta functions for the matter sector are
$g^2$ independent, the asymptotically free trajectory is the one with $\xi_2$ constant
and equal to its fixed-point value.
%

\subsection{$(P\!>\!1)$-scaling solutions}
\label{subsec:P>1_with_relevant}
%
\begin{figure}[!t]
\begin{center}
 \includegraphics[width=0.4\textwidth]{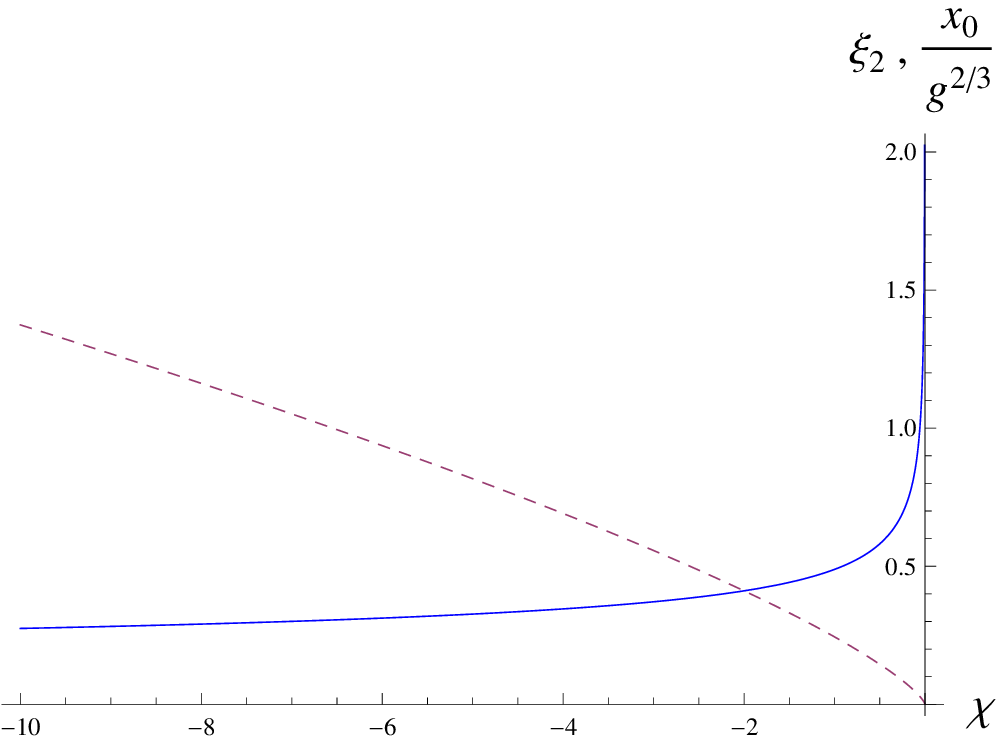}
 \includegraphics[width=0.4\textwidth]{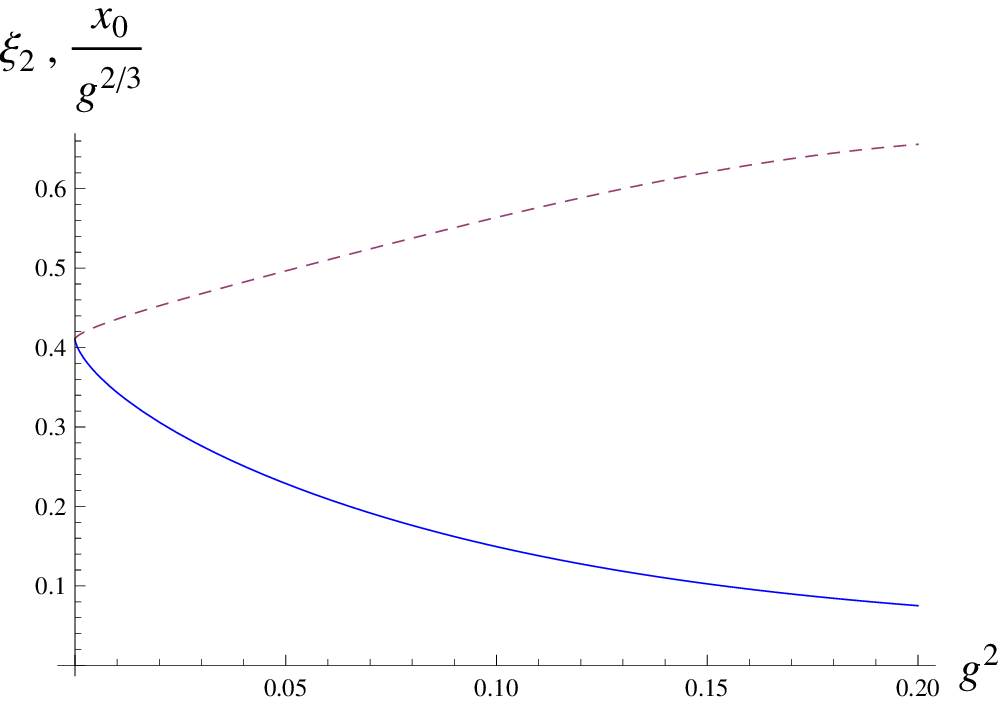}
 \caption{ Roots of the fixed-point equations $\beta_{x_0}=0$ and
   $\beta_{\xi_2}=0$ in terms of the gauge-rescaled dimensionless
   field expectation value $x_0/g^{2/3}$ (solid) and the
   gauge-rescaled coupling $\xi_2$ (dashed) as functions of
   $\chi=\xi_3/g^{2/3}$ and $g^2$ for the $P=2$ scaling solution in
   the effective-theory approximation with $N_p=2$.  As example
   values, the upper panel uses $g^2=10^{-9}$, and the lower panel
   $\chi=-2$.  }
\label{fig:P2_FP}
\end{center}
\end{figure}

For $P>1$, we replace $g^{2(1-P)}$ with the inverse $g^{2(P-1)}$ as
the appropriate variable for the Taylor expansion.  For $1<P<2$ this
is also the leading term. To leading-order, the $\beta$ functions in
the effective-field theory approximation with $\Np=2$ read for $1<P<2$
\begin{eqnarray}
\beta_{x_0}&=&-2x_0+ g^{2(P-1)}\frac{9}{16\pi^2x_0^2\xi_2}+O(g^{4(P-1)}), \label{eq:x0Pg1l2}\\
\beta_{\xi_2}&=&g^{2(P-1)}\left(\frac{9}{8\pi^2 x_0^3}
+\frac{9\xi_3}{16\pi^2 x_0^2\xi_2}\right)
+O(g^{4(P-1)}).\ \ \ \label{eq:xi2Pg1l2}
\end{eqnarray}
If $P=2$ the leading variable $g^{2(P-1)}$ merges with $g^2$, and the
simplest approximation we can inspect is an expansion to linear order in
$g^2$
\begin{eqnarray}
\beta_{x_0}&=&-2x_0+ g^{2}\left(\frac{9}{16\pi^2x_0^2\xi_2}
-\frac{11 x_0}{12\pi^2}\right), \label{eq:x0P2}\\
\beta_{\xi_2}&=&g^{2}\left(
\frac{9}{8\pi^2x_0^3}+\frac{11 \xi_2}{6\pi^2}
+\frac{9\xi_3}{16\pi^2 x_0^2\xi_2}\right)
.\ \label{eq:xi2P2}
\end{eqnarray}
For $P>2$, both $g^{2P}$ and $g^{2(P-1)}$ are larger than $g^2$ at
small coupling, such that only the leading term in $g^2$ is relevant
for the UV behavior.  The lowest-order approximation is, however, not
sufficient in the present truncation, since the only \fgFP\
would yield $x_0=\xi_2=0$. At next-to-leading order, we encounter a
term linear in $g^{2(P-1)}$ in the $\beta$ functions:
\begin{eqnarray}
\beta_{x_0}&=&-2x_0 -g^{2}\frac{11 P x_0}{24\pi^2}
+ g^{2(P-1)}\frac{9}{16\pi^2x_0^2\xi_2}\ ,\label{eq:x0Pg2}\\
\beta_{\xi_2}&=&g^{2}
\frac{11 P\xi_2}{12\pi^2}+g^{2(P-1)}\left(\frac{9}{8\pi^2 x_0^3}
+\frac{9\xi_3}{16\pi^2 x_0^2\xi_2}\right),\ \ \label{eq:xi2Pg2}
\end{eqnarray}
which reproduces the correct fixed-point values for $g\to 0$, if compared
  to the non-Taylor-expanded full form.

In summary, these weak-coupling $\beta$ functions feature \fgFPs\
of $x_0$, and $\xi_2$ at finite values of the gauge coupling for all
values of $P>1$, provided that $\xi_3$ scales like $\xi_3=\chi
g^{-2(P-1)/3}$ with negative $\chi$. The leading order $g^2$
dependence of these fixed points is given by
\begin{eqnarray}
x_0 &=& \frac{1}{2}\sqrt{\frac{3}{\pi}}
\frac{g^{2(P-1)/3}}{(-\chi)^{1/4}}, \quad\text{for}\,\,P>1, \label{eq:xoFPPg1}\\
\xi_2 &=& \frac{1}{4}\sqrt{\frac{3}{\pi}}
(-\chi)^{3/4}\ \label{eq:xi2FPPg1}.
\end{eqnarray}
We observe that $x_0$ vanishes as $g^{2(P-1)/3}$ towards the UV,
however the dimensionless field expectation value at the same time
runs to infinity, $\kappa\sim g^{-2(2P+1)/3}$. A difference to the
previous cases is that $\xi_3$ must scale like a negative power of
$g^2$ which implies a somewhat less suppressed $\lambda_3$.  Yet,
$\lambda_3$ is still more suppressed than $\lambda_2$, since
$\lambda_2\sim g^{4P}\xi_2$ while $\lambda_3\sim g^{2(8P+1)/3}$.  The
full $\chi$ and $g^2$ dependence of the \fgFPs\ as
  obtained from the non-Taylor-expanded full form of the $\beta$
  functions is shown in Figs.~\ref{fig:P2_FP} for $P=2$; all other
cases with $P>1$ are rather similar.

\section{Functional renormalization group}
\label{sec:FRG}

\subsection{Flow equation for the nonabelian Higgs model}

The effective-field theory method used in the previous section is
based on several explicit or implicit assumptions which need to be
checked in order to claim the existence of asymptotically free scaling
solutions in nonabelian Higgs models: explicitly, we assumed that the
flow equation of the highest coupling $\lambda_{\Np+1}$ of our
polynomial operator expansion of the effective potential
\Eqref{eq:efffieldpot} exhibits solutions with the desired
properties. While we have argued that this can always be arranged for
by our expansion technique to any desired order, an implicit
assumption underlying this construction is that the resulting
all-order expansion $\Np\to\infty$ (or a resummed version thereof)
yields a physically legitimate potential $U(\phi)$.

In particular, we demand for a globally defined regular potential for
all field amplitudes $\phi^\dagger \phi\geq 0$. For instance, it is
well known and widely studied that the criterion of global existence
of the fixed point potential singles out physical solutions of the
fixed-point equations such as the Wilson-Fisher fixed point in O($N$)
models below $d=4$ dimensions, see, e.g.,
\cite{Hasenfratz:1985dm,Morris:1998da} and
App.~\ref{app:WF}. We also demand for global stability, i.e., the
fixed point potential should be bounded from below. This criterion is,
for instance, at the center of ongoing discussions about the possible
existence of interacting fixed points in scalar theories in higher
dimensions
\cite{Fei:2014yja,Fei:2014xta,Percacci:2014tfa,Herbut:2015zqa,Mati:2016wjn,Eichhorn:2016hdi}. A
third, equally important criterion is the requirement of large-field
self-similarity or polynomial boundedness of the scaling solutions
near the fixed points \cite{Morris:1998da,Bridle:2016nsu,Rosten:2010vm}. This
criterion guarantees the quantization of critical exponents associated
with the scaling directions. Giving up this criterion would, for
instance, give rise to a continuum of critical exponents and
associated asymptotically free trajectories in pure scalar O($N$)
models in $d=4$ dimensions
\cite{Halpern:1994vw,Halpern:1995vf,Periwal:1995hw,Bonanno:2000sy,Gies:2000xr},
going beyond conventional quantum field theory standards
\cite{Morris:1996nx,Halpern:1996dh,Bridle:2016nsu}.

For a discussion of our asymptotically free trajectories in the light of
these criteria, we need to study the renormalization of the full
scalar potential, i.e., we need a $\beta$ functional for the
potential. This is provided by the functional renormalization group,
which can be formulated in terms of a flow equation for a
scale-dependent effective action $\Gamma_k$. The Wetterich equation  \cite{Wetterich:1992yh},
\begin{equation}\label{flowequation}
	\partial_t\Gamma_k[\Phi]
        =\frac{1}{2}\mathrm{STr}\{[\Gamma^{(2)}_k[\Phi]+R_k]^{-1}(\partial_t R_k)\}\, .
\end{equation}
determines the evolution of the effective action in the space of
action functionals as a function of a running scale $k$ at which
fluctuations are regularized by means of a regulator function
$R_k$. For permissible choices, this function $R_k$ acts
simultaneously as an IR mass-like regulator in the exact
field-dependent propagator $[\Gamma^{(2)}_k[\Phi]+R_k]^{-1}$, and as
multiplicative UV regulator through the momentum-shell-defining
functional $\partial_t R_k$, for details see
\cite{Aoki:2000wm,Berges:2000ew,Pawlowski:2005xe,Gies:2006wv,Delamotte:2007pf,Rosten:2010vm,Braun:2011pp}.

For our study of asymptotically free trajectories, we consider an
approximation of the full effective action spanned by the following
operators,
\begin{align}\label{eq:truncation}
\Gamma_k = \int d^dx &\Big[ U(\rho)+Z_{\phi}(D^{\mu}\phi)^{\dagger}(D_{\mu}\phi)\\
&+\frac{Z_{W}}{4}F_{\mu \nu}^iF^{i\mu \nu} +\frac{Z_{W }}{2\alpha} G^iG^i
-\bar{c}^i\mathcal{M}^{ij}c^j \Big]\, .\nonumber
\end{align}
where the effective potential $U$ and the wave-function
renormalizations $Z_{\phi,W}$ are $k$ dependent. Also the parameter
$\alpha$ could be considered as scale dependent, but we choose to work
in the Landau gauge $\alpha\to 0$ which is known to be a fixed-point
of the RG \cite{Ellwanger:1995qf,Litim:1998qi}. The quantization can
most conveniently be performed in a background-field $R_\alpha$ gauge. 
The definition of this gauge requires splitting the scalar field 
into the bare expectation value $\bar{v}$ and
the fluctuations $\Delta\phi$ about the expectation value
\begin{equation}\label{fluctandvev}
\phi^a=\frac{\bar{v}}{\sqrt{2}}\hat{n}^a+\Delta\phi^a, \quad
\Delta\phi^a=\frac{1}{\sqrt{2}}(\Delta\phi_1^a + i\Delta\phi_2^a),
\end{equation}
where $\hat{n}$ is a unit vector ($\hat{n}^{\dagger}_a\hat{n}^a=1$)
defining the direction of the expectation value in fundamental space.
A label ${\hat{n}}$ in place of a fundamental index denotes the
contraction with the unit vector ${\hat{n}}$ (or ${\hat{n}}^\dagger$,
depending on the position of the index). For vanishing background
field, the gauge fixing can then be written as
\begin{equation}\label{eq:gaugefixingcond}
G^i(W)=\partial_\mu W^i_\mu + i\alpha
\bar{v}\bar{g}(T^i_{{\hat{n}}\check{a}}\Delta\phi_1^{\check{a}}+iT^i_{{\hat{n}}a}\Delta\phi_2^a)=0, \quad \check{a}\neq\hat{n}
\end{equation}
where the component $\Delta \phi_1^{\hat{n}}$ is not included in the
sum over $\check{a}$, such that only the would-be Goldstone-boson
directions are involved, excluding the radial mode. In the presence of
a background field $\overline{W}_\mu$, the partial derivative $\partial_\mu$ acting on
the gauge field is replaced by a background covariant derivative
$\overline{D}_\mu\equiv D_\mu(\overline{W})$. This gauge condition gives rise to the gauge-fixing term in \Eqref{eq:truncation} as well as the Faddeev-Popov ghost term featuring the Faddeev-Popov operator
\begin{equation}\label{eq:FadeevPopovoperator}
\mathcal{M}^{ij}=-\partial^2\delta^{ij}-\bar{g}f^{ilj}\partial_{\mu}W^{l\mu}
+\sqrt{2}\alpha \bar{v}\bar{g}^2T_{{\hat{n}}\check{a}}^iT^j_{\check{a}b}\Delta\phi^b,
\end{equation}
given here for vanishing background and again excluding $a=\hat{n}$ in
the sum over $\check{a}$. Focusing on asymptotically free behavior,
it suffices to study the running of the gauge sector
perturbatively. For extracting the flow of the gauge coupling, it is
convenient to use the background-field method \cite{Abbott:1980hw}
which relates the running of the coupling to the wave function
renormalization of the background field $\eta_{\overline{W}}$, such
that the renormalized coupling and its $\beta$ function are given by
\begin{equation}
\beta_{g^2} = \pat g^2 = (d-4+ \eta_{\overline{W}}) g^2, \quad g^2=\frac{\bar{g}^2}{Z_{\overline{W}}k^{4-d}}, \label{eq:dimlesshg}
\end{equation}
where we have used the anomalous dimension $\eta_{\overline{W}}$ of the background
field. The anomalous dimensions of the fields are defined by
\begin{equation}
 \eta_{\overline{W}}=-\partial_t\log Z_{\overline{W}} ,\,\,\,
 \eta_W=-\partial_t\log Z_W ,\,\,\, \eta_{\phi}=-\partial_t \log
 Z_{\phi}.
\end{equation}
For the remainder of the paper, we identify the anomalous dimensions
of the background field and the fluctuation field,
$\eta_W=\eta_{\overline{W}}$, which maintains the one-loop exactness
of all flow equations and only introduces minor errors in the
higher-loop terms contained in the threshold functions, see below,
which become exceedingly irrelevant in the asymptotically free region. 

In order to specify the flow equations for the potential and the
anomalous dimension, it is useful to introduce several auxiliary
quantities.  Let us start with the mass matrix of the gauge bosons as
a consequence of a nontrivial (unrenormalized) minimum $\bar{v}^2/2$
of the potential $U(\rho)$,
\begin{equation}
\label{gaugemass}
\bar{m}^{2\ \ ij}_W=\frac{1}{2}Z_\phi \bar{g}^2 \bar{v}^2 \{T^i,T^j\}_{{\hat{n}}{\hat{n}}}.
\end{equation}
The mass matrix can be diagonalized by an appropriate basis in adjoint space, 
\begin{equation}
\label{gaugemassdiag}
\bar{m}^{2\ \ ij}_W=\bar{m}^2_{W,i}\delta^{ij}\quad\text{(no sum over $i$)}\, .
\end{equation}
The scalar mass matrix reads
\begin{equation}
\bar{m}_\phi^{2\,ab}=\bar{v}^2U''\left(\frac{\bar{v}^2}{2}\right)\hat{n}^a\hat{n}^{\dagger b}\, .\label{eq:scalarmass}
\end{equation}
In a diagonalizing basis, we have
$\bar{m}_\phi^{2\,ab}=\bar{m}_{\phi,a}^2\delta^{ab}$ (no sum over
$a$), with vanishing eigenvalues for the would-be Goldstone modes
corresponding to the broken generators in this gauge. The flow
equations can most conveniently be written in terms of renormalized
dimensionless quantities, such as the dimensionless potential in terms
of the dimensionless field invariant,
\begin{equation}
u(\tilde{\rho})=k^{-d}U(Z_{\phi}^{-1}k^{d-2}\tilde{\rho}),\quad
\tilde{\rho}=\frac{Z_\phi \rho}{k^{d-2}}. \label{eq:dimlessu}  
\end{equation}
The correspondingly dimensionless expectation value $\kappa$ already
used in the preceding section then reads
\begin{equation}
\kappa=\frac{Z_\phi
  \bar{v}^2}{2k^{d-2}}=\tilde{\rho}_{\min}, \label{eq:kappa1}
\end{equation}
where the wave function renormalization has been properly included
now. The analog polynomial expansions of the dimensionless potential
in the symmetric or broken regimes then read
\begin{equation}\label{eq:uexp}
u=\sum_{n=1}^{N_\text{p}}\frac{\lambda_{n}}{n!}\tilde{\rho}^n\, ,\quad 
\text{or}\,\,\,u=\sum_{n=2}^{N_\text{p}}\frac{\lambda_{n}}{n!}(\tilde{\rho}-\kappa)^n\,,
\end{equation}
where, in comparison with Eqs.~\eqref{eq:efffieldpot}
and~\eqref{eq:efffieldpotSSB}, the wave function renormalizations are
included now for the definition of the couplings $\lambda_n$.  

A crucial role is played by dimensionless renormalized mass parameters
\begin{equation}
\mu^2_{W,i}=\frac{\bar{m}^2_{W,i}}{Z_W k^2},\quad \mu^2_{\phi,a}=\frac{\bar{m}^2_{\phi,a}}{Z_\phi k^2}\,. 
\label{eq:defmu}
\end{equation}
Most of the analysis presented in this work will refer to the example
of a gauge group SU($N=2$).  In this case
\begin{equation}
\mu_W^2=\frac{1}{2} g^2 \kappa, \quad \mu_{\text{H}}^2=2
\lambda_2\kappa \ . \label{eq:dimlessmuSU2}
\end{equation}
Long-range observables are best described in terms of dimensionful
renormalized quantities, which are then easily obtained by
\begin{equation}
m_W^2=\mu_W^2 k^2, \quad m_{\text{H}}^2=\mu_{\text{H}}^2 k^2 \ . \label{eq:massesSU2}
\end{equation}
An analogous relation holds for the  dimensionful renormalized expectation value
\begin{equation}
v= \sqrt{ 2 \kappa}\, k^{(d-2)/2} \equiv Z_\phi^{1/2} \bar{v}\ . \label{eq:vev}
\end{equation}
The RG flow equations for the present system have been derived in
\cite{Gies:2013pma} for an even larger system including chiral
fermions. Taking over the results for the sector of the nonabelian
Higgs model, the flow equation of the scalar potential and the
equation for the scalar anomalous dimension read
\begin{widetext}
\begin{eqnarray}
\label{floweq:potential}
\partial_t u &=& -d u + (d-2 + \eta_\phi)\tilde\rho u' +  2 v_d \Big\{ 
(d-1)\sum_{i=1}^{\dG}l_{0}^{(\mathrm{G})d}\left(\mu_{W,i}^2 (\tilde\rho)\right)
+(2N-1)l_0^{(\mathrm{B})d}\left( u' \right) + l_0^{(\mathrm{B})d}\left(u' + 2 \tilde\rho u'' \right) \Big\} \\
\eta_{\phi}&=&\frac{8 v_d}{d}\Big\{
\tilde\rho(3 u'' +2\tilde\rho u''')^2 m_{2,2}^{(\mathrm{B})d}(u'+2\tilde\rho u'',u'+2\tilde\rho u'')
+(2N-1)\tilde\rho u''^2 m_{2,2}^{(\mathrm{B})d}(u',u')\nonumber\\
&&-2g^2 (d-1) \sum_{a=1}^{N}\sum_{i=1}^{\dG} T^i_{{\hat n}a}T^i_{a{\hat n}} \, 
l_{1,1}^{(\mathrm{BG})d}\left( u', \mu_{W,i}^2 \right) 
+(d-1)\sum_{i=1}^{\dG} \frac{\mu_{W,i}^4}{\tilde\rho} \left[ 
2 a_1^d\left(\mu_{W,i}^2 \right) + m_2^{(\mathrm{G})d}\left(\mu_{W,i}^2 \right) \right]\Big\}\Big|_{\tilde{\rho}=\tilde{\rho}_{\text{min}}}
\label{floweq:etaphi}
\end{eqnarray}
\end{widetext}
where $\mu_{W,i}^2 (\tilde\rho)$ are defined as functions of the full
scalar field in analogy with
Eqs.~\eqref{gaugemass},\eqref{gaugemassdiag}, reducing to the
dimensionless gauge boson renormalized masses for $\tilde\rho=\kappa$.
Furthermore, we have used the abbreviation $v_d=1/(2^{d+1}\pi^{d/2}
\Gamma(d/2))$, e.g., $v_4=1/(32\pi^2)$, and $N$ and $N^2-1$ denote
the dimension of the fundamental and adjoint representations,
respectively. The threshold functions $l$, $m$, $a$, represent the
various regularized loop contributions involving the full RG-improved
propagators, with the superscripts B and G indicating contributions
from scalar and gauge boson fluctuations, respectively. All threshold
functions depend on generally field-dependent mass-like arguments;
they vanish for large argument and approach finite constants for zero
arguments. The threshold functions have an additional dependence on the
anomalous dimensions of the fluctuating fields signifying the RG
improvement. In the present work, we use the threshold functions
arising from a piece-wise linear regulator
\cite{Litim:2000ci,Litim:2001up} for simplicity,
\begin{eqnarray}
l_0^{(\mathrm{B/G})d}(\omega) &=& \frac{2}{d}\frac{1-\tfrac{\eta_{\phi/\mathrm W}}{d+2} }{1+\omega} \, ,\nonumber \\
l_{1,1}^{(\mathrm{BG})d}(\omega, \omega) &=& \frac{2}{d}
\frac{2-\frac{\eta_\phi+\eta_\mathrm W}{d+2}}{(1+\omega)^3}  \, ,\nonumber\\
m_{2,2}^{(\mathrm{B})d}(\omega, \omega) &=& \frac{1}{(1 + \omega)^4} = m_2^{(\mathrm{G})d}(\omega) \, ,\nonumber\\
a_1^d(\omega) &=& \frac{1- \tfrac{\eta_\mathrm W}{d}}{d-2}\frac{1}{(1 + \omega)^3} \, .\label{eq:threshold}
\end{eqnarray}
For integral representations for general threshold functions, see the
appendix of \cite{Gies:2013pma}.

From \Eqref{floweq:potential}, we can extract the $\beta$ functions
for the effective vertices $\lambda_n$ by projecting both sides of
this flow equation onto the coefficients of the potential expansion
Eq.~\eqref{eq:uexp}. As a matter of principle, we would have to
distinguish the scalar anomalous dimension for the would-be Goldstone
modes from that of the radial mode in the symmetry-broken regime. For
simplicity, we ignore this difference. As the Goldstone modes as such
would not propagate in unitary gauge, we compute the scalar anomalous
dimension $\eta_\phi$ in the broken regime by projecting the flow onto
the radial scalar operators.

Finally, we need the anomalous dimension of the (background) gauge
field $\eta_{\overline{W}}$ for the running of the gauge coupling,
cf. \Eqref{eq:dimlesshg}. For almost all results deduced below, the
simple one-loop perturbative result would be sufficient. For reasons
of consistency with the scalar sector, we also include the
corresponding threshold behavior. As discussed in detail in
\cite{Gies:2013pma}, slightly different definitions of the gauge
coupling can be used in the symmetry-broken regime arising from
different legitimate choices of the relative orientation between the
scalar expectation value in fundamental space and the background color
field in adjoint space. In the following, we use the so-called minimal
decoupling option, for which the anomalous dimension for SU($N=2$), in
$d=4$ Euclidean dimensions reads
\begin{eqnarray}\label{floweq:gauge}
  \eta_{\overline{W}}&=&\frac{-g^2}{48 \pi^2}\Big(22N \,
  \mathrm L_W(\mu_{W,i}^2)- \mathrm L_\phi(\mu_{\phi,a}^2)\Big),\nonumber
\end{eqnarray}
The threshold functions satisfy $L_{W,\psi,\phi}(0)=1$, such that the
standard universal one-loop $\beta$ function for the gauge coupling is
obtained in this limit which corresponds precisely to the deep
Euclidean regime.  For generic arguments instead, they read for the
piece-wise linear regulator,
\begin{align}\label{floweq:gaugeLSU2}
\mathrm L_W(\mu_{W}^2)&=\frac{1}{44}\left(21+\frac{21}{1+\mu_{W}^2}+2\right) ,
&\!\! \mu_{W}^2&=\frac{g^2\kappa}{2},\nonumber\\
\mathrm L_\phi(\mu_{\text{H}}^2)&=\frac{1}{2}\left(1+\frac{1}{1+\mu_{\text{H}}^2}\right) ,
&\!\! \mu_{\text{H}}^2&=2\lambda_{2}\kappa .
\end{align}
This concludes our presentation of the functional RG flow for the
theory space spanned by the effective action $\Gamma_k$ of
\Eqref{eq:truncation}. The polynomial expansion of
\Eqref{floweq:potential} together with the $\beta$ function of the
gauge sector directly reproduces all $\beta$ functions for the
couplings $\lambda_n$ and its descendants used in the preceding section. 

\subsection{Gauge-rescaled flows}

We have characterized the scaling trajectories discovered before
within a perturbative or an effective-field-theory setting by their
UV-asymptotic scaling behavior of the scalar interactions with the
gauge-coupling. For instance, the $P$-scaling solutions are those enjoying
the property that $\lambda_2\sim g^{4P}$ in the far UV. According to
\Eqref{eq:xinP}, this generalizes to all higher couplings and thus
suggests to define a gauge-rescaled dimensionless potential,
\begin{equation}
f(x)= u(\trho)\Big|_{\trho= g^{-2P}x} = k^{-4} U\!\left( \rho \right)\Big|_{\rho= g^{-2P}Z_\phi^{-1}{k^2} x},\label{eq:deff}
\end{equation}
being a function of the gauge-rescaled scalar field
\begin{equation}
x=g^{2P}\trho=g^{2P}\frac{ Z_\phi}{k^2}\rho, \quad \rho=\phi^\dagger \phi. \label{eq:grescal}
\end{equation}
The flow equation for $f(x)$ follows straightforwardly from
\Eqref{floweq:potential},
\begin{eqnarray}
\partial_t f &=& \beta_f\equiv -4\, f+(2 + \eta_\phi-P \eta_W)x f' \label{floweq:fpotential}\\
&&+\frac{1}{16\pi^2} \Big\{3 \sum_{i=1}^{\dG} l_{0\mathrm T}^{(\mathrm{G})4}\left(g^{2(1-P)}\omega_{W,i}^2 (x)\right)\nonumber\\
&&+(2N-1)l_0^{(\mathrm{B})4}\! \left(g^{2P}f' \right)+ l_0^{(\mathrm{B})4}\!\left(g^{2P}(f' + 2 x f'') \right)\!
 \Big\},\nonumber 
\end{eqnarray}
where $\omega_{W,i}^2(x)=g^{2(P-1)}\mu_{W,i}^2$; e.g., for SU(2), we
simply have $\omega_{W,i}^2=x/2$ for any $i=1,2,3$.  A polynomial
expansion of $f(x)$ as well as \Eqref{floweq:fpotential} either about
$x=0$ or about a nontrivial minimum $x_0$ straightforwardly reproduces
the $\beta$ functions for the couplings $\xi_n$ as well as the minimum
$x_0$ as used above for various values of $P$. More concretely, we
have considered the truncation
\begin{equation}
f(x)= \frac{\xi_2}{2!}(x-x_0)^2+\frac{\xi_3}{3!}(x-x_0)^3\ ,
\end{equation}
in the previous section, resulting in a specific scaling of $x_0$ and
$\xi_3$ with respect to $g^2$ at fixed $\xi_2$, which is summarized in
Tab.~\ref{tab:polynomial_g2_scaling}.
\begin{table}[!t]
\begin{center}
\resizebox{4cm}{1cm}{
    \begin{tabular}{| l | l | l |}
    \hline
    $P$			&$x_0$		&$\xi_3$ \\ \hline
    $(0,1/2]$		&$g^{2P}$	&$g^{2P}$ \\ \hline
    $[1/2,1]$		&$g^{2(1-P)}$	&$g^{2(1-P)}$ \\ \hline
    $[1,\infty)$	&$g^{2(P-1)/3}$	&$g^{-2(P-1)/3}$\\ \hline
    \end{tabular}
}
 \caption{Asymptotic $g^2$-scaling of the polynomial couplings at fixed $\xi_2$ as a function of $P$.}
 \label{tab:polynomial_g2_scaling}
 \end{center}
\end{table}
In other words, we have identified an asymptotically free trajectory
for each choice of the two-parameter family $(P,\xi_2)$. The scaling
behavior of Tab.~\ref{tab:polynomial_g2_scaling} entails that the
minimum $x_0$ of $f(x)$ always vanishes asymptotically in the UV, the
curvature $\xi_2=f''(x_0)$ attains a finite nonvanishing value, while
the fate of $\xi_3=f'''(x_0)$ depends on $P$. For instance, the
dependence of $x_0$ on $(P,\xi_2)$ and its asymptotic behavior with
$g$ can be made explicit as follows:
\begin{equation}
 x_0=\left\{ \begin{array}{ll}
             g^{2P} \frac{3}{32\pi^2} & \text{for}\,P\in(0,1/2)\\
             g\left(\frac{3}{32\pi^2}+\frac{9}{128\pi^2\xi_2}\right) & \text{for}\, P=1/2 \\
             g^{2(1-P)}\frac{9}{128\pi^2\xi_2} & \text{for}\, P\in(1/2,1)\\
		\frac{9}{32\pi^2\xi_2(2+x_0)^2} & \text{for}\, P=1 \\
		g^{2(P-1)/3}\ \frac{1}{2\xi_2^{1/3}}\left(\frac{3}{2\pi}\right)^{2/3} & \text{for}\, P\geq 1
            \end{array}
            \right.
\label{eq:summary_x0_of_xi2}
\end{equation}
In the following, we aim at substantiating these properties studying
the full flow of $f(x)$ without recourse to an expansion in the scalar
field $x$. Our strategy is similar to the previous sections: we are
looking for a \fgFPh\ behavior of the flow of $f(x)$ even at finite
values of $g^2$, such that the full model approaches asymptotic
freedom in the UV. The corresponding \fgFPh\ potential is defined by a
vanishing $\beta_f$ functional for $f(x)$ even at finite $g^2$.

\subsection{Weak-coupling analysis}

In our preceding studies, we observed that such a \fgFP\ as a function
of the coupling $g^2$ represents a good estimate for the
asymptotically free trajectory. The difference between a true RG
trajectory and the \fgFP\ is controlled by the coupling $g^2$ itself
and vanishes in the limit $g^2\to0$, see, e.g.,
Fig.~\ref{fig:P3o8_AF_trajectory}. This allows us to construct the
\fgFPh\ potential in terms of an expansion of the fixed point equation
$\pat f(x)=0$ in powers of the gauge coupling. By keeping only the
leading $g^2$-dependence, we expect to get a portrait of the
asymptotically free trajectory which accurately describes the UV
asymptotics of the latter.

This type of construction guarantees that we hit a genuine fixed point
at $g^2=0$ also in the scalar sector. At finite $g^2$, the
\fgFPh\ condition $\beta_f=0$ can be interpreted as the definition of
the marginally-relevant perturbation associated with switching on the
gauge coupling. Denoting the leading $g^2$-dependent terms inside
$\beta_f$ by $\delta\beta_f$, we are thus interested in solving the
\fgFPh\ condition
\begin{equation}
 \beta_f\big|_{g^2=0}+\ \delta\beta_f=0\ . \label{eq:fgFPcond}
\end{equation}
For $P\neq 1$ the zeroth order coincides with the canonical terms (up to a field-independent additive constant)
\begin{equation}
 \beta_f\big|_{g^2=0}=-4f(x)+2xf'(x)\ , \quad P\neq 1 \label{eq:deltabetaneq1}
\end{equation}
while the leading $g^2$ dependence is given in
Tab.~\ref{tab:leading_g2_dep}, being corroborated in the next
sections.
\begin{table}[!t]
\vskip-3mm
\begin{equation*}
{\renewcommand{\arraystretch}{1.68}
\begin{array}{| l | l |}    
    \hline
    P & \delta \beta_f \\ \hline
    (0,1/2)\ \ \ & -g^{2P}\left(\frac{f'(x)}{8\pi^2}+\frac{x f''(x)}{16\pi^2}\right) \\ \hline
    1/2  & -g \left(\frac{9 x}{64\pi^2} +\frac{f'(x)}{8\pi^2}+\frac{x f''(x)}{16\pi^2}\right) \\ \hline
    (1/2,1)  & - g^{2(1-P)}\frac{9 x}{64\pi^2} \\ \hline
    (1,2)  & \frac{3(8g^{2(P-1)}+x)}{32\pi^2 (2g^{2(P-1)}+x)}\\ \hline
    [2,\infty)  & \frac{3(8g^{2(P-1)}+x)}{32\pi^2 (2g^{2(P-1)}+x)}+g^2\frac{11\, P}{24\pi^2}x f'(x)\\
    \hline
\end{array}}
\end{equation*}
\vskip-3mm
\caption{Asymptotic $g^2$-dependence of $\beta_f$ for small $g^2$ depending on the values of $P$.}
 \label{tab:leading_g2_dep}
\end{table}
For $P=1$ the zeroth-order features also a term due to
gauge-bosons loops
\begin{equation}
\beta_f\big|_{g^2=0}=-4f(x)+2xf'(x)+\frac{3}{32\pi^2}\frac{8+x}{2+x}\ , \quad P=1. \label{eq:deltabeta1}
\end{equation}
The leading dependence of $\delta\beta_f$ on the gauge coupling is of
order $g^2$, and it multiplies an extensive correction to the $\beta$
function which we do not display here but can straightforwardly be
obtained by expansion.  

We emphasize that the zeroth-order result for the \fgFPh\ condition
is a first-order linear ordinary differential equation (ODE) for any $P$ that
allows for a one-parameter family of solutions $f_*$, depending on $P$
and an integration constant $\xi$.  These solutions are
\begin{equation}
f_*(x)=\left\{ \begin{array}{ll}
\xi x^2 & \text{for}\, P\neq1 \\
\xi x^2-\left(\frac{3}{16 \pi}\right)^2 \left[2x+
  x^2\log\left(\frac{x}{2+x}\right) \right] & \text{for}\, P=1
\end{array}\label{eq:FPpotential}
\right. 
\end{equation}
and $\xi_2=2\xi$. Upon
inclusion of the leading-order correction $\delta \beta_f$, the
\fgFPh\ condition \eqref{eq:fgFPcond} becomes a second order ODE. As
detailed below, this does not introduce a further free parameter such
that the same counting of free parameters as for the zeroth order
persists also to higher orders. The solutions to \Eqref{eq:fgFPcond}
define an approximation to marginally relevant trajectories
which parametrically depend also on $g^2$.  Clearly, by taking the
$g^2\to 0$ limit of the latter potentials one should recover the
zeroth-order expressions above, such that these describe the first
corrections to exact scaling along the marginally relevant
asymptotically free trajectories.  In the following we will give
details of the shape of the leading-order $g^2$-dependent potentials.

\subsubsection{$(P\!<\!1)$-scaling solutions}
\label{subsec:P<1_potentials}

In order to stay on the level of elementarily integrable equations, we
keep only the leading power in $g^2$ for the case $P=1/2$. This is
indeed sufficient for our purposes. As verified in
App.~\ref{app:leadingpower}, the same leading-order
effective-field-theory analysis leads to the correct \fgFPh\ values
for $x_0$ and $\xi_2$ as discussed in Subsect.~\ref{subsec:P=1/2}, and
introduces only errors at order $\mathcal{O}(g^4)$ for the
higher-order coupling $\lambda_3$. The same pattern with
parametrically quantifiable error estimates also holds for $P<1/2$ as
well as for $P\in(1/2,1)$.

Hence, we expand $\beta_f$ to linear order in $g^{2P_{\mathrm L}}$,
with $P_{\mathrm L}=P$ or $(1-P)$, for $P\leq1/2$ or $P\geq1/2$
respectively, and set the result to zero. This yields a linear
differential equation which is solved by
{\renewcommand{\arraystretch}{1.4}
\begin{equation}
 f(x)= \left\{ \begin{array}{ll}
   \xi x^2-\xi \frac{3}{16\pi^2}g^{2P} x & \text{for}\,P\in(0,1/2)\\
   \xi x^2-\frac{3(3+8\xi)}{128\pi^2}g\ x & \text{for}\, P=1/2 \\
   \xi x^2-\frac{9}{128\pi^2}g^{2(1-P)} x & \text{for}\, P\in(1/2,1)
 \end{array}
 \right. . \label{eq:fofxP<1/2}
\end{equation}
}
By means of these parametrizations one can easily reproduce the
position of $x_0$ as a function of $g^2$ and the curvature at the minimum, $\xi_2=2\xi$,
that we have found in Sec.~\ref{subsec:P=1/2}, \ref{subsec:P<1/2_with_relevant}  
and \ref{subsec:P>1/2_with_relevant}, which is summarized in \Eqref{eq:summary_x0_of_xi2}.

For $P\leq 1/2$, the linear ODE 
for the \fgFPh\ potential
is of second order, and has therefore 
a two-parameter family of solutions.
We restricted our analysis to a one-parameter sub-class labeled by $\xi$, 
because the remaining solutions contain exponentially growing pieces,
and thus do not correspond to physical
trajectories~\cite{Bridle:2016nsu}.
To be precise, the large-$x$ behavior of the
discarded solutions for $P\leq 1/2$ is
\begin{equation*}
f(x)\widesim[1.8]{x\to \infty}\frac{g^{2P}}{2^{23}\pi^{10}x^4}\ e^{\frac{32\pi^2 x}{g^{2P}}}+O(g^{4P}),
\end{equation*}
and thus they do not satisfy the criterion of uniform convergence \cite{Bridle:2016nsu}.

\subsubsection{$(P\!=\!1)$-scaling solutions}
\label{subsec:P=1_potentials}

Our study of the $P=1$-scaling solutions in
Sec.~\ref{subsec:P<1/2_with_relevant} has shown that not only $\xi_2$
but also $x_0$ and $\xi_3$ attain nontrivial $g^2$-independent finite
values in the far UV.  This is compatible with the \fgFPh\ potential found
in \Eqref{eq:FPpotential}.  Indeed it is straightforward to check that
the values of $x_0$ and $\xi_3$, as functions of $\xi_2=2\xi$, that
can be extracted from this analytic potential exactly agree with those
computed from the polynomial beta-functions of
Sec.~\ref{subsec:P<1/2_with_relevant}.  Thus, the upper panel of
Fig.~\ref{fig:P1_FP} already implicitly displayed these functions.  To
compute the equivalent of the lower panel of Fig.~\ref{fig:P1_FP} in
the present functional approach, i.e., the scaling of the potential to
asymptotic freedom alongside $g^2$, we need to consider the leading
correction to scaling, which is encoded in the linearization of
$\beta_f$ with respect to $g^2$.  The latter is a lengthy expression
as in the polynomial truncations, and the resulting ODE for the
\fgFP\ is difficult to treat analytically. We display some
numerical solutions below.

At this point, a remark is in order: so far, we have abstained from a
standard technique of fixed-point analysis as often used for
Wilson-Fisher type fixed points within the functional RG approach. This analysis would proceed in terms
of a linearization of the flow equations with respect to the
functional perturbation $\delta{f}(x)=f(x)-f_*(x)$.  In fact, such a
linearized analysis would not be consistent for a similar reason as
has been discussed recently for nonpolynomial interactions in scalar
field theory in \cite{Bridle:2016nsu}.  For instance for the case
$P=1$, equating the linearized beta function of $\delta{f}(x)$ to zero
yields a first-order ODE that corresponds to the traditional
marginality condition for a small perturbation about the fixed point.
One solution is the $x^2$-perturbation expected from the usual
marginally irrelevant quartic coupling.  The only other solution is
the marginally relevant direction associated to the gauge coupling, of
the form ${g^2} \Delta f(x)$.  Here $\Delta f(x)$, for large $x$,
behaves like $x^2\log(2/x)$ times a coefficient that is
positive. Hence, it would be tempting to conclude that this
perturbation gives rise to bounded potentials only for ${g^2}<0$ and
it is therefore unphysical.  However, such a conclusion would not be
well founded because $\Delta f$ grows faster than $ f_*$ itself for
large fields, thus signaling the breakdown of the linearized
analysis.  Indeed, that this seeming stability problem is a fallacy of
the linearization becomes obvious from the asymptotic behavior of the
marginally relevant $f(x)$ for large $x$, as is discussed below.

\begin{figure}[!t]
\begin{center}
 \includegraphics[width=0.4\textwidth]{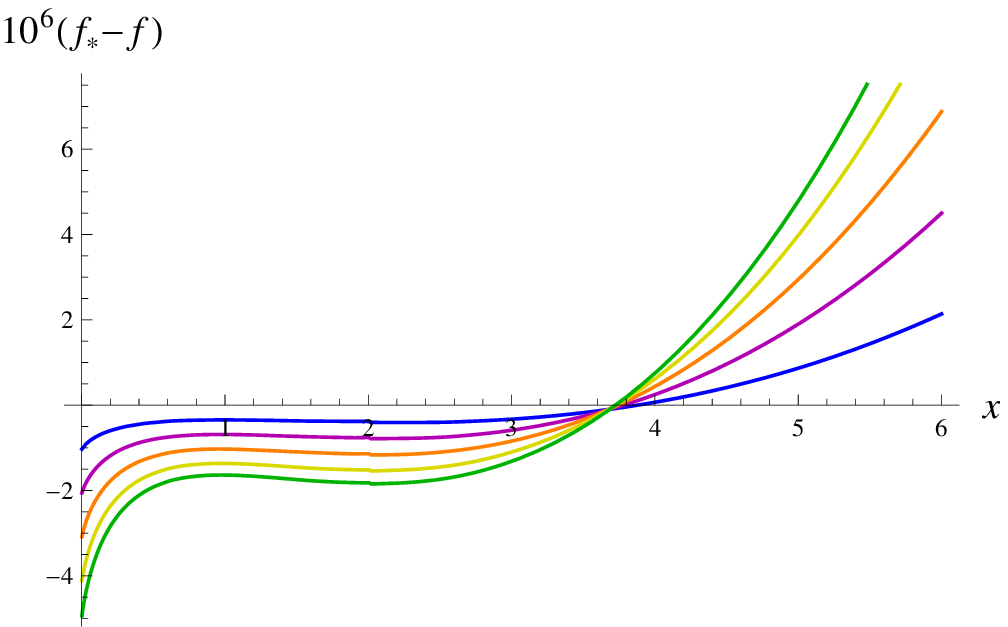}
 \includegraphics[width=0.4\textwidth]{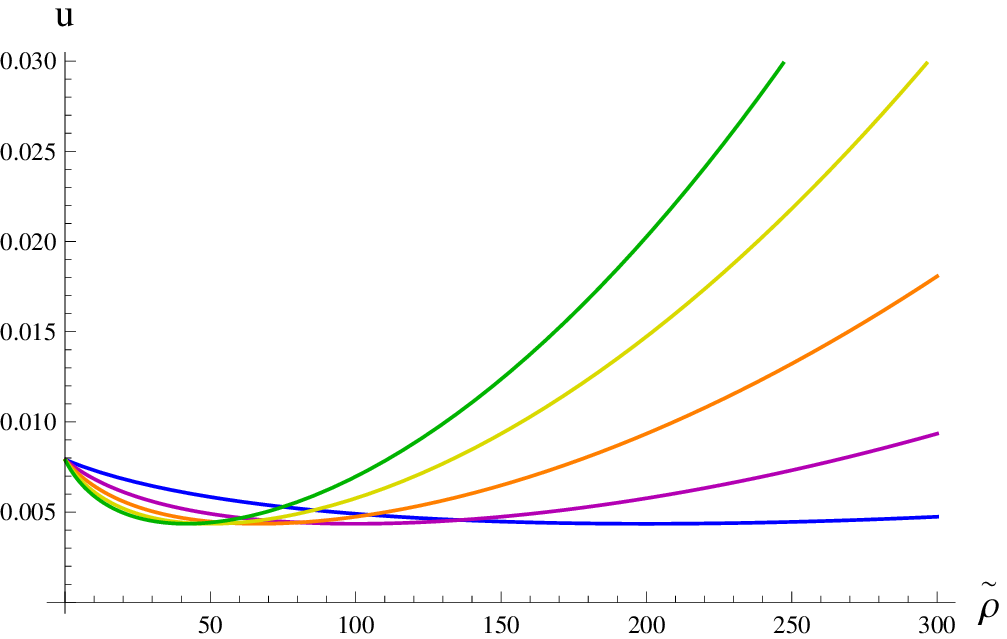}
 \caption{Upper panel: the difference between the \fgFPh\ potential
   $f_*(x)$ and the marginally-relevant potential $f(x)$ of the
   SU($N=2$) model in a weak-coupling analysis, both at $(P=1,\xi\simeq2\times 10^{-4})$, that is
   at fixed ${x_0}=2$, for increasing values of $g^2$ from blue to
   green (smaller to larger) $(g^2 \in \{0.01, 0.02, 0.03, 0.04,
   0.048\})$.  Lower panel: the corresponding conventional
   renormalized dimensionless potentials $u(\trho)$.  Both plots are
   produced ignoring $\eta_\phi$.  }
\label{fig:numerical_marginal_perturbation}
\end{center}
\end{figure}

\begin{figure}[!t]
\begin{center}
 \includegraphics[width=0.4\textwidth]{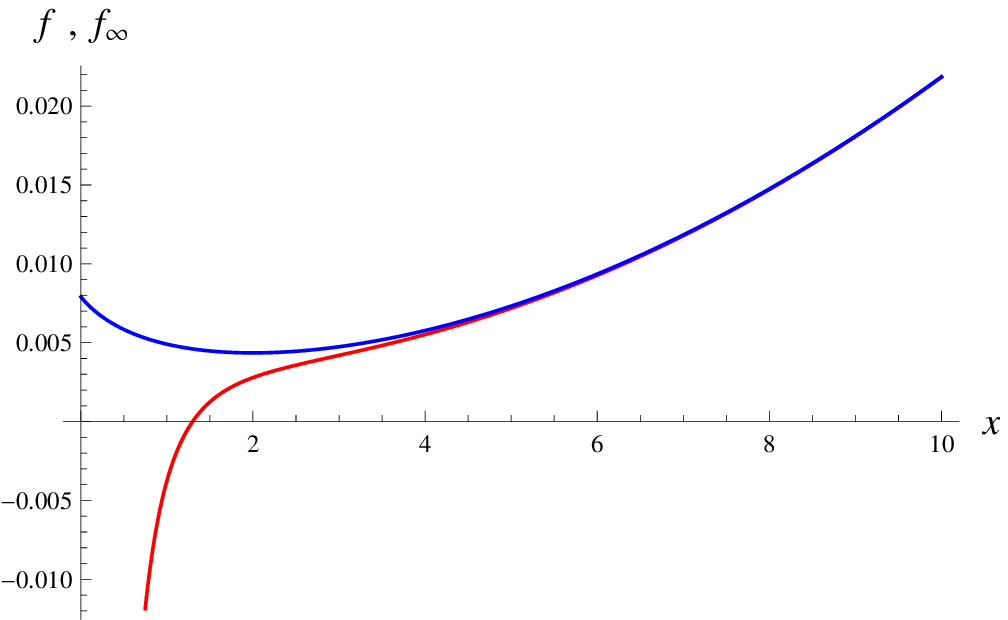}\hskip 1cm
 \includegraphics[width=0.4\textwidth]{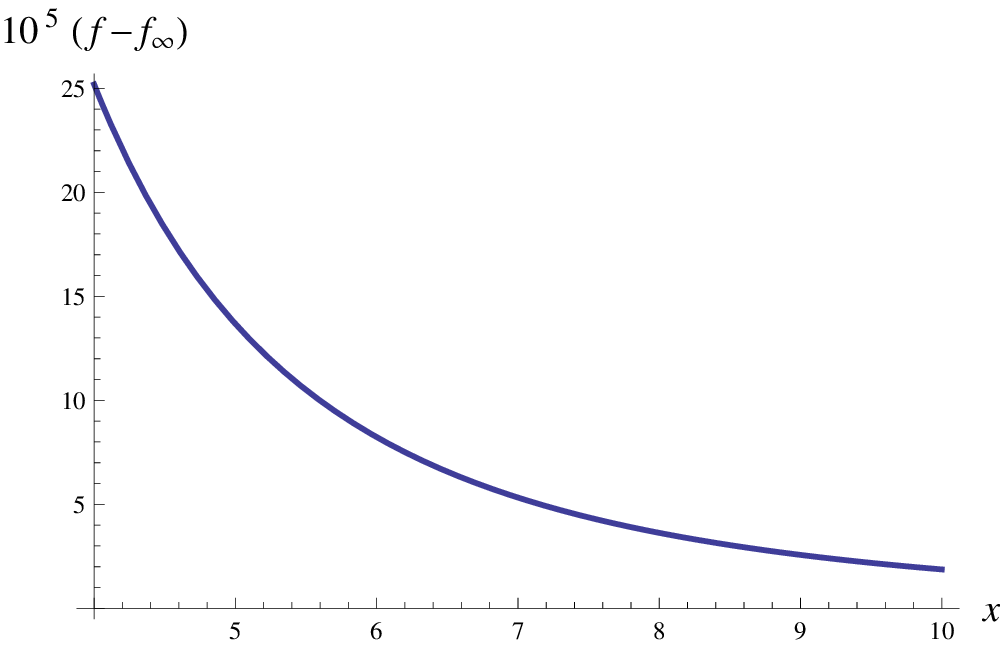}
 \caption{Upper panel: the numerical solution $f(x)$ (found by
   shooting from $x_0$) and the analytic large-field-asymptotic
   expansion result $f_{\infty}(x)$ of \Eqref{eq:fgFPcond} for the
   potential along the marginally-relevant trajectory of the $P=1$
   scaling solution of the SU($N=2$) model.  Both curves correspond to
   ${x_0}=2$ and $g^2=10^{-2}$.
   Lower panel: the difference between the two very same functions.  Both
   plots are produced ignoring $\eta_\phi$. }
\label{fig:marginal_and_asymptotics}
\end{center}
\end{figure}

Let us now continue the analysis of the $g^2$-dependence of $f(x)$
along the marginally relevant asymptotically free trajectory by
numerical means.  By neglecting $\eta_\phi$ for simplicity, we solve
the ODE \Eqref{eq:fgFPcond} numerically by shooting both from $x_0$
and from large field.  In the former case, $x_0$ can be used as a free
parameter labeling the asymptotically free trajectories, like $\xi$ at
$g^2=0$.  One of the boundary conditions is provided by the
requirement $ f'( {x_0})=0$, while the other one can be parametrized
by the value of $f''( {x_0})$.  We know that the \fgFPh\ condition has
exactly a one-parameter family of solutions for $g^2=0$.  Hence we
expect that this remains true also for any fixed value of $g^2\neq0$,
as we already observed for $P<1$. A closer global inspection of the
full \fgFPh\ equation performed below provides evidence for the
existence of at least one solution corresponding to a value of $ f''(
{x_0})$ that, for small $g^2$, is close to the \fgFPh\ value.  Such a
solution to \Eqref{eq:fgFPcond} is plotted in
Fig.~\ref{fig:numerical_marginal_perturbation} (lower panel) with the
asymptotic approach to the \fgFP\ depicted in more detail in the upper
panel.  Figure \ref{fig:marginal_and_asymptotics} shows a comparison
to an analytical large-field asymptotics $f_\infty(x)\sim
x^{2+\mathcal{O}(g^2)}$ determined from \Eqref{eq:fgFPcond},
and given in Eqs.~(\ref{eq:weak_g2_asymptotics_1}),~(\ref{eq:weak_g2_asymptotics_2}).
%

\subsubsection{$(P\!>\!1)$-scaling solutions}
\label{subsec:P>1_potentials}

The lowest-order analytic approximation of the polynomial beta
functions for $1<P<2$, as presented in
Sec.~\ref{subsec:P>1_with_relevant}, was obtained by expanding to
linear order in $g^{2(P-1)}$, and neglecting any other
$g^2$-dependence.  If we do the same with the beta function of the
scalar potential, the leading $g^2$-dependent term in $\beta_f$ is
$9g^{2(P-1)}/(16\pi^2 x)$  and by integrating the \fgFPh\ 
condition we obtain the following effective potential
\begin{equation}
f(x)=\xi x^2+g^{2(P-1)}\frac{3}{32\pi^2 x}\ . \label{eq:fppotentialPg1}
\end{equation}
Clearly this potential has a nontrivial minimum for nonvanishing
$g^2$, and we can easily extract the polynomial couplings at this minimum,
which are $\xi_2=6\xi$ and
\begin{eqnarray}
x_0&=&g^{2(P-1)/3}\ \frac{1}{2\xi_2^{1/3}}\left(\frac{3}{2\pi}\right)^{2/3}\\
\xi_3&=&-g^{-2(P-1)/3}\ 4\left(\frac{2\pi\xi_2^2}{3}\right)^{2/3} \label{eq:x0Pg1}
\end{eqnarray}
and perfectly agree with Eqs.~\eqref{eq:xoFPPg1} and
\eqref{eq:xi2FPPg1} found in Sec.~\ref{subsec:P>1_with_relevant}.
Furthermore, one can also deduce that higher derivatives at the
minimum have alternating signs and scale like $\xi_n\sim
g^{-2(n-2)(P-1)/3}$, that is $\lambda_n\sim g^{2n(2P+1)/3+4(P-1)/3}$.
Notice that this potential is singular at $x=0$, which does not
inhibit to extract physical interaction vertices at the nontrivial
minimum.  Nevertheless, analyzing the origin of this pole helps us to
get a better approximation of the full potential.

The singular term in the beta function is produced by expanding the
contribution from gauge loops (2nd line of \Eqref{floweq:fpotential})
in powers of $g^{2(P-1)}$.  In fact, a general feature of $P>1$ is the
appearance of a negative power of $g^2$ in the beta function of the
potential, such that an expansion in powers of $g^{2(P-1)}$ is the same
as expanding in $1/x$.  Yet, the gauge loop contribution has finite
$g^2\to 0$ and $x\to 0$ limits.  This suggests that we should keep the
whole $g^{2(P-1)}$-dependence of the beta function, that is the whole
gauge loop, which for the linear regulator reads
\begin{equation}
\delta \beta_{f}(x)=\frac{3(8g^{2(P-1)}+x)}{32\pi^2 (2g^{2(P-1)}+x)}\ . \label{eq:deltabetaPg1}
\end{equation}
In this way, \Eqref{eq:fgFPcond} leads to the following
Coleman-Weinberg-like potential
\begin{equation}
f(x)= \xi x^2-\left(\frac{3}{16 \pi}\right)^2 \left[\frac{2x}{g^{2P_\mathrm L}}+
\frac{x^2}{g^{4P_\mathrm L}}  \log\left(\frac{x}{2g^{2P_\mathrm L}+x}\right) \right]
\label{eq:fppotentialPg1gl}
\end{equation}
where $P_\mathrm L=P-1$.  This time the potential and its first
derivative have a finite $x\to 0$ limit, but the higher derivatives
become singular.  Also, as for the simpler parametrization obtained
by Taylor expansion in $g^{2(P-1)}$, the $g^2\to 0$ and $x\to 0$
limits do not commute.  In fact, such singularities of the
higher-derivatives are generic for Coleman-Weinberg potentials for
$x\to 0$, while the physical correlation functions extracted by taking
derivatives at the nontrivial minimum stay finite.

\begin{figure}[!t]
\begin{center}
 \includegraphics[width=0.4\textwidth]{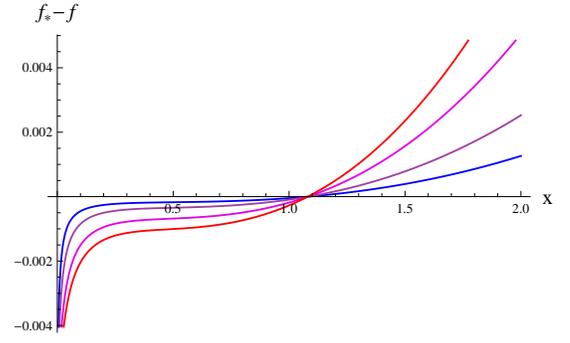}
 \caption{
The difference between the zeroth order approximation $f_*$ of \Eqref{eq:FPpotential} and the finer 
approximation $f$ of \Eqref{eq:fpPg1eq5} for the
scalar potential along the marginally relevant trajectory with $P=2$ and $\xi=1$.
Here $g^2\in\{5\ 10^{-3}, 10^{-2}, 2\ 10^{-2}, 3\ 10^{-2}\}$ from
blue (flatter) to red (steeper).
}
\label{fig:P3_FP}
\end{center}
\end{figure}

Following Sec.~\ref{subsec:P>1_with_relevant}, we expect that a linear
term in $g^2$ becomes at least as important as the contribution of the
gauge loops from $P=2$ on.  The only piece of $\beta_f$ which is
linear in $g^2$, comes from the $\eta_\mathrm W$ entering in the
canonical scaling term of the flow equation for $f$, as a consequence
of the rescaling of the field with a power of $g^2$.  Therefore, the
inclusion of this just term results in a shift of the dimensionality of the
SU$(2)$ invariant $x$ from $2$ to
\begin{equation}
d_{x}=2+g^2\frac{11\, P}{24\pi^2}\ . \label{eq:dx}
\end{equation}
Let's address first the case $P>2$, where $g^2$ is the leading
power. If we retain only this term, the solution of \Eqref{eq:fgFPcond} is
simple,
\begin{equation*}
f(x)= \xi x^{4/d_x},
\end{equation*}
exhibiting again a logarithmic singularity at the origin.  In this
leading-power approximation, the nontrivial minimum found by the
effective-field theory method is not visible.  Indeed, in
Sec.~\ref{subsec:P>1_with_relevant} we already observed that by
considering only the linear terms in $g^2$ we were not able to
describe the full \fgFPs\ for $P>2$ without the next-to-leading term
$\sim g^{2(P-1)}$.  In the present analysis, this is tantamount to considering
\begin{equation}
\delta \beta_{f}(x)=\frac{9}{16\pi^2 x}
+g^2\frac{11\, P}{24\pi^2}x f'(x) \label{eq:fpPg1eq1}
\end{equation}
whose integration leads to
\begin{equation*}
f(x)= \xi x^{4/d_x}+g^{2(P-1)}\frac{9}{16 \pi^2 (4+d_x) x}. \label{eq:fpPg1eq2}
\end{equation*}
Expanding this form to linear order in $g^2$ and $g^{2(P-1)}$, we find
\begin{equation}
f(x)= \xi x^{2}+g^{2(P-1)}\frac{3}{32 \pi^2 x}-g^2 \xi \frac{11\,
  P}{24\pi^2} x^2 \log(x)\ . \label{eq:fpPg1eq3}
\end{equation}
By means of the latter expression, we can again extract the
corresponding polynomial couplings upon expansion about the minimum
$x_0$.  There are at least two minima for $x_0$. One of them is
independent of $\xi$ and diverges as the gauge coupling vanishes,
since it is related to the two terms multiplied by $\xi$ in the
previous expression.  The other one is instead well described by
dropping the last logarithmic term for small gauge coupling, and thus
retaining the same simple potential that we found for $1<P<2$. The
fact the the asymptotic description of the scalar interactions in the
two intervals $1<P<2$ and $P\geq 2$ is the same, matches with what we
have already found in Sec.~\ref{subsec:P>1_with_relevant}.  Again, for
a more accurate description of the potential, we can retain
the complete gauge loop, yielding
\begin{equation}
\delta \beta_{f}(x)=\frac{3(8g^{2(P-1)}+x)}{32\pi^2 (2g^{2(P-1)}+x)}
+g^2\frac{11\, P}{24\pi^2}x f'(x)\ . \label{eq:fpPg1eq4}
\end{equation}
Thus, the equation defining $f$ becomes the same as for $1<P<2$, apart
for the effective change in field-dimensionality.  Then the solution
then takes the more general form
\begin{eqnarray}
f(x)&=& \xi x^{4/d_x} -2\left(\frac{3}{16 \pi}\right)^2 \left[ 
\frac{4\pi/d_x}{\sin(4\pi/d_x)}\left(\frac{x}{2 g^{2(P-1)}}\right)^{4/d_x} \right.\nonumber\\
& & -\left. {}_2 F_1\!\left(1;-\frac{4}{d_x};1-\frac{4}{d_x}, -\frac{x}{2g^{2(P-1)}}\right)\right]
\label{eq:fpPg1eq5}
\end{eqnarray}
Upon expansion to linear order in $g^{2(P-1)}$, we recover the
simplified expressions discussed before. The approach of
\Eqref{eq:fpPg1eq5} to the Gau\ss{}ian fixed point written as the
zeroth order result $f_*$ for the \fgFPh\ condition,
cf. \Eqref{eq:FPpotential}, is shown in Fig.~\ref{fig:P3_FP} for the
case $P=2$, $\xi=1$ and various values of the gauge coupling.

%

\subsection{Linearized flow near the Gau\ss ian fixed point}
\label{sec:linearized_flow}

The standard way of characterizing the properties of the RG flow in
the neighborhood of a fixed point is through linearization with
respect to the displacement from the fixed point.  Here we discuss how
the analysis of such a linearized flow of the dimensionless potential
$u(\tilde\rho)$ relates to the construction of scaling solutions for
$f(x)$.  It should first be emphasized that while a fixed-point
condition for $u(\tilde\rho)$ differs from one for $f(x)$
because of the supplementary boundary conditions, yielding
different scaling phenomena, the information encoded in the
differential equation, i.e. the flow equations, for $u$ and $f$ is one
and the same.  Hence, the UV asymptotics of the flows of the potential
$f(x)$ should be inside the flow equation for $u(\tilde\rho)$ in
vicinity of the Gau\ss ian fixed point.  Making the meaning of 'vicinity' more precise
requires a discussion of subtleties.

The analysis of the stability properties of the Gau\ss ian fixed point can be
reduced to an eigenvalue problem if one sets up an approximate
description of the RG flow in terms of a linear operator. The standard
construction proceeds by separating the RG-time dependence from the
field-dependence \cite{Morris:1998da}
\begin{eqnarray}
g^2(t)=\epsilon e^{-\theta t}\delta g^2\\
u(\tilde\rho ;t)=\epsilon e^{-\theta t}\delta u(\tilde\rho) \label{eq:lin1}
\end{eqnarray}
i.e. considering $\delta g^2$ and $\delta u(\tilde\rho)$ as
$t$-independent, and expanding the flow equations for $g^2$ and $u$ to
linear order in $\epsilon$.  This provides an eigenvalue system for
$\delta g^2$ and $\delta u$, where the eigenvalue $\theta$ controls
the RG evolution along the corresponding eigendirection, since
$\partial_t \left( g^2(t), u(\tilde\rho ;t)\right)=-\theta
\left( g^2(t), u(\tilde\rho ;t)\right)$.  As $\beta_{g^2}\propto g^4$,
eigenvectors with $\theta\neq 0$ must have $\delta g^2=0$ and
correspond to the usual purely-scalar Gau\ss ian eigenperturbations
$\delta u(\tilde\rho)= \tilde\rho^n$ with $\theta_n=4-2 n$,
\ $n\in\mathbb{N}$.  The $n=4$ case gives us the marginal $\phi^4$
interaction, which is marginally irrelevant in the standard
  analysis.  The other marginal direction descends from keeping
$\delta g^2\neq0$, and it corresponds to a scalar potential solving
the marginality condition
\begin{equation}
-\frac{9}{64\pi^2}\trho\delta g^2-4 \delta u(\tilde\rho) +\left(2\tilde{\rho}-\frac{1}{8\pi^2}\right)\delta u'(\tilde\rho)-\frac{\tilde\rho}{16\pi^2}\delta u''(\tilde\rho)=0, \label{eq:lin2}
\end{equation}
implying up to field-independent constants
\begin{equation}
\delta u(\tilde\rho)=\zeta\tilde{\rho}^2
-\left(\frac{48\zeta+18\delta g^2}{256\pi^2}\right)\tilde{\rho}, \label{eq:lin3} 
\end{equation}
where $\zeta$ is an integration constant.  The other integration
constant of this second order linear ODE needs to be set equal to zero
otherwise the potential would have an exponential growth for large
$\tilde\rho$, which violates self-similarity of the linearized RG
flow~\cite{Morris:1998da} (see the discussion in Sec.~\ref{subsec:P<1_potentials}). 
Here the only constraints on $\zeta$ come
from self-consistency of the linearized analysis, i.e. it should 
be sufficiently small, and from the stability of the corresponding potential,
i.e. it should be positive. At the level of the eigenvalue analysis
$\zeta$ is also $t$-independent by construction, on the same footing
as $\delta g^2$ and $\delta u$.

Beyond linearization, we already know that this perturbation cannot be
exactly marginal.  Indeed the eigenvalue problem simply illustrates
the local properties of the RG flow. By consistency the solution given
above should describe the leading asymptotics of the running scalar
potential along the marginally relevant trajectory governed by the
gauge coupling.  Hence, we can replace the infinitesimal displacements
from the fixed point $\delta u$ and $\delta g^2$ with their running
counterparts $u$ and $g^2$, and require that the corresponding
potential vanishes when $g^2$ approaches zero.  This is possible only
if $\zeta$ equals zero or if it is an appropriate function of $g^2$.
The first option is clearly not acceptable since it does not lead to
stable potentials as in the analysis of~\cite{Salam:1978dk}, and
furthermore it clashes with the freedom in the choice of $\zeta$
allowed by the linearized analysis.  On the other hand assuming
$\zeta$ to be a function of $g^2$ one can comply with all the
requirements listed so far, as long as $\zeta(g^2)>0$ and
$\lim_{g^2\to0}\zeta(g^2)=0$.

Since we are still confining ourselves to the UV asymptotics, we can
parametrize the small-$g^2$ behavior of this function by a simple
power law $\zeta(g^2)=\xi g^{4P}$, with $P>0$ and a sufficiently small
$\xi>0$.  For the consistency of the linearized analysis yielding
\Eqref{eq:lin3} for $\delta u$, we need to constrain $P\leq1/2$;
otherwise $u(\tilde\rho)$ would become of the same order of higher
powers of $g^2$ that have been neglected in the linearization of the
flow.  Indeed, by translating the above formula for the linearized
potential in terms of rescaled quantities $x=g^{2P}$ and
$f(x)=u(\tilde\rho)$ one obtains exactly the next-to-leading
parametrizations of the $(P\leq1/2)$-scaling solutions presented in
Sec.~\ref{subsec:P<1_potentials}.  The $P>1/2$ scaling solutions are
beyond the scope of the present conventional linearized analysis,
since one needs to consistently take into account higher powers of
$g^2$ inside $\beta_u$.  Unfortunately, if one sticks to the variables
$\tilde\rho$ and $u(\tilde\rho)$ the next-to-leading power of $g^2$
appearing inside $\beta_u$ is $g^4$ and this comes together with
nonlinearities and more explicit field-dependencies. This renders the
$\beta$ function too complicated for an analytic functional treatment.
However, if $u(\tilde\rho)$, its minimum and its derivatives scale
like definite powers of $g^2$, one does not need to take into account
all of the $O(g^4)$-terms inside $\beta_u$, and an intermediate
approximation order becomes available between $g^2$ and $g^4$.  This
is precisely what the rescaling in the definition of $x$ and $f(x)$
accounts for.

\subsection{Global analysis of the \fgFPh\ potentials}
\label{sec:beyond_polynomial_truncations_numerical}

An analysis of the flow for the full potential $f(x)$ (or $u(\trho)$)
requires to consider the global properties of this partial
differential equation in field space. So far, we have performed this
analysis either in an effective-field-theory spirit (polynomial
expansion in $x$ or $\trho$) or in terms of a weak-coupling expansion,
each coming with its own limitations. In the following, we improve on
both by considering the functional \fgFPh\ condition $\partial_t
f(x)=0$ at a given finite value of $g^2$. As argued in general above
and verified explicitly in various limits, the \fgFPh\ condition
describes a locus of points in theory space which gets closer and
closer to the true asymptotically free trajectory for small gauge
coupling. For increasing coupling, the solution to $\partial_t f(x)=0$
as a function of this coupling $g^2$ may still represent an acceptable
approximation of the true RG trajectory. The technical advantage of
solving the \fgFPh\ condition clearly is that it corresponds to an ODE
instead of the PDE of the full flow in $x$ and $t$.

From the global viewpoint in field space, the expansions so far are
not only expected to have a limited range of validity. By expanding
the beta functional of $f(x)$ in powers of $g^2$ or $x$, we also
modify the nonlinear structure of the flow equation, most importantly
removing movable or fixed singularities. Indeed, these singularities
of the flow equation are known to lead to rather selective criteria
for the existence and physical eligibility of fixed-point solutions
\cite{Hasenfratz:1985dm,Morris:1998da,Bervillier:2007tc,Codello:2012ec,Demmel:2015oqa,Dietz:2015owa,Vacca:2015nta,Hellwig:2015woa,Borchardt:2015rxa,Litim:2016hlb}. Hence, the results obtained so far still have to
pass the test of global existence and eligibility, as the expansions
used so far might decisively change the behavior of $f$ for large
fields or close to the origin. For a significant test, we need to
maintain the singularity structure of the flow equations, and
construct again \fgFPh\ potentials $f(x)$.

For the following numerical analysis, we neglect the dependence of the
anomalous dimensions on the expectation value of the scalar
$\sim\kappa$. This corresponds to dropping the scalar loops in the
expressions of the anomalous dimensions and keeping only the gauge
loops; (technically, this corresponds to setting $x_0=0$ or $\kappa=0$
inside $\eta_W$ and $\eta_\phi$).  For $P>1/2$ this is fully justified
at weak coupling as the terms dropped are of higher order than the
leading order $\sim g^2$. For $P< 1/2$, the expectation value
$\kappa=3/(32\pi^2)$ is numerically small, also justifying the
approximation. The latter also holds for $P=1/2$ except for
non-generically small values of $\xi_2\ll 1$ which we exclude from the
following analysis. In summary, this approximation is quantitatively
well justified, and, most importantly, does not significantly alter
the functional structure of the flow equation.  The remaining analysis of
the solutions still requires numerical methods, since the \fgFPh\ equation
is of second order and highly nonlinear.

\subsubsection{Shooting from the minimum}
\label{sec:shooting_ from_minimum}

Let us start by numerically constructing the solution to the
\fgFPh\ equation $\partial_t f=0$ for fixed values of $g^2$ first by
shooting from the minimum. For a given value of $P$, we expect the
existence of a one parameter family of solutions labeled, for
instance, by the parameter $\xi$ above. In fact, this problem at first
sight seems very similar to that of constructing fixed point solutions
for the scalar potential at Wilson-Fisher-type fixed points. We
review the latter for completeness in App.~\ref{app:WF}, where we
demonstrate that the techniques used in the following can reproduce
known results to a high accuracy. The \fgFP\ condition can be
rewritten into a conventional ODE form,
\begin{eqnarray}
f''&=& -\frac{1}{64\pi^2} \frac{e(f,f';x)}{x\, s(f,f';x)}, 
\label{eq:ffnd}\\
&& s(f,f';x)=-4f+(2+\eta_\phi-P\eta_W) x f',
\end{eqnarray}
where the numerator function $e$ depends nonlinearly on its arguments
and can be straightforwardly computed from
\Eqref{floweq:fpotential}. Both, numerator $e$ and denominator
function $s$ depend on $g^2$. We integrate the ODE by starting from
the minimum $x=x_0$ towards both smaller values (shooting inwards
$x\to 0$) and larger values (shooting outwards $x\to \infty$). The
initial conditions,
\begin{equation}
f'(x=x_0)=0, \,\,\, \text{and}\,\, \sigma:=f''(x=x_0), \label{eq:init}
\end{equation}
a priori form a 2-parameter set $(x_0,\sigma)$ of solutions. As in the
well-known Wilson-Fisher case (App.~\ref{app:WF}), the right-hand side
of \Eqref{eq:ffnd} has both a fixed singularity at $x=0$, as well as a
movable singularity at the point $x$, where $s(f,f';x)=0$. Shooting
outwards, the movable singularity is generically hit at some value
$x=x_{\text{s+}}>x_0$, $s(f,f';x_{\text{s+}})=0$. The existence of a
regular solution requires to fix one of the parameters, e.g.,
$\sigma$, such that also the numerator vanishes
$e(f,f';x_{\text{s+}})=0$ and the solution can go in a regular fashion
through $x_{\text{s+}}$ and be continued to $x\to\infty$.

In the Wilson-Fisher case, the fixed singularity at $x=0$ implying the
condition $e(f,f';x=0)=0$ also fixes the second parameter and thus
only a discrete set of solutions exists, yielding a quantization of
both physical fixed points as well as critical exponents. As discussed
in App.~\ref{app:WF}, this solution can be found by tuning the
remaining parameter, say $x_0$, to its critical value. 

This is precisely the point, where the present nonabelian Higgs model
in $d=4$ differs from the pure scalar case (in $d=3$): we have
numerically not been able to satisfy the condition $e(f,f';x=0)=0$ for
curing the fixed singularity at $x=0$ for a wide range of numerically
explored parameter regions $(x_0,\sigma)$. Whereas this seems like a
reason for concern, it comes actually not unexpectedly, as the
weak-coupling solutions studied above also partly exhibited
singularities in the higher-derivatives of $f(x)$. In fact, we find
that the potential $f(x)$ and its first derivative $f'(x)$ stay
numerically well-behaved down to rather small values of $x$, see
Fig.~\ref{fig:large_VS_minimum_shooting} . Very close to
$x=0$, the fixed singularity in $f''(x)$ eventually contaminates
standard integration algorithms and the numerical integration
stops. This behavior of a stable solution for $f$ and $f'$ and a
divergence in higher-derivatives is rather independent of the initial
conditions at $x=x_0$. Again, this is in contrast to the Wilson-Fisher
case, where singularities in higher derivatives as well as the
potential itself appear and depend strongly on initial conditions if
the criterion $e(f,f';x=0)=0$ is missed.

\begin{figure}[!t]
\begin{center}
 \includegraphics[width=0.4\textwidth]{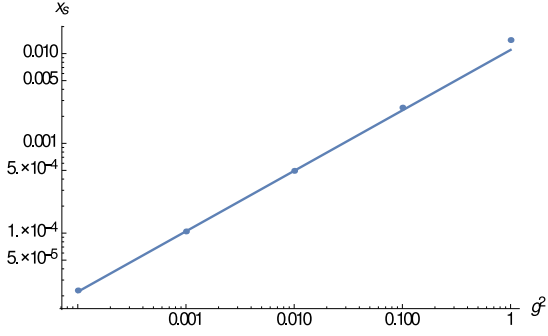}
 \caption{Width of the singular regime induced by the fixed
   singularity at $x=0$ as a function of $g^2$. The width is estimated
   by the criterion $f'''(x_{\text{s--}})=1$. For $P=1/2$, the full
   numerical data (points) are well approximated by a power law,
   $x_{\text{s--}} \sim (g^2)^{\alpha}$ with $\alpha\simeq2/3$ (line),
   suggesting that the \fgFPh\ potential approaches asymptotic freedom
   globally on field space.}
\label{fig:xs}
\end{center}
\end{figure}

Still, singularities also in higher derivatives would represent a
problem if they persisted in the limit $g^2\to 0$, as this could
contradict asymptotic freedom. In order to test this, we have
determined the onset of the singularity near $x\simeq0$ in the
numerical solution as a function of $g^2$. As a criterion, we have
computed the position $x_{\text{s--}}$ where $f'''(x_{\text{s--}})=1$,
which is still in the region where $f(x)$ and $f'(x)$ behave regularly
and slowly varying, whereas the higher derivatives start to vary
rapidly. An estimate of $x_{\text{s--}}$ is shown in Fig.~\ref{fig:xs}
for the case $P=1/2$ for a wide range of $g^2$ values. A fit to the
data at small coupling confirms that the singular regime $x\lesssim
x_{\text{s--}}$ shrinks to zero for $g^2\to 0$. Numerically, we find
$x_{\text{s--}} \sim (g^2)^{\alpha}$ with $\alpha\simeq2/3$ for $P=1/2$. A similar behavior is
observed also for other values of $P$.

\begin{figure}[!t]
	\begin{center}
		\includegraphics[width=0.4\textwidth]{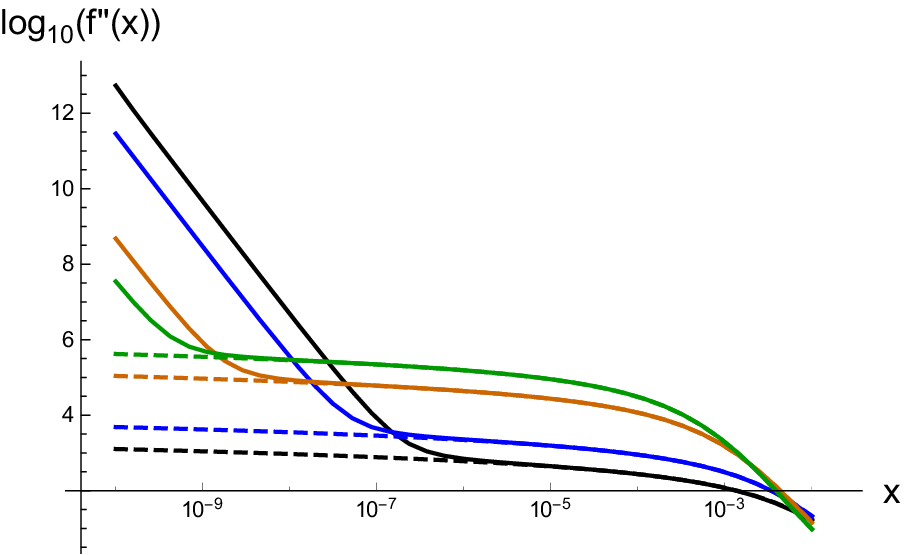}
		\includegraphics[width=0.4\textwidth]{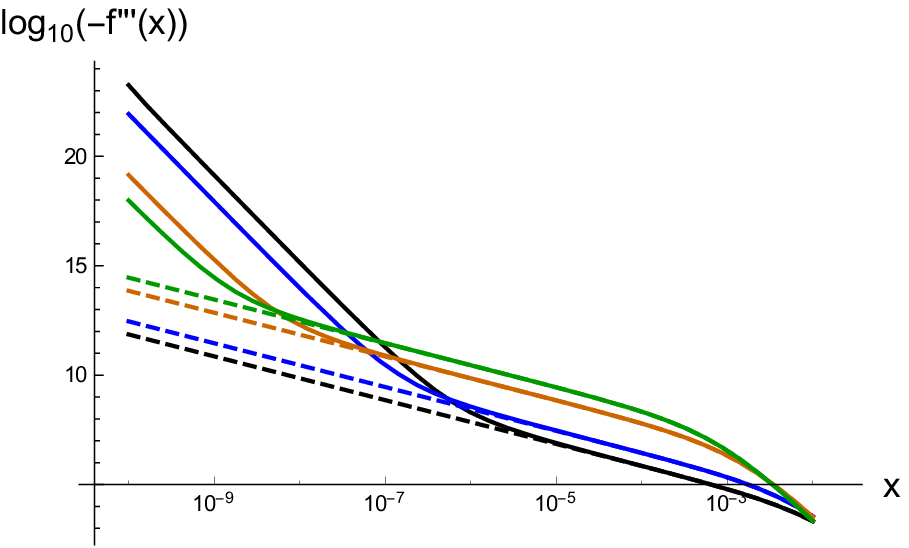}
		
		\caption{
			$P=2$. 
			Comparison between the numerical solution (full curves) from shooting
			from the minimum, and the analytic approximation 
			in~\Eqref{eq:fpPg1eq5} (dashed curves), 
			for $\xi=1$ and  
			$g^2\in\{10^{-2}, 5\ 10^{-3}, 10^{-3}, 5\ 10^{-4}\}$ from
			the lower-right (black) to the upper-right (green) curves. 
		}
		\label{fig:diverging_at_zero}
	\end{center}
\end{figure}

In fact, for $P\geq1$ the description of this singular behavior close
to $x=0$ seems within reach of the weak-coupling expansion, because
the corresponding \fgFPh\ potentials exhibit logarithmic structures,
leading to divergences in their higher order derivatives.  It is thus
interesting to compare these features to those of the full numerical
solution.  Since the small field behavior is only weakly sensible to
the position of the movable singularities at $x>x_0$, we do not need
to fine-tune the parameters $(x_0,\xi_2)$ to their \fgFPh\ values as
boundary conditions of the numerical integration.  For definiteness,
we insert values according to the leading weak-gauge-coupling \fgFPh\
relation of~\Eqref{eq:summary_x0_of_xi2}.  For instance, at $P=2$ we
can use the approximation in~\Eqref{eq:fpPg1eq5}, and relate
$\xi_2=6\xi$ as in the \fgFPh\ case, even if $g^2$ is not tiny.  The
comparison between this analytic form and the numerical solution is
shown in Fig.~\ref{fig:diverging_at_zero}, where a discrepancy
is visible in a neighborhood of the origin. This failure of the
weak-coupling expansion is reasonable, since it relies on the 
assumption that the argument of the gauge-thresholds $g^{2(1-P)}x$ is
big, which excludes values of $x$ much smaller than $g^{2(P-1)}$.
Indeed, these plots show that this region progressively shrinks as
$g^2$ decreases, such that the range of applicability of the analytic
approximation gets larger.

We take these numerical studies as evidence that global solutions to
the \fgFPh\ equation exist. These solutions form a one-parameter
family for each value of $P$ which can be parametrized, e.g., by the
position of the minimum $x_0$ (alternatively, this could be rephrased
in terms of a parameter $\xi$ as before). These solutions develop
singularities in their higher derivatives near $x=0$. This singular
regime vanishes with $g^2\to0$ such that the \fgFPh\ potential
approaches asymptotic freedom globally on field space.

\begin{figure}[!t]
\begin{center}
 \includegraphics[width=0.47\textwidth]{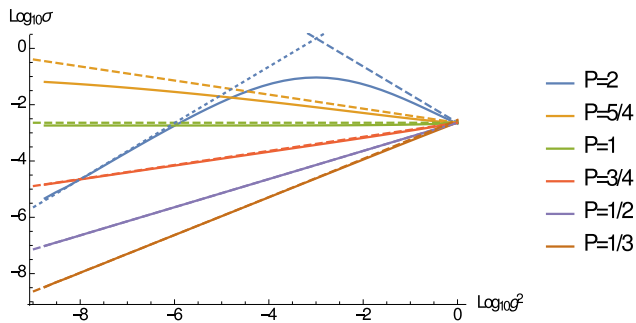}
 \caption{Comparison of the scaling of the critical value of the
   initial condition $\sigma$ from outward shooting for the gauge
   coupling $g^2$ for a given fixed value of $x_0=0.1$ and various
   values of $P$. The full numerical data (solid lines) are compared
   to the effective-field-theory estimates after identifying $\sigma$
   with the polynomial expansion coefficient $\sigma=\xi_2$. The
   dashed lines show the estimates of \Eqref{eq:compb} (for
   $P=1/2,3/4,1,5/4,2$) and \Eqref{eq:compc} (for $P=1/3$); the dotted
   line shows the estimate of \Eqref{eq:compa} for $P=2$ (the
   corresponding line for $P=5/4$ is outside the plot region).}
\label{fig:comp}
\end{center}
\end{figure}

As a further check, we can compare these numerical findings with the
approximate results obtained in the effective-field-theory setting in
the weak-coupling limit. If the numerical solutions obtained by
shooting correspond to those of the effective-field-theory analysis,
the minimum $x_0$ and the critical value for $\sigma$ fixed by outward
shooting should be tightly related for a given value of the gauge
coupling $g^2$. A comparison with the expansion of the potential in
\Eqref{eq:Uofphiexp} implies to identify $\sigma$ with the expansion
coefficient of $f(x)$ at the minimum, $\sigma\equiv\xi_2$. From the
effective-field-theory analysis, we hence expect in the weak-coupling limit
\begin{eqnarray}
\sigma&=&\frac{9}{32\pi^2} \frac{g^{2(P-1)}}{x_0^3}, \quad \text{for}\,\, P>1, \label{eq:compa}\\
\sigma&=&\frac{9}{128\pi^2} \frac{g^{2(1-P)}}{x_0}, \quad \text{for}\,\, 1/2<P<1, \label{eq:compb}\\
\sigma&=&\frac{9}{128\pi^2} \frac{g^{2(P-1)}}{x_0-\frac{3}{32\pi^2} g^{2P}}, \quad \text{for}\,\, P<1/2, \label{eq:compc}
\end{eqnarray}
where \Eqref{eq:compa} summarizes Eqs.~\eqref{eq:xoFPPg1} and
\eqref{eq:xi2FPPg1}, \Eqref{eq:compb} follows from \Eqref{eq:FPP1},
and \Eqref{eq:compc} arises from an analogous analysis of
\Eqref{eq:x0Pless05} if the next-to-leading order $\sim g^{2(P-1)}$ is
included. The result of the comparison of our numerical studies with
these analytical forms is shown in Fig.~\ref{fig:comp} as a function
of $g^2$ for fixed $x_0=0.1$ for various values of $P$. The analytical
estimates Eqs.~\eqref{eq:compb} and \eqref{eq:compc} are shown as
dashed lines and agree with the full numerical curves for all the
depicted values of $P\leq1$ and small values of $g^2$. Interestingly,
\Eqref{eq:compb} also approximates the full numerical solution for
$P\gtrsim 1$ for $g^2\lesssim 1$, while the proper analytic form
\Eqref{eq:compa} (dotted line) takes over for sufficiently small,
$g^2\ll 1$ (for $P=5/4$, this asymptotics sets in for only much
smaller values of $g^2$ and thus is not visible in this plot). 
We
conclude that the numerical analysis of the full \fgFPh\ equation
agrees very well with the effective-field-theory analysis in the
corresponding validity limit, lending support to the existence of
asymptotically free trajectories in the nonabelian Higgs model.

%

\subsubsection{Shooting from large fields}
\label{sec:shooting_ from_infinity}

Another option to extract information from the \fgFPh\ condition
$\partial_t f=0$ beyond the weak-field, weak-coupling expansion is to
compute the large-$x$ asymptotic expansion of the \fgFPh\ potential. This
can be done for any $P$ and without neglecting subleading powers of
$g^2$; in fact, no expansion in $g^2$ is involved in the following.
The result is given in Eqs.~(\ref{eq:general_asymptotics_large_x})
and~(\ref{eq:asexp1bis}).
For the present discussion it suffices to refer to the leading term of
this expansion, which is 
\begin{align}
f(x)&\widesim[1.8]{x\to \infty}f_{\infty}(x) 
= \xi_\infty\  x^{N_\infty}+\dots\\
N_{\infty}&=4/d_x\ ,\quad d_x=2+\eta_\phi-P\eta_W\label{eq:asexp1}
\end{align}
where $\xi_\infty$ is an arbitrary integration constant, while the
subleading terms depend on $P$, $g^2$, $\eta_\phi$, $\eta_W$ and
$\xi_\infty$ itself. 
Since $N_\infty$ is always positive the scalar potential is stable as
long as $\xi_{\infty}\geq 0$.

For the following analysis, we can again ignore the $x_0$ dependence
of the anomalous dimensions. This is consistent with the large-field
expansion, as the anomalous dimensions depend only on $x_0$ which is
considered as much smaller than the expansion point $x$, $x_0\lll x$.
Remarkably, in the approximation of neglecting the contribution of
scalar loops to the anomalous dimensions, the latter are both negative
and controlled by $g^2$. This allows for the case
$0<P<\eta_\phi/\eta_W$ which corresponds to $d_x<2$ and $N_\infty>2$.
Yet, already in the one-loop approximation such range of $P$ values is
quite restricted, being $P<27/86$.  The contribution of the scalar
loops to $\eta_\phi$ are positive and are expected to further restrict
this bound.  Quite in general, instead, AF requires $d_x$ to be a
monotonically increasing function of $P$, such that by choosing $P$
arbitrarily large we can make $N_\infty$ arbitrarily close to zero.

The analytic large-field asymptotic expansion (and its first derivative) can be used
as a boundary condition at a large field value $x=x_M$,
for numerical integration of the \fgFPh\ equation towards the origin~\cite{Morris:1994ki}.
This procedure can be called `shooting', since it involves an arbitrary
coefficient $\xi_\infty$. The criterion for fixing this free parameter
is usually provided by requirements on the behavior of the solutions 
at $x=0$, which are absent in the present context. 
Hence, we expect to be able to construct a one-parameter family of
solutions labeled by $\xi_\infty$, which can then be related to the
position of the nontrivial minimum $x_0$ or to the curvature $\xi_2$.
For consistency, one needs to make sure that the chosen $x_M$ is big
enough, for instance, by checking that the values of $x_0$, $\xi_2$, $\xi_3\dots$
are stable against an increase of $x_M$. 
This procedure can be repeated for any $g^2$, and $x_M$ must be
adjusted correspondingly.

Such an analysis is important for testing the stability of the
potential, for which a method capable of probing the large-field region is
mandatory.  For $P\geq 1$ the weak coupling expansion is one such
tool, since one can solve the ODE without requiring $x$ to be small.
This is because the only nontrivial explicit $x$-dependence of
$\beta_f$ appears in the gauge loop, through the combination
$g^{2(1-P)}x$, such that small $g^2$ and large $x$ play the same role.
For $P<1$ instead, the weak coupling expansion is not applicable when
$g^2$ is kept fixed and $x$ is taken arbitrarily large.  Also shooting
from the minimum is not helpful, since the corresponding solution
always hits a movable singularity for $x\gtrsim x_0$.

\begin{figure}[!t]
\begin{center}
 \includegraphics[width=0.4\textwidth]{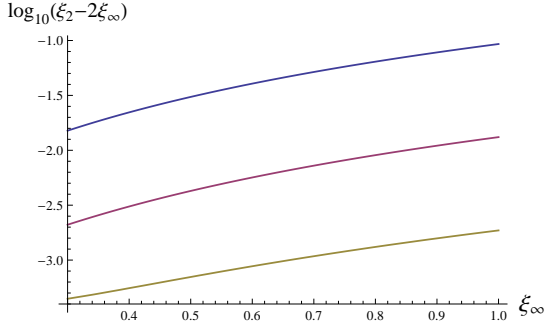}
 \caption{
$P=1/2$. 
Difference between $\xi_2$ as a function of $\xi_\infty$,
obtained from the large-field shooting,  and  $2\xi_\infty$.
The three curves refer to $g^2\in\{10^{-1}, 10^{-2}, 10^{-3}\}$ from
the upper (blue) to the lower (yellow) one.
}
\label{fig:Ponehalf_xi2_minus_2xiinfty}
\end{center}
\end{figure}

\begin{figure}[!t]
\begin{center}
 \includegraphics[width=0.4\textwidth]{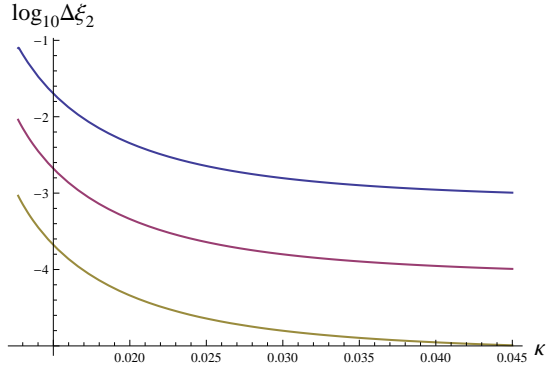}
 \caption{
$P=1/2$. 
Difference between the two functions $\xi_2(\kappa)$ (here $\kappa=x_0/g$) 
obtained from the large-field shooting and from the
leading-order weak coupling expansion, i.e. \Eqref{eq:summary_x0_of_xi2}. 
The three curves refer to $g^2\in\{10^{-1}, 10^{-2}, 10^{-3}\}$ from
the upper (blue) to the lower (yellow) one.
In the plotted interval of $x_0$ values, $\xi_2$ varies approximately from $0.05$ to $2$.
}
\label{fig:Ponehalf_deltaxi2_of_kappa}
\end{center}
\end{figure}

For definiteness, let us discuss in detail the $P=1/2$ case.
The corresponding weak-$g^2$ expansion in \Eqref{eq:fofxP<1/2}
provides two crucial relationships: one between the large-field behavior of the potential
($\xi_\infty=\xi$) and the curvature at the minimum ($\xi_2=2\xi$), the other
between this curvature and the position of the minimum.
These are fulfilled also by the present numerical solutions to an accuracy controlled by the gauge coupling,
 see Figs. \ref{fig:Ponehalf_xi2_minus_2xiinfty}
and  \ref{fig:Ponehalf_deltaxi2_of_kappa} respectively. 
In the $g^2\to 0$ limit,
the deviations
are roughly of the same order of magnitude as $g^2$.

\begin{figure}[!t]
	\begin{center}
		\includegraphics[width=0.4\textwidth]{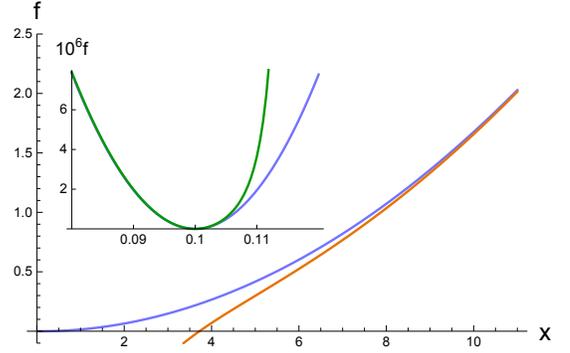}
		\caption{
			$P=1/2$, $g^2=0.3$, $x_0=0.1$. 
			Comparison between the numerical solution constructed by shooting from large fields (blue, upper) and the corresponding analytic asymptotic behavior (orange, lower).
			Inset: the numerical solution by shooting from the minimum (green, upper)
			and the one by shooting from large fields (blue, lower).
		}
		\label{fig:large_VS_minimum_shooting}
	\end{center}
\end{figure}

One can also compare the two solutions obtained by the two shooting procedures
from a large field value $x_M$ or from the minimum $x_0$.
The two solutions have 
different ranges of
applicability: while the former is accurate for the whole inner region $x<x_M$,
the latter is appropriate
for $x\lesssim x_0$, and quickly breaks down
towards larger values of $x>x_0$.
Hence, we compare the two solutions in the domain of applicability
of the latter.
Since we provide different kinds of boundary conditions in the two cases,
and since at non-tiny $g^2$ there is a small difference between
$\xi_\infty$ and $\xi_2$, we need to tune the parameters in such a way that
the produced solutions have the same value of $x_0$.
These are compared to each other and to the analytic large-field asymptotics
at a given value of $g^2$ in Fig.~\ref{fig:large_VS_minimum_shooting}.
This confirms the compatibility of the two approaches.

\subsubsection{Large-$N$ approximation}
\label{sec:largeN}

The matrix
$\mu_W^{2\ \ ij}(\tilde\rho)=\tilde{\phi}^\dagger\{T^i,T^j\}\tilde{\phi}$
can be diagonalized such that it reduces to
$\mu_W^{2\ \ ij}(\tilde\rho)=\mu_{W,i}^2(\rho)\delta^{ij}$.  For
SU($N$) and $\tilde{\phi}$ pointing into the $N^2-1$-direction
  in fundamental color space, one has
\begin{equation}
\mu_{W,i}^2=\tilde{\rho}\left\{ \begin{array}{ll}
             0 & \ \ \text{for}\ \  i\in[1,N(N-2)]\\
             1/2 &\ \  \text{for}\ \ i\in[(N-1)^2,N^2-2]\\
		1-1/N &\ \  \text{for}\ \ i=N^2-1
            \end{array}
            \right. .
\label{eq:mu_i_SU(N)}
\end{equation}
We have explicitly verified for various gauge groups that the
  spectrum does not depend on the direction of $\tilde{\phi}$ in the
  Cartan. The multiplicities in this spectrum scales with $N^2,N,N^0$
  respectively, such that the first line in \Eqref{eq:mu_i_SU(N)}
  seems to dominate in the large-$N$ limit. However, the combination
$T^i_{\hat{n}a}T^i_{a\hat{n}}$ that enters $\eta_\phi$ agrees with the
the Casimir invariant $C_{\text{F}}=(N^2-1)/(2N)\to N/2$ and thus is
in fact dominated by the second line of \Eqref{eq:mu_i_SU(N)}. In
order to extract the large-$N$ limit of the flow equations, a
rescaling is convenient
\begin{equation}
 \tilde\rho\to
N\tilde\rho\ , \quad u\to N u \ ,\quad g^2\to g^2/N. \label{eq:Nrescale}
\end{equation}
The flow equation in the $N\to\infty$ limit then  simplifies to
\begin{eqnarray}
\label{floweq:potential_largeN}
\partial_t u &=& -d u + (d-2 + \eta_\phi)\tilde\rho u' 
+4 v_d l_0^{(\mathrm{B})d}\left( u' \right) \nonumber\\
&&+4 v_d (d-1)l_{0}^{(\mathrm{G})d}\left(g^2 \tilde\rho/2\right)\\
\eta_{\phi}&=&\frac{8 v_d}{d}\Big\{(d-1) g^2 \Big[-l_{1,1}^{(\mathrm{BG})d}\left( u', g^2 \tilde\rho/2 \right)\nonumber\\
&&+g^2\tilde\rho \Big( a_1^d\left(g^2 \tilde\rho/2 \right) +\frac{1}{2} m_2^{(\mathrm{G})d}\left(g^2 \tilde\rho/2\right)\Big)
\Big]\nonumber\\
&&+2\tilde\rho u''^2 m_{2,2}^{(\mathrm{B})d}(u',u') \Big\}\Big|_{\tilde{\rho}=\tilde{\rho}_{\text{min}}}
\label{floweq:etaphi_largeN}
\end{eqnarray}
%
%
%
%
%
%
%
The dependence of $\eta_\phi$ on $\tilde{\rho}_{\text{min}}=\kappa$,
together with the general expectation that $\kappa\neq 0$ holds
because of the contribution of the gauge loops, makes the analysis of
this system harder than a corresponding large-$N$ limit for purely
scalar models or Yukawa models. To simplify the system, we use the
observation that the finite values of $\kappa$ exert a small
quantitative influence on the flows in the weak coupling regime. As in
the preceding analysis, we hence set $\kappa=0$ in both anomalous
dimensions $\eta_\phi$ and $\eta_W$. As a consequence, these anomalous
dimensions become simple functions of $g^2$ only. For instance, the
one-loop expressions are re-obtained to lowest order (including the
rescaling \eqref{eq:Nrescale})
\begin{equation}
\eta_\phi=-\frac{3}{16\pi^2}g^2\ , \quad \eta_W=-\frac{11}{24\pi^2}g^2\ .
\label{eq:Netas}
\end{equation}
For instance, the full expression for $\eta_\phi$ in the limit
$\kappa\to 0$ consists of only the first line
of~\Eqref{floweq:etaphi_largeN}, which most importantly contains the
$B'$ term in the one-loop beta-function of the quartic coupling
(cf. \Eqref{eq:1lbetal}) as well as its RG improvement due to the
presence of $\eta_\phi$ and $\eta_W$ inside
$l_{1,1}^{(\mathrm{BG})d}\left(0,0\right)$.  Let us introduce the
notation
\begin{equation}
y=\frac{1}{2}g^2\tilde\rho\ ,\quad n(y)=u(\tilde\rho)\ .
\end{equation}
Then the flow equation for the potential function $n$ is a first order
PDE which can be recast into the form of two inhomogeneous ODEs by the
method of characteristics. By denoting $\tau(y)=n'(y)$, for the
linear regulator these read

\begin{eqnarray}
\label{floweq:ODEtau}
-\eta_W g^2 \frac{\mathrm{d}\tau}{\mathrm{d}g^2}&=& (2-\eta_\phi+\eta_W)\tau
+\frac{6}{32\pi^2}\frac{1-\eta_W/6}{(1+y)^2}\\
-\eta_W g^2 \frac{\mathrm{d}y}{\mathrm{d}g^2}&=& (2+\eta_\phi-\eta_W)y
-\frac{2}{32\pi^2}\frac{1-\eta_\phi/6}{(1+g^2 \tau/2)^2}\quad\quad
\label{floweq:ODEy}
\end{eqnarray}
where both $\tau$ and $y$ have to be interpreted as functions of $g^2$
only. The desired potential function $\tau(y)$ is obtained from the
previous system as the solution corresponding to the initial condition
$\tau(y)|_{g^2=g^2_\Lambda}=\tau_\Lambda(s)$, representing the
bare potential, and $y(g^2=g^2_\Lambda)=s$. Because of the difference
between $s$ and $y$ at $g^2\neq g^2_\Lambda$, the freedom to specify
the function $\tau_\Lambda(s)$ can be translated into a freedom in
determining the running of the potential $\tau(y)$ w.r.t. the RG time
$g^2$.  This is another way of recognizing the freedom to choose a
boundary condition for the integration of the RG equations.  The goal
of this subsection is to explicitly show how the latter allows for the
construction of $(P,\xi)$-scaling in the UV.

Unfortunately, this system of ODEs does not offer a straightforward
explicit solution, though it could be solved numerically rather
straightforwardly. For reasons of analytical insight, we perform
another physically motivated simplification: the second term on the
right-hand side of~\Eqref{floweq:ODEy} arises from the would-be
Goldstone modes. Owing to the Higgs mechanism, these modes are
actually not propagating degrees of freedom, but remain visible here
only because of our choice of using a Landau-type gauge. In
particular, such contributions would be absent in unitary gauge which
is more adapted to the physical degrees of freedom in the Higgs
phase. Therefore, we drop these contributions in~\Eqref{floweq:ODEy},
which
decouples the two solutions and makes the latter equation trivially
solvable. It is interesting to note that this line of argument
together with the large-$N$ limit removes the scalar-loop
contributions to the flow of the scalar sector altogether. Hence, we
expect the conventional triviality problem of perturbative nonabelian
Higgs models to be alleviated in the current limit anyway.

In this limit, we can again go back to the PDE for the potential
function $n(y)$, such that \Eqref{floweq:potential_largeN} reads
\begin{eqnarray}
\eta_W g^2 \frac{\partial n}{\partial g^2}&=&
 -d n + (d-2 + \eta_\phi-\eta_W)y n' \nonumber\\
&&+4 v_d (d-1)l_{0}^{(\mathrm{G})d}\left(y\right) \label{eq:flowofn}.
\end{eqnarray}
Here and in the following, we use the lowest-order form
\Eqref{eq:Netas} for the anomalous dimensions and ignore the
dependence of the threshold function on $\eta_W$.  This makes
\Eqref{eq:flowofn} accessible to
straightforward integration.  The solution is a linear combination of
the corresponding \fgFPh\  solution and of the solution of the
homogeneous flow equation.  For the linear regulator in $d=4$, it reads
\begin{eqnarray}
n(y,g^2)&=&\frac{3}{32\pi^2}\left(-y+y^2\log\left(1+\frac{1}{y}\right)\right)\nonumber\\
&&+4 y^2 Q\!\left(\frac{1}{g^2}+\frac{11}{48\pi^2}\log(y)\right),
\label{eq:nsol}
\end{eqnarray}
where $Q$ is an arbitrary function that can depend on $y$ and $g^2$
only through its argument. From here, we can recover the
$(P,\xi)$-scaling solutions from different choices of $Q$.  First, we
go over from the variable $y$ to the gauge-rescaled field
$x=2g^{2(P-1)}y$ as used before, identifying $f(x)=n(y)$,
\begin{eqnarray}
f(x,g^2)&=&\frac{x^2}{g^{4(P-1)}} Q\!\left(\frac{1}{g^2}+\frac{11}{48\pi^2}\log\left(\frac{x}{2g^{2(P-1)}}\right)\right)\\
&&\!\!\!\!\!\!\!\!\!\!\!\!\!\!\!\!\!\!\!\!\!\!+\frac{3}{32\pi^2}\left(-\frac{x}{2g^{2(P-1)}}+\frac{x^2}{4g^{4(P-1)}}\log\left(1+\frac{2g^{2(P-1)}}{x}\right)\right).\nonumber
\label{eq:Nfsol}
\end{eqnarray}
We then look for finite limits of $x$ and $f$ when $g^2\to 0$.  

If $P\neq 1$ the inhomogeneous solution corresponding to the
  second line becomes trivial in the $g^2\to 0$ limit (even though it
this is not regular at $x=0$).  In order to have a finite nontrivial
potential $f$ then we need to demand for the homogeneous solution
(first line) to remain different from zero.  This is the case if
$Q(z)\sim\xi z^{2(1-P)}$ when $z\to+\infty$, such that $f(x)\sim\xi
x^2$ when $g^2\to 0$.  

If $P=1$ it is sufficient to choose a function that approaches a
constant $Q(z)\sim\xi $ for $z\to+\infty$ in order to recover the
corresponding \fgFPh\ solution; note that the different numerical
factors in front of the inhomogeneous piece compared to the $N=2$
solutions arises from subleading terms of the $1/N$ expansion. We
conclude that we rediscover the family of $(P,\xi)$-scaling solutions
towards asymptotic freedom also in the large-$N$ limit.

We observe that only the asymptotic behavior of $Q$ for large argument
is fixed by the requirement to realize asymptotic freedom in the UV.
Apart from this asymptotic behavior, $Q$ remains largely undetermined
in the present approximation. Whereas this freedom does not affect the
approach to asymptotic freedom, the shape of $Q$ at finite argument
can take influence on the IR behavior. For instance, if  $Q$
exhibited an exponential behavior for small argument,
\begin{equation}
Q(z)=c e^{\frac{48}{11}\pi^2 z}=c \frac{2x}{g^{2(P-1)}}e^{\frac{48\pi^2}{11g^2}}
=2c g^{2}\tilde\rho k^2, \label{eq:Qmass}
\end{equation}
this would correspond to a dimensionful relevant component, i.e., a
nonvanishing dimensionful mass, which increases logarithmically with
the gauge coupling $g^2$ towards the IR. Within the limitations of the
current large-$N$ analysis, it is difficult to judge whether the
freedom to choose $Q$ (beyond the asymptotics which controls
asymptotic freedom) is merely an artifact of the approximations or
whether it can parametrize further relevant components as in
\Eqref{eq:Qmass}.

\section{IR flows and mass spectrum}
\label{sec:UVIR}

Our analysis of the UV behavior of the model allowed us to classify
the asymptotically free trajectories which are attracted by the
Gau\ss{}ian fixed point at high energies. In total, we found a
four-parameter family of these trajectories which we label by
$(P,\xi,c_\Lambda,g_\Lambda^2)$. Whereas $P$ and $\xi$ are fixed
parameters for each trajectory, the parameters $c_\Lambda$ and
$g_\Lambda^2$ quantify the magnitude of the relevant component and the
gauge coupling, respectively, at a reference scale $\Lambda$. This
reference scale is not physical, as a change of $\Lambda$ can be 
compensated by a corresponding change of $c_\Lambda$ and $g_\Lambda^2$
as dictated by the renormalization flow along the trajectory.

Once, the trajectory is chosen in terms of these parameters, the
theory is fixed and all long-range observables can, in principle, be
predicted. Technically, this corresponds to solving the full coupled
PDE/ODE system of our truncation from $\Lambda$ down to the IR
scales. For the initial conditions for this PDE/ODE system, we need to
relate the parameters $(P,\xi,c_\Lambda,g_\Lambda^2)$ to the
corresponding functions/variables. In particular, we have to fix the
full potential $u_\Lambda(\rho)$ at the initialization scale. We have
argued that this can be approximated by the solution of a
\fgFPh\ equation. This approximation becomes better for smaller gauge
coupling: the parametrizations obtained in the previous sections are
trustworthy if the $P$-dependent powers of $g_\Lambda$ appearing in
Tab.~\ref{tab:leading_g2_dep} are small. Hence, this suggests to fix
the initial conditions at scales $\Lambda$ much larger than any IR
scale (e.g., the Fermi scale, or the scale $\Lambda_{\mathrm{SU}(N)}$
where the gauge coupling grows large). Though using the
\fgFPh\ solution may introduce small errors in the irrelevant
components as compared to the true scaling solution, RG universality
guarantees that these errors are washed out by the RG flow towards IR
scales. In practice, this suggests to compute the IR flow
approximately in terms of a simple polynomial expansion of the
potential. In fact, the reliability of such an expansion for a
description of Fermi scale observables has been verified for a variety
of standard-model-like (gauged)-Higgs-Yukawa models in
\cite{Gies:2013fua,Gies:2014xha,Eichhorn:2015kea,Borchardt:2016xju,Jakovac:2015kka}.

However, there is an apparent clash: on the one hand, we can apply
truncated polynomial (effective-field-theory-like) approximations for
the IR flow of the potential; on the other hand, our asymptotically
free UV scaling solutions require boundary conditions that fix
higher-order couplings to gauge-rescaled dimensionless ratios and that
give rise to a global existence of the potential in field space. For
instance in the effective-field-theory setting, we have discovered the
asymptotically free trajectories, provided that the ratios
$\xi_2=\lambda_2/g^{4P}$ and
\begin{equation}
\chi=\lambda_3\left\{ \begin{array}{ll}
             g^{-8P} & \text{for}\,P\in(0,1/2]\\
             g^{-2(1+2P)} & \text{for}\, P\in[1/2,1]\\
		g^{-2(1+8P)/3} & \text{for}\, P\geq 1
            \end{array}
            \right. , 
\label{eq:chi_to_lambda3_genP}
\end{equation}
etc. approach nonvanishing constants in the UV. It is obvious that this
property cannot meaningfully persist along the RG flow towards the IR:
for instance, it would imply that the dimensionful couplings could
diverge in the IR, e.g., the six-point vertex would scale as
$g^{6P+2P_3(P)}\chi/k^2$ for $k\to 0$, where $2 P_3(P)$ is the
$P$-dependent power given in the second column of
Tab.~\ref{tab:polynomial_g2_scaling}. This would correspond to a
substantial deviation from Wilsonian power counting, being in strong
contradiction to the anticipated vicinity of the Higgs sector to the
Gau\ss{}ian fixed point. 

Therefore, we expect the scaling conditions such as
\Eqref{eq:chi_to_lambda3_genP} to be satisfied in the UV whereas
Wilsonian scaling should hold in the IR. The quantitative details of
the transition between the different scaling regimes are governed by
the full PDE. In the present section, we aim at a simple estimate for
the flow between the different regimes. For this purpose, we stay
within the polynomial effective-field-theory setting and model the
behavior of a higher-order coupling; it turns out that the results for
the long-range observables only show a mild dependence on the details
of this modeling.

It is worthwhile to compare the parameter fixing to that of
conventional perturbation theory. In the latter case, the couplings
would be fixed in terms of the long-range observables of Higgs mass
$m_{\text{H}}$, gauge boson mass $m_W$ and vacuum expectation value
$v$. These translate into renormalization conditions for, say
$\lambda$, $g^2$, and $\kappa$ at a certain fixing scale
$\Lambda$. For the asymptotically free trajectories, we have one
additional parameter. Say, we fix this additional one in terms of a
concrete choice for $P$. Next, we choose some (small) value of
$g_\Lambda^2$. The asymptotically free scaling potential can then be
constructed from the \fgFPh\ condition in conjunction with the choice
for the third parameter $\xi$. Now, in the simplest effective-field
theory approximation including a $\phi^6$ term, we can trade $\xi$ for
the ratio $\chi$ of \Eqref{eq:chi_to_lambda3_genP} which approaches a
constant in the deep UV. The choice of the four parameters is then
completed by adding a value $c_\Lambda$ for the relevant direction. In
practice, this value has a strong influence on the value of the Fermi
scale, i.e., the value of the vacuum expectation value $v$ in units of
$\Lambda$.  This choice of parameters fixes all initial conditions;
e.g., in the effective-field-theory setting to order $\lambda_3$, also
$\kappa$ and $\lambda_2$ are fixed by the \fgFPh\ conditions at the
scale $\Lambda$ in this manner.

\begin{figure}[!t]
\begin{center}
 \includegraphics[width=0.4\textwidth]{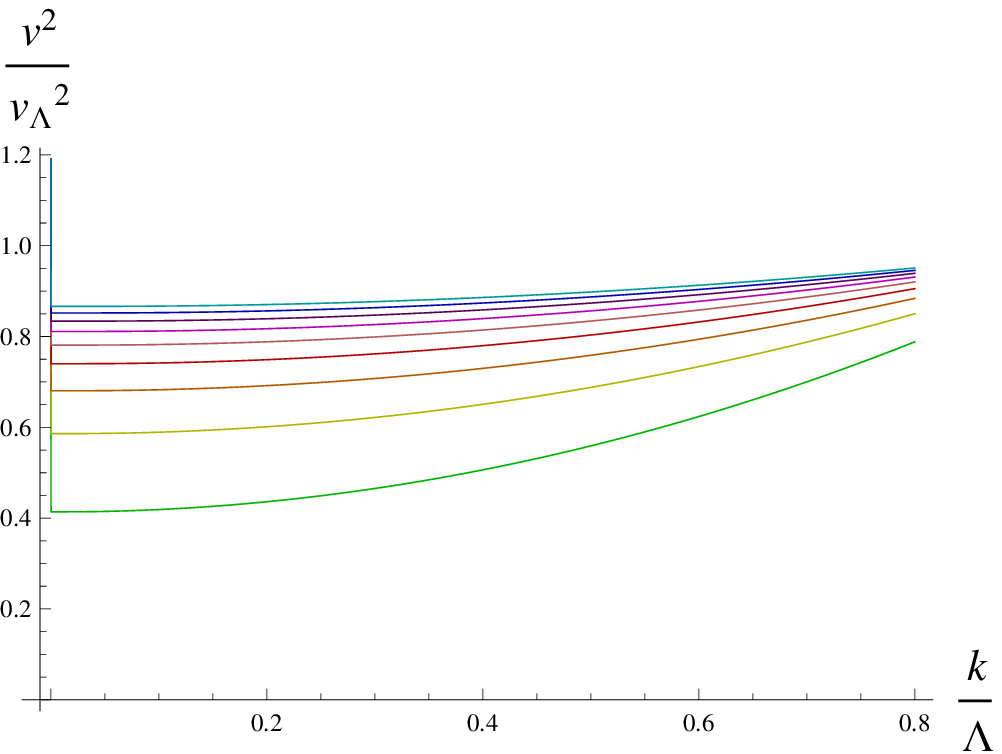}\hskip 1cm
\includegraphics[width=0.4\textwidth]{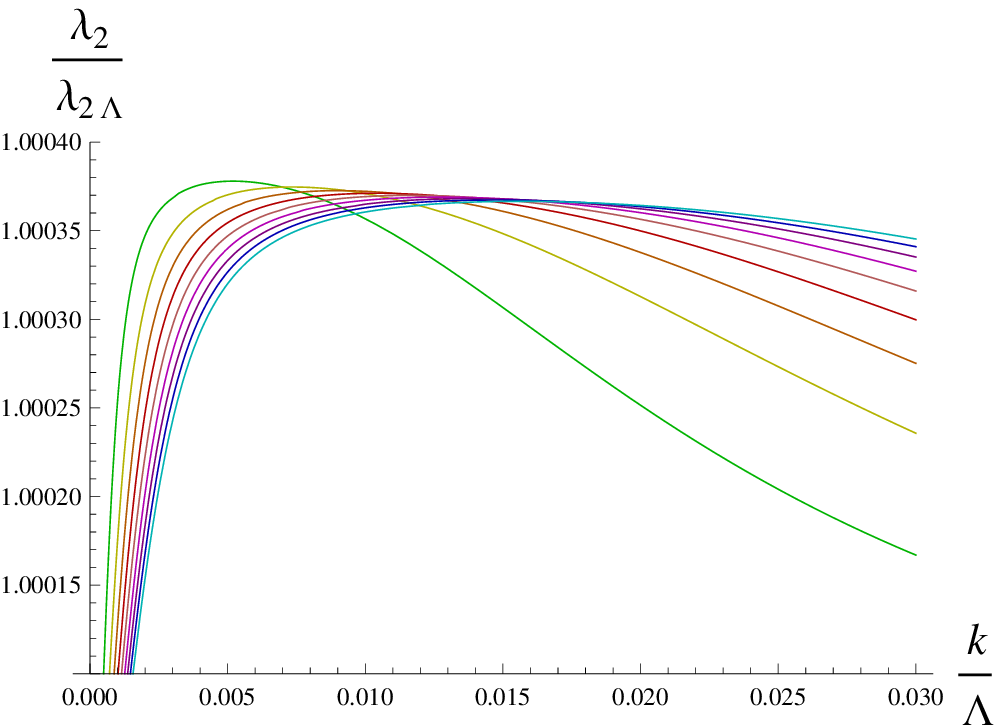}
 \caption{Flows of the dimensionful renormalized field expectation value $v^2$ and quartic
   coupling as functions of $k/\Lambda$ for $P=1/2$.  The initial data
   is defined by $g^2_\Lambda=10^{-4}$ and the choice $\chi=-10^{-3}$,
   which implies $\xi_2=9.98\times 10^{-4}$. The relevant component
   $c_\Lambda$ is chosen such that $\frac{-2 g_\Lambda
     c_\Lambda}{10 \lambda_{2,\Lambda}}$ is an integer number ranging
   from 1 (green, steeper curve) to 9 (cyan, flatter curve).  All
   flows are computed in the effective-field-theory approximation with
   $N_p=2$ including anomalous dimensions. We observe the physical
   decoupling behavior of massive modes at the scale
   $\mathcal{O}(1-10\%\Lambda)$. As an oversimplification, $\chi$ is
   kept constant at all scales inducing an unphysical behavior at low
   scales $\sim\mathcal{O}(0.1\%\Lambda)$.}
\label{fig:P1o2_fixchi_smallg}
\end{center}
\end{figure}

Figure~\ref{fig:P1o2_fixchi_smallg} shows a set of example flows for 
$P=1/2$.
Here, we have chosen $g^2_\Lambda=10^{-4}$ and naively kept
$\chi=-10^{-3}$ fixed over all scales. The latter choice implies
$\xi_2=9.98\times 10^{-4}$ (fixing $\lambda_{2,\Lambda}$), and the
relevant component $c_\Lambda$ is varied such that $\frac{-2
  g_\Lambda c_\Lambda}{10 \lambda_{2,\Lambda}}$ is an integer
number ranging from 1 to 9. Even for this unphysical case of keeping
$\chi=-10^{-3}$ constant over all scales, we observe a freeze-out
behavior of the dimensionful renormalized expectation value $v$ (upper panel) and quartic
coupling (lower panel) towards the infrared at a scale
$k_{\text{F}}=\mathcal{O}(1-10\%\Lambda)$ for sufficiently big $c_\Lambda$. This
indicates the generation of the Fermi scale and the Higgs mass, and
marks the decoupling of all massive modes. Only for much lower scales
$\sim\mathcal{O}(0.1\%\Lambda)$, the artificial choice of a constant
$\chi$ spoils the decoupling behavior towards the deep IR, triggering
an unphysical strong flow of $\lambda_2$ and of $v$. This signals the
break down of the oversimplifying approximation of keeping $\chi$
constant on all scales, destabilizing the physical decoupling regime
in the deep IR.

\begin{figure}[!t]
\begin{center}
\includegraphics[width=0.4\textwidth]{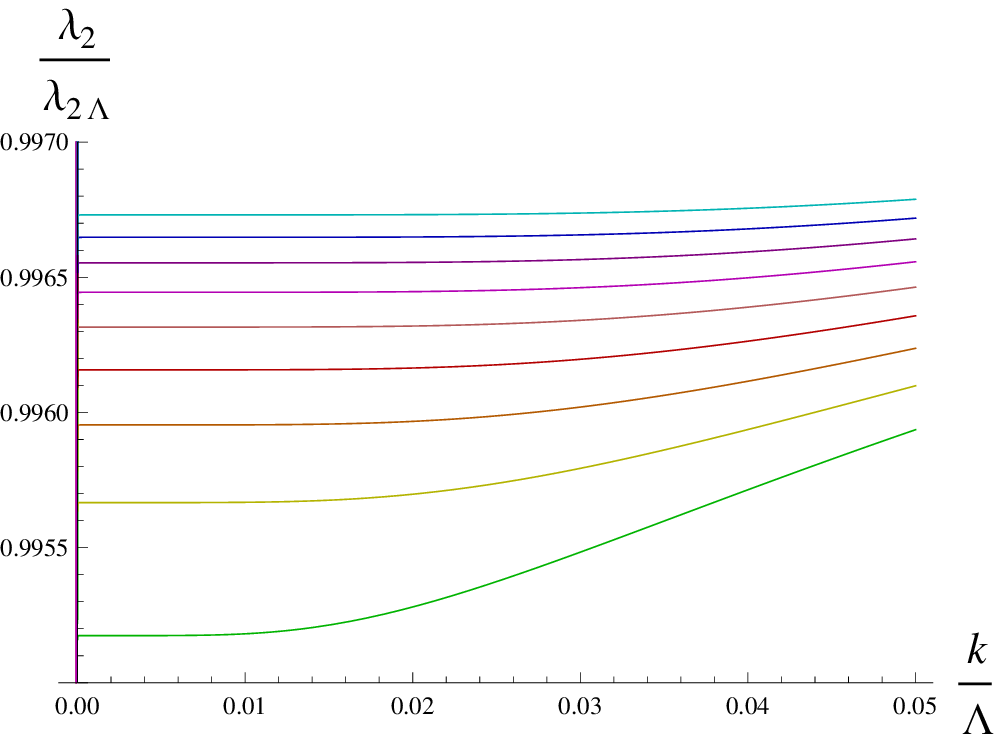}
\includegraphics[width=0.4\textwidth]{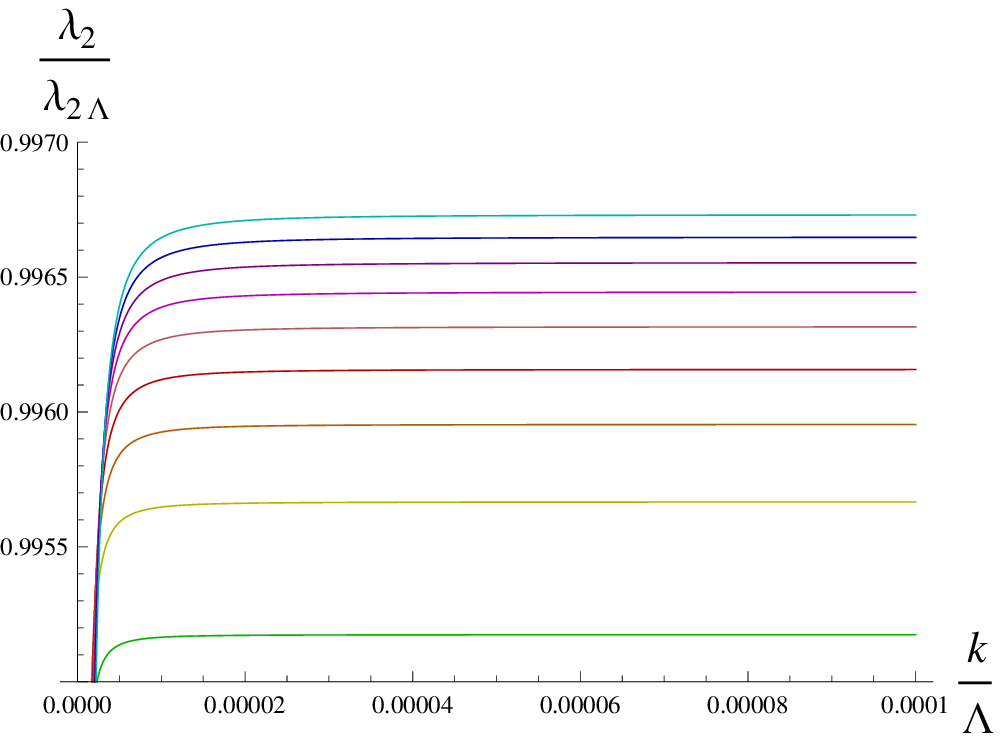}
 \caption{RG flow of the quartic coupling as functions of $k/\Lambda$
   for $P=1/2$. The initial data and the same effective-field-theory
   approximation is chosen as in Fig.~\ref{fig:P1o2_fixchi_smallg},
   but $\chi$ is set to zero along the flow, leading to a more stable
   decoupling regime. The deep IR instability caused by the artificial
   would-be Goldstone bosons appears at much lower scales and can well
   be separated from the long-range physics at decoupling.}
\label{fig:P1o2_chizero_smallg}
\end{center}
\end{figure}

A simple approximation to model the UV to IR transition region is to
switch from $\chi$=const. to a fixed renormalized dimensionful
coupling $\sim\lambda_3$ after the onset of the physical freeze-out
behavior. A simple choice is even to set $\chi=0$ near
decoupling, which is enough to achieve the freeze-out of $\lambda_3$. 
Resulting flows for the simplest choice with
  $\chi=0$ for $k\leq\Lambda$ are shown in
Fig.~\ref{fig:P1o2_chizero_smallg} , again for $P=1/2$ and the same
initial conditions as before.
The decoupling regime is now more stable (upper panel), facilitating
to read off the physical long-range observables; for instance, the
Higgs mass would correspond to $m_{\text{H}}=v\lambda_2$. Only in the
very deep IR, yet another pathological IR behavior sets in, which
artificially drives $\lambda_2\to 0$ (lower panel). This artifact is
caused by the spurious presence of propagating would-be Goldstone
bosons in the Landau gauge used here \cite{Gies:2014xha}. The dominating positive
contribution of these modes to the scalar anomalous dimension drives
the flow artificially in the very deep IR. This artifact would be
absent in the unitary gauge.

\begin{figure}[!t]
\begin{center}
 \includegraphics[width=0.4\textwidth]{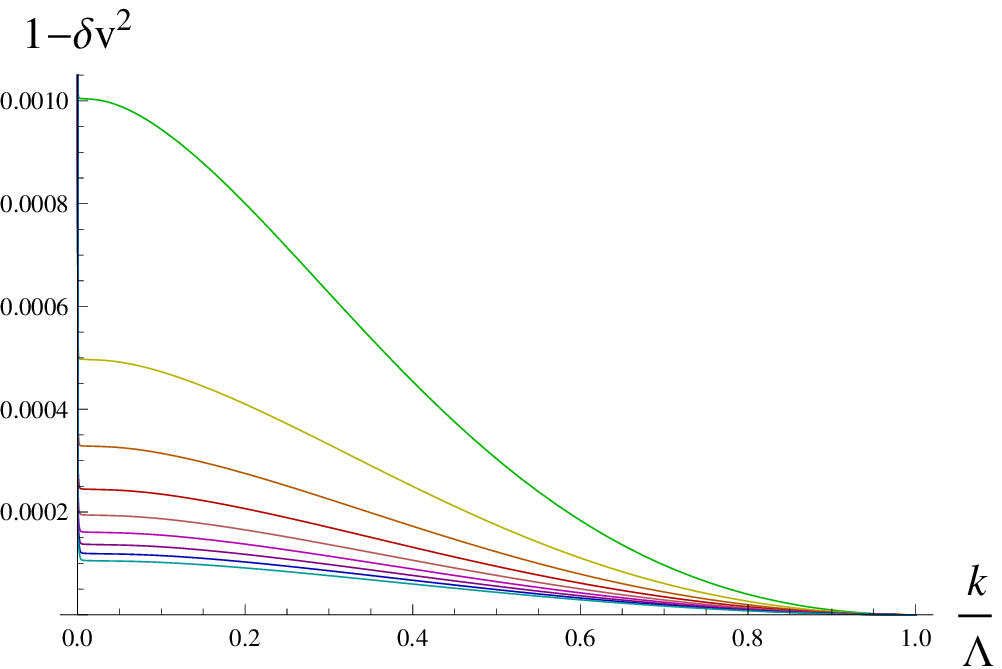}\hskip 1cm
\includegraphics[width=0.4\textwidth]{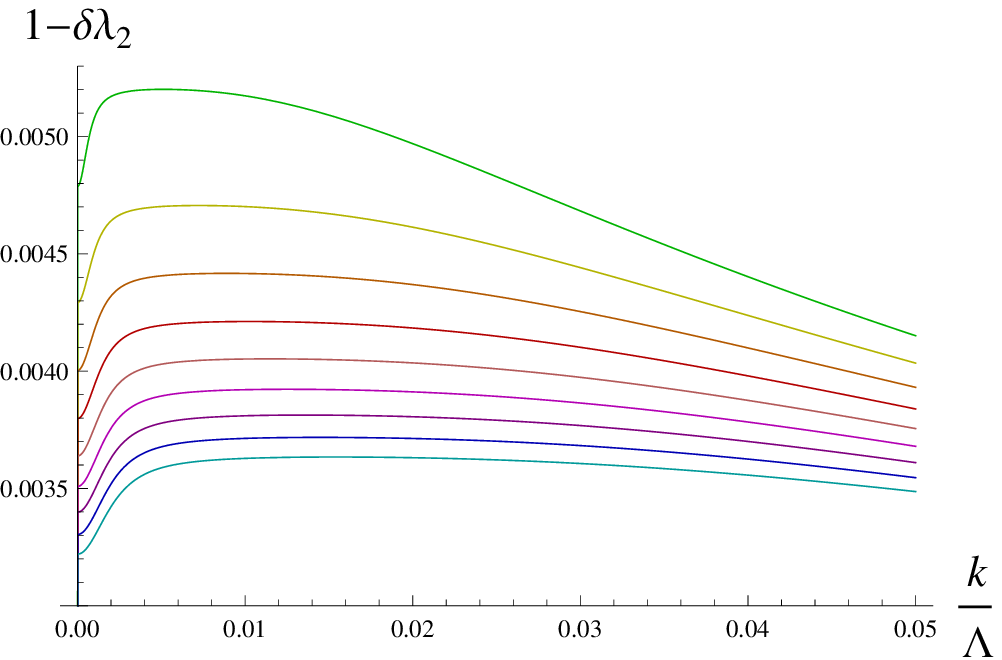}
 \caption{Quality check of the effective-field-theory flows: $\delta
   v^2$ and $\delta \lambda_2$ denote the ratios between the
   corresponding quantities obtained using the $\chi\!=\!0$ and the
   constant $\chi\!\neq\!0$ flows at $P=1/2$,
   cf. Figs.~\ref{fig:P1o2_fixchi_smallg} and
   \ref{fig:P1o2_chizero_smallg}. All other parameters are as in the
   preceding figures.}
\label{fig:P1o2_fixorzerochi_smallg}
\end{center}
\end{figure}

The difference between the extreme choices of $\chi=$const.~and
$\chi=0$ for the flows is shown in
Fig.~\ref{fig:P1o2_fixorzerochi_smallg} for the expectation value and
the quartic coupling.
The quantitative difference in the physical decoupling
regime is on the per mille level and decreases for a larger relevant
component and a smaller initial $g^2_\Lambda$.  Of course, any smooth
modeling of the transition regime from constant $\chi$ to vanishing
$\chi$ is equally possible, but is expected to lead to even smaller
differences.

In the examples shown in
Figs.~\ref{fig:P1o2_fixchi_smallg},~\ref{fig:P1o2_chizero_smallg},~\ref{fig:P1o2_fixorzerochi_smallg}
the separation of scales between initialization and decoupling is
rather small, as encoded in the ratio $k_{\text{F}}/\Lambda\approx
10^{-2}$. Still, it is already visible that the details about the
treatment of the nonrenormalizable interaction $\lambda_3$ have a tiny
effect on the IR spectrum, as long as the chosen boundary conditions
allow to end up in a Higgs phase. This corroborates the expectation
that boundary conditions compatible with asymptotic freedom in the UV
and a Higgs phase in the IR can also preserve the IR-irrelevance of
nonrenormalizable interactions.  For phenomenological applications the
separation of scales is expected to be larger, implying that the
effect of different boundary conditions for nonrenormalizable
operators will even be smaller.

Still, the task to compute phenomenologically viable flows is far from
straightforward, as these would require the decoupling scale
$k_{\text{F}}$ to be near the Fermi scale. At that scale, the gauge
coupling should approach the physical value $g^2_{\text{F}}\approx
(80/123)^2$ (as dictated by the $W$ boson mass relative to the vacuum
expectation value), which is not small. As the \fgFPh\ condition
facilitates an accurate determination of the asymptotically free
trajectory only for small gauge couplings, the flow of the
trajectories has to be followed over many orders of magnitude, say
from the Planck scale to the Fermi scale. In addition, the relevant
component at the Planck scale has to be fine-tuned very accurately to
trigger decoupling near the Fermi scale. This is a manifestation
of the standard hierarchy problem in the present setting. Computing
such a functional flow for a full potential over many orders of
magnitude represents a viable challenge for modern FRG PDE solvers
\cite{Borchardt:2016pif}.

Here, we will follow a simpler pragmatic approach: we assume that RG
flows from the asymptotically free trajectories, including suitable
small relevant perturbations, ending up in the Higgs phase in
the IR do exist. The preceding studies represent simple examples for such
flows. In order to separate the UV regime from the Fermi scale, the
parameter $c_\Lambda$ for the relevant direction has to be very small
at a high UV scale $\Lambda$. In fact, it has to have a negligible
influence on a large part of the flow towards the IR until $c_k$
becomes sufficiently large at a cross-over (CO) scale $k_{\text{CO}}$
where it kicks the system off the logarithmically slow running in the
UV fixed point regime. In this way, the relevant direction triggers
the approach to decoupling at $k_{\text{F}}$. As the relevant
perturbation increases with critical exponent $\theta_2=2$, the
cross-over scale $k_{\text{CO}}$ will already be comparable to
$k_{\text{F}}$. At the cross-over scale, the four UV parameters
$(\xi,P,g^2_\Lambda,c_\Lambda)$ can be mapped by the RG flow onto a
set of four other suitable parameters, e.g., the values of the
couplings in the effective-field-theory description $(\lambda_{2,
  \text{CO}},\lambda_{3,\text{CO}},g^2_{\text{CO}},\kappa_{\text{CO}})$. In
principle, also the values of all higher order parameters are
determined by the UV parameters; however, the full PDE would have to
be solved accurately for a quantitative estimate. Nevertheless, as
long as the flow ends up in the IR Higgs regime where Wilsonian power
counting becomes applicable, the precise details do not matter but are
washed out by the RG flow.

If so, this suggests to stay within the effective-field-theory
viewpoint and set up the IR flow at a fiducial cross-over scale
$k_{\text{CO}}$. Though it is difficult to relate the precise initial
data at $k_{\text{CO}}$ to the four UV parameters
$(\xi,P,g^2_\Lambda,c_\Lambda)$, we can at least estimate them by
solving the \fgFPh\ condition for the scalar sector with
$g^2_{k_\text{CO}}=g^2_{\text{CO}}$, and by superimposing a suitable
relevant component with parameter $c_{\text{CO}}$. In other words, we
model our ignorance about the effect of the relevant component on the
relations among higher order couplings by suddenly and discontinuously
switching on the relevant component at $k_{\text{CO}}$. 

\begin{figure}[!t]
\begin{center}
 \includegraphics[width=0.4\textwidth]{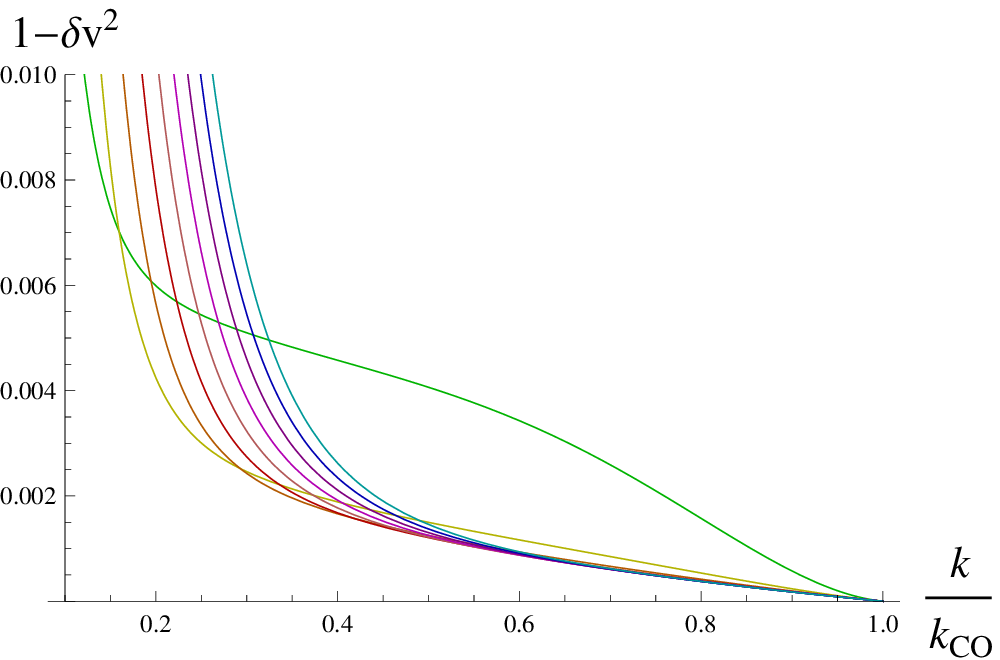}\hskip 1cm
\includegraphics[width=0.4\textwidth]{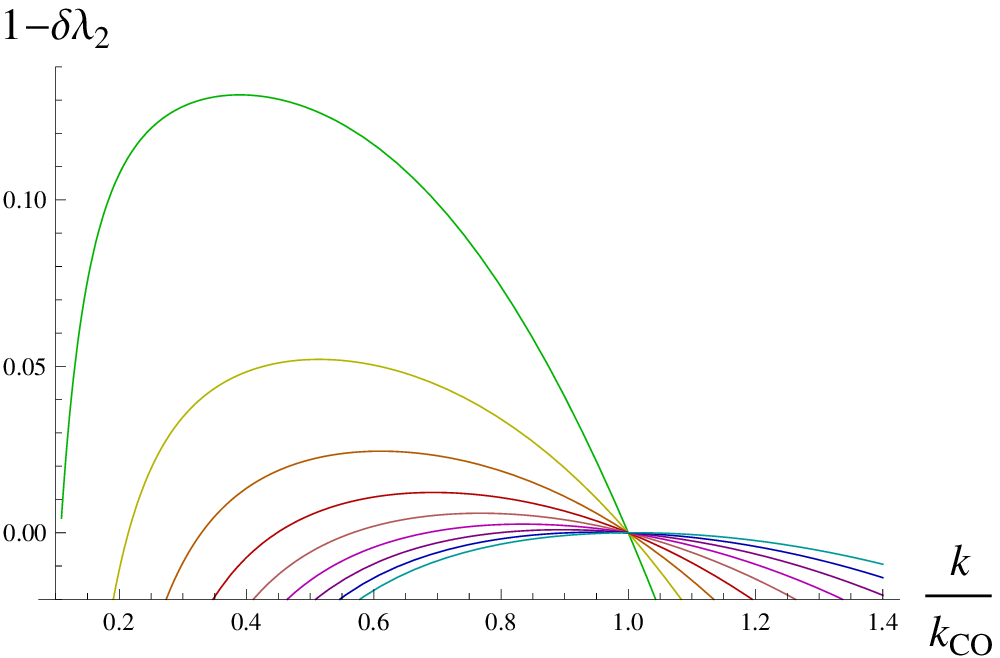}
 \caption{Self-consistency check of the effective-field-theory flows
   in the IR at stronger gauge coupling: $\delta v^2$ and $\delta \lambda_2$
   denote the ratios between the corresponding quantities obtained
   using the $\chi\!=\!0$ and the constant $\chi\!\neq\!0$ flows at
   $P=1/2$. The initialization is performed at
   $g^2_{\text{CO}}=(80/123)^2-10^{-2}$, all other parameters and
   approximations are as in the preceding figures, i.e., the different
   lines correspond to a relevant component $c_{\text{CO}}$ such that
   $\frac{-2g_{\text{CO}} c_{\text{CO}}}{10 \lambda_{2,{\text{CO}}}}$
   is an integer number ranging from 1 (green, steeper curve) to 9
   (cyan, flatter curve).}
\label{fig:P1o2_fixchi}
\end{center}
\end{figure}

As a self-consistency check of this procedure, we can again study the
sensitivity of the results in the decoupling regime to the details of
how we treat the higher-order interactions, as we did before in the
weak-coupling case. For this purpose, we now initialize the flow at
the fiducial $k_{\text{CO}}$ at a coupling value $g_{\text{CO}}^2$
slightly below the desired $g_{\text{F}}^2=(80/123)^2$ and compute the
flow towards decoupling in the $\chi=$const. as well as the
$\chi=0$ approximation. All other parameters are chosen as
before. Also for this bigger value of the initial gauge coupling, we
observe that adding a relevant component is sufficient to drive the
system towards decoupling in both approximations. The resulting
differences for the field expectation value and the quartic coupling
are shown in Fig.~\ref{fig:P1o2_fixchi}.
These differences are again found to be comparatively small though
somewhat larger than before because of the larger gauge coupling. We
conclude that different choices of boundary conditions for the running
of $\lambda_3$, even those that would artificially violate its
Wilsonian IR irrelevance such as for $\chi=$const., have only a minor
effect on the properties at decoupling, even in the case that the
initialization and decoupling scale are separated by less than one
order of magnitude, and even for a larger value of the
gauge coupling.

Within this simple effective-field-theory setting, we can now
construct an approximate mapping of the four UV parameters
$(\xi,P,g^2_\Lambda,c_\Lambda)$ onto the long range observables. As
the gauge coupling is fixed by the desired IR value
$g_{\text{F}}^2=(80/123)^2$ ($W$ boson mass), and the relevant
component is determined by ending up with a vacuum expectation value
at the Fermi scale, $v_{\text{F}}\simeq 246$GeV, a variation of the
parameters $P$ and $\xi$ is expected to shift the Higgs mass.

In our approximate treatment, some care is required for the
determination of the magnitude of the relevant component at the
cross-over scale $k_{\text{CO}}$ parametrized by
$c_{\text{CO}}$. Self-consistency of our cross-over picture requires
$|c_{\text{CO}}|$ to be sufficiently small such that it is justified to
ignore the relevant component at larger scales. On the other hand,
$|c_{\text{CO}}|$ should be sufficiently large such that it triggers to
flow towards the decoupling regime. For instance, if $|c_{\text{CO}}|$
would be chosen too small in our approximation, the physical
decoupling could overlap with the artificial IR flow of the would-be
Goldstone bosons which do not properly decouple in the Landau gauge in
our approximation.

\begin{figure}[!t]
\begin{center}
 \includegraphics[width=0.4\textwidth]{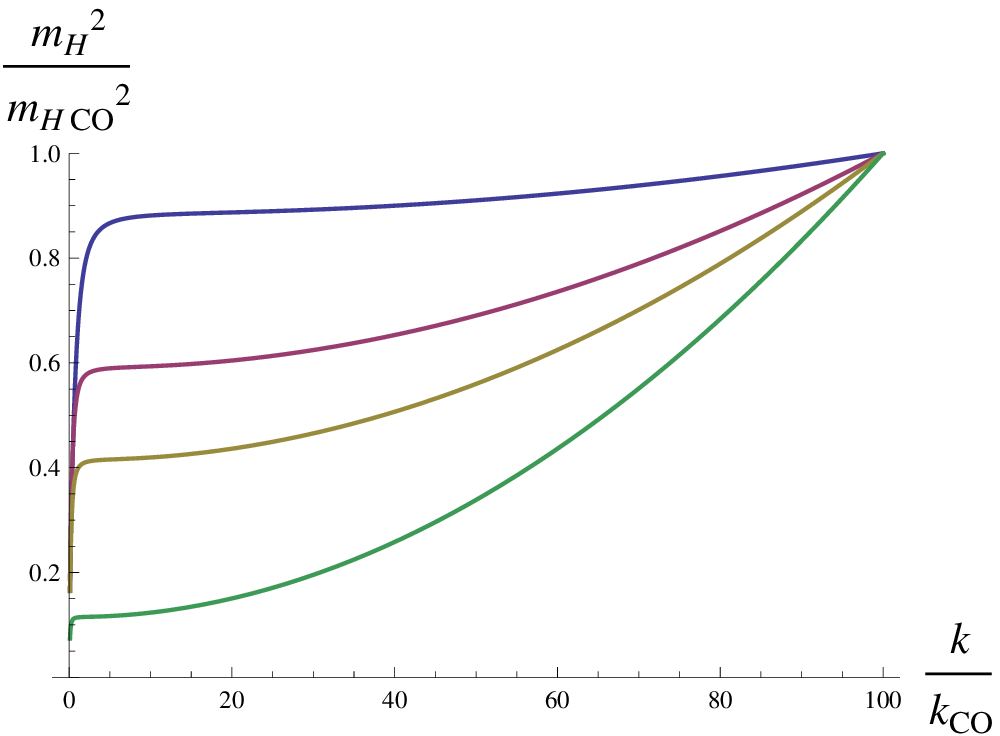}\hskip 1cm
\includegraphics[width=0.4\textwidth]{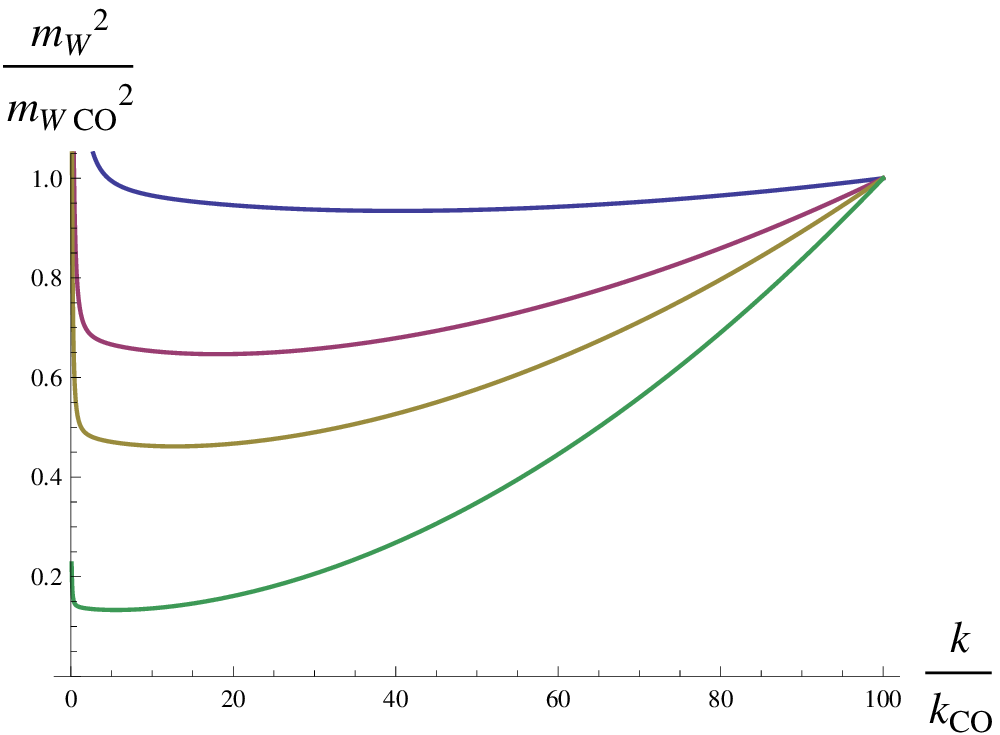}
 \caption{The running squared Higgs and $W$ mass, normalized to those at
   $k=k_{\text{CO}}$, for $P=1/2$ and $\xi=0.2$.  Both flows are
   initialized with $g^2_{\text{CO}}=(80/123)^2-10^{-2}$, by using the
   analytic approximations of the potential $f(x)$ along the AF
   trajectories obtained in Sec.~\ref{sec:FRG}, and adding a relevant
   component $+c_{\text{CO}}x$ with $-c_{\text{CO}}\in\{0.001,0.005,
   0.01, 0.05\}$ from green (deeper) to blue (flatter).  These flows
   are obtained from a $N_p=4$ effective-field-theory approximation.}
\label{fig:P1o2_optimal_relevant}
\end{center}
\end{figure}

This gauge insufficiency also has the consequence that the decoupling
is not visible in the gauge coupling, which is driven by gauge modes
even in the deep IR in our approximation. This contaminates the
freeze-out behavior of the flow of the $W$ mass and creates a minimum
in $m_{W}^2$ as a function of $k$ instead of the plateau behavior
which we observe for the Higgs mass $m_{\text{H}}$ and signals
decoupling. If $|c_{\text{CO}}|$ is chosen too large, the minimum in
$m_{W}^2$ occurs at larger $k$ values than the plateau of
$m_\text{H}^2$, such that we cannot unambiguously identify the point
of decoupling.  This is shown in Fig.~\ref{fig:P1o2_optimal_relevant}
for the $P=1/2$ case, where the running masses are exhibited for
various choices for $c_{\text{CO}}$. 

In summary, we observe that there is an interval of acceptable values
of $c_{\text{CO}}$ for any choice of $P$ and $\xi$ which is consistent
with the physical and technical requirements listed above. This
interval spans approximately an order of magnitude.  Inside this
interval we prefer small values of $|c_{\text{CO}}|$, as it
  simplifies the identification of the decoupling scale. For the sake of inspecting
several $P$ and $\xi$ values at once, $c_{\text{CO}}=-0.01$ turns out
to be a reasonable choice. By fixing this value to a constant, the
physical Fermi scale in turn implicitly fixes the cross-over scale,
such that only $P$ and $\xi$ remain as parameters.

\begin{figure}[!t]
\begin{center}
 \includegraphics[width=0.5\textwidth]{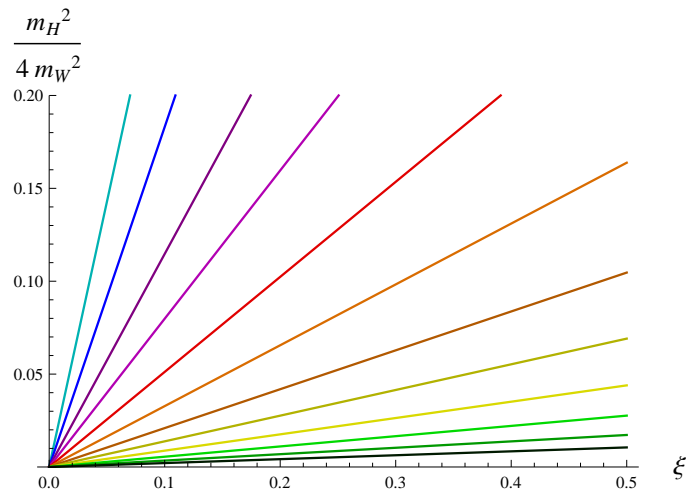}
 \caption{The squared ratio between the Higgs-mass and twice the $W$ mass,
at decoupling, as a function of $\xi$ and for
various values of  $P=i/4$, $i$ taking integers values from 1
(cyan, steeper) to 12 (green, flatter).  For both flows, the
initialization is performed at $g^2_{\text{CO}}=(80/123)^2-10^{-2}$,
by using the analytic approximations of the potential $f(x)$ along the
AF trajectories obtained in Sec.~\ref{sec:FRG} and adding a relevant
component $+c_{\text{CO}}x$ with $c_{\text{CO}}=-0.01$.  These flows
are obtained from an $N_p=4$ conventional polynomial truncation.}
\label{fig:HiggstotwoW}
\end{center}
\end{figure}

With all these prerequisites, we are now ready to explore the
properties of the mass spectrum. For this, we first estimate the
initial data at the cross-over scale from the analytic
parametrization of the asymptotically free trajectories obtained in
Sec.~\ref{sec:FRG}. Having fixed the Fermi scale as well as the $W$
boson mass by the choice of the gauge coupling at $k_{\text{CO}}$, we
can then study the dependence of the Higgs mass on the parameters $P$
and $\xi$. The ratio of Higgs to $W$ boson mass is shown in
Fig.~\ref{fig:HiggstotwoW}. We observe that each of the two parameters
can be used to tune the Higgs mass. This is similar to perturbation
theory where $\lambda_2$ governs the Higgs mass. Hence, $P$ and $\xi$
are not uniquely fixed by the knowledge of the mass spectrum, since
for each $P$ one can find a suitable $\xi$.  Indeed, these numerical
scans show that the qualitative dependence of the mass ratio is well
described by the same ratio at initialization, which in the cross-over
picture is the same as the ratio at the fixed-point regime. We find
\begin{equation}
\frac{m^2_{\mathrm H}}{4m^2_{\mathrm W}}\Big|_{\text{CO}}
=\frac{\lambda_{2{\text{CO}}}}{g^2_{\text{CO}}}
=2\xi \ g^{2(2P-1)}_{\text{CO}}\ .\label{eq:massratio}
\end{equation}
As a consequence, the theory is not fully specified by the mass
spectrum and the Fermi scale contrary to the perturbative setting, but
an additional higher-order operator has to be measured in order to
determine the asymptotically free trajectory that can guarantee UV
completion of the theory.

We close this section with the comment that the present estimate of IR
observables is close to the spirit of a perturbative estimate of the
spectrum. In particular, the present treatment is blind to
nonperturbative bound-state effects triggered by the gauge
sector. This becomes obvious from the fact that $c_{\Lambda}$ or
$c_{\text{CO}}$ could be chosen such that the system seemingly stays
in the perturbatively massless phase. In fact, such a phase does not
exist, as the gauge sector eventually grows strong towards the IR. As
a consequence, the perturbatively massless phase actually corresponds
to a QCD-like phase, with a massive spectrum of glueballs and
corresponding gauge-Higgs bound states. The properties of and
distinction between these phases has recently been under intense
investigation \cite{Maas:2012tj,Maas:2013aia,Maas:2014pba}. A
particular interesting question is as to whether the non-perturbative
effects can exert a strong influence also on the Higgs side of the
phase transition, e.g., in terms of providing a lower bound on the
Higgs mass \cite{Maas:2014pba}, or even reshape the spectrum in
comparison with a pure perturbative reasoning
\cite{Maas:2015gma,Maas:2016ngo}.

\section{Conclusions}
\label{sec:conc}

Our work demonstrates the construction of asymptotically free
renormalization group trajectories in nonabelian Higgs models. We
emphasize that we have identified these trajectories also for those
models, where low-order perturbation theory does not exhibit
asymptotic freedom. Whereas most of our analysis is performed at weak
coupling, the essential difference to standard perturbation theory
arises from the fact that we carefully pay attention to boundary
conditions to be imposed on the renormalized action.

While perturbation theory is largely insensitive to boundary
conditions and merely implicitly assumes that suitable boundary
conditions exist, the importance of boundary conditions for the
existence of scaling solutions is well known from the study of
interacting fixed points in critical phenomena. Our work can be viewed
as an extension of this concept to \fgFPs\ which solve a fixed-point
condition for the scalar potential at finite but fixed gauge
coupling. The corresponding scalar \fgFPh\ potential yields an
estimate of asymptotically free RG trajectories of the model becoming
accurate in the limit of vanishing gauge coupling.

The set of asymptotically free models for a given gauge group SU($N$)
has four parameters which is one more than for a perturbative
analysis. In the latter, nonabelian Higgs models have a relevant
parameter (mass parameter of the scalar potential), the marginally
relevant gauge coupling and the marginally irrelevant Higgs
self-coupling. In our construction the relevant mass-like parameter
persists and the marginally relevant gauge coupling goes along with a
nontrivial perturbation of the scalar potential determined by the
\fgFPh\ conditions. The latter, however,  
can be solved in terms of two
parameters $P,\xi>0$ for which we have not found any further
constraint. They may be viewed as exactly marginal in an RG language.

While we have presented indications for this scenario first in the
languages of simple low-order perturbation theory and effective field
theory, the picture unfolds both more quantitatively and conceptually
with the method of functional renormalization. This technique
facilitates to extract global information about
the scalar potential by means of analytic asymptotic
expansions and numerical shooting methods. Most importantly, this
function passes all standard tests of being polynomially
(uniformly) bounded and self-similar for large field values. As an interesting feature which
is different from pure scalar models for critical phenomena, we
observe a logarithmic non-analyticity of the \fgFPh\ potential at the
origin in field space where the potential and its first derivative
still stay finite. This non-analyticity implies that the
asymptotically free trajectories globally belong to a different
functional space than that spanned by polynomial interactions -- even
though many features of these trajectories are visible in polynomial
effective-field-theory like approximations.

The formulation of the property of asymptotic freedom as a standard RG fixed-point condition
allows to perform a standard Wilsonian RG
relevance classification of perturbations about scaling solutions. In
the weak-coupling regime, they exhibit a standard (quantized) behavior
of the critical exponents. As a consequence, our asymptotically free
trajectories may solve triviality problems but -- within a perturbative
analysis of the mass spectrum -- they appear to feature the same
properties with respect to a separation of hierarchies as the
perturbative standard model.

The functional RG has also helped to further corroborate the evidence
for such asymptotically free theories in the large-$N$ limit, where an
even larger set of parameters in the form of a whole function seems
admissible in order to find asymptotically free trajectories. While
this might be an artifact of the large-$N$ limit, we have been able to
rediscover unambiguously the family of trajectories parametrized by
$P$ and $\xi$, again corresponding to boundary conditions for the solution
of the differential RG equations.

Whereas we have concentrated on pure nonabelian Higgs models, viewed
as a key ingredient of the standard model or of suitable sectors of
grand unified models, 
first signatures of the existence of the
trajectories described here had already been found before in a gauged
Yukawa model in \cite{Gies:2013pma}. That previous study had in fact
been inspired by the search for \textit{asymptotically safe}
trajectories, potentially being triggered by a quasi-conformal running
of the field expectation value \cite{Gies:2009hq,Gies:2009sv}. While
the latter may still be a viable ingredient, our present results
illustrate that such a property is not necessary for the
construction of asymptotically free trajectories. As far as asymptotic
safety is concerned, a wide class of gauged Higgs-Yukawa models has
recently been identified that allows for a controlled determination of
a non-trivial UV limit in terms of an interacting fixed point
\cite{Litim:2014uca,Bond:2016dvk,Codello:2016muj,Molgaard:2016bqf,Bajc:2016efj}.

Finally, we have estimated the long-range properties of theories that
arise from our asymptotically free trajectories, and identified
initial/boundary conditions giving rise to a conventional Higgs
phase. Whereas solving a full functional flow from a high UV scale
down to the Fermi scale remains a technical challenge, our estimates
give access to the particle spectrum analogous to perturbation
theory. We observe that the range of particle masses remains
unconstrained because of the freedom encoded in the parameters $P$ and
$\xi$. For a realistic application of such asymptotically free
trajectories to standard model physics, the inclusion of chiral
fermion degrees of freedom is, of course, mandatory.

\acknowledgments

We thank J\"org J\"ackel, Axel Maas, Christof Wetterich and Omar Zanusso for
interesting discussions, and Stefan Rechenberger, Ren\'{e}
Sondenheimer, and Michael Scherer for collaboration on related
projects. We acknowledge support by the DFG under grants
No. GRK1523/2, and Gi 328/5-2 (Heisenberg program).

\appendix

\section{Lowest-order weak-coupling effective field theory}
\label{app:leadingpower}

In the effective field theory analysis in Sect.~\ref{sec:EFT1}, we
consider the $\beta$ functions fully in the spirit of effective field
theory including all couplings of a given truncation as they would
arise from corresponding Feynman diagrams. By contrast, the functional
RG analysis in Sect.~\ref{sec:FRG} is mostly based on a weak-coupling
analysis, sorting the $\beta$ functions in powers of the gauge
coupling $g$ and disregarding higher orders. This seems to imply a
clash between the two approaches, as the functional RG analysis
ignores terms that have been fully taken into account in the effective
field theory analysis. 

In this Appendix, we show that the lowest-order weak-coupling analysis
can also be performed in the effective-field theory case. Though this
is a rather crude approximation, results for the \fgFPh\ values of the
potential minimum $x_0$ and the rescaled coupling $\xi_2$ are
maintained, and the error for the higher-order couplings is
parametrically quantifiable. The order of the following results
parallels that of the main text.

We start with the case $P=1/2$ discussed in
Subsect.~\ref{subsec:fg}. There the $\beta$ function in
\Eqref{subsec:fg} contains terms of order $g$ and $g^2$. Keeping both
already suffices in order to unfold the gauge-rescaling pattern of the
higher order terms. However, if we strictly impose a weak-coupling
scheme, the leading-order analogue would simply read $\beta_{\xi_2}=-g
D \xi_3$. Hence, $\xi_2\sim\mathcal{O}(g^0)$ is still a \fgFP\ if
$\xi_3=0$. With $\xi_3=\mathcal{O}(g)$ being the all-order result,
this lowest-order result implies an error of $\mathcal{O}(g^4)$ for
the six-point coupling $\lambda_3$.

This observation generalizes to the generalized scaling solutions
presented in Subsect.~\ref{subsec:gss}: the $\beta$ function for
$\xi_2$ to leading power in $g$ reads (cf. \Eqref{eq:mussnochschnellrein}),
\begin{equation}
\beta_{\xi_2} = - D g^{2P}\xi_3\ . \label{eq:mnsr2}
\end{equation}
Solving the \fgFPh\ equation for $\xi_3$ at arbitrary $\xi_2$, we find
the simple solution $\xi_3=0$.  This simple leading-order estimate
entails an error of $O(g^{8P})$ in the determination of $\lambda_3$
along the asymptotically free trajectory.

Let us now turn to the more general case that includes the relevant
direction, i.e. a possibly nontrivial field expectation value. We
start again with the $P=1/2$ case as discussed in
Subsect.~\ref{subsec:P=1/2}. To lowest order in the coupling, the
$\beta$ functions for $x_0$ and $\xi_2$ read,
cf. Eqs.~\eqref{eq:x0P05} and \eqref{eq:xi2P05},
\begin{eqnarray*}
\beta_{x_0}&=&-2x_0+ g \left(\frac{3}{16\pi^2} +\frac{9}{64\pi^2\xi_2}\right)+O(g^2)\\
&=&g \left(\frac{3}{16\pi^2}-2\kappa +\frac{9}{64\pi^2\xi_2}\right)+O(g^2)\\
\beta_{\xi_2}&=&g\left(-\frac{\xi_3}{16\pi^2}+\frac{9\xi_3}{64\pi^2\xi_2}\right)+O(g^2)
\end{eqnarray*}
The \fgFPh\ solution reproduces the correct value for $\kappa$, and
suggests $\xi_3=0$ as before.  Thus, the degree of accuracy of this
approximation, as far as couplings beyond $\phi^4$ are concerned,
would be the same as in the corresponding approximation without the
relevant direction.

This pattern generalizes to $P<1/2$ as discussed in
Subsect.~\ref{subsec:P<1/2_with_relevant}: the lowest-order
approximation does not affect $\beta_{x_0}$ in \Eqref{eq:x0Pless05},
so that the \fgFP\ for $\kappa=3/(32\pi^2)$ remains correct at that
order. The corresponding $\beta$ function for $\xi_2$ of
\Eqref{eq:xi2Pless05} simplifies to
$\beta_{\xi_2}=-g^{2P}\xi_3/(4\pi^2)$ with the solution $\xi_3=0$,
implying an error for $\lambda_3$ of $O(g^{8P})$. The same pattern
holds also for the lowest-order calculation for the case
$1/2<P<1$. The \fgFPh\ values of \Eqref{eq:FPP1} persist, whereas the
result $\xi_3=0$ to lowest order would represent an error of order
$O(g^{4P+2})$ in the determination of $\lambda_3$ along the
asymptotically free trajectories.

For $P\geq1$, the main text in Subsect.~\ref{subsec:P=1_with_relevant}
and subsequent sections is already devoted to a leading-order
analysis, yielding non-trivial values for $\xi_3$ at the \fgFP. In
conclusion, the effective-field theory and functional RG analysis are
partly complementary, but fully agree with each other in the
regime of overlapping applicability.

\section{Singularities, boundary conditions and free parameters of fixed point solutions}
\label{app:WF}

The determination and properties of fixed point solutions of
functional RG equations is a widely studied subject. These so-called
fixed functionals have been constructed for a wide variety of theories
with different methods, see,
e.g. \cite{Hasenfratz:1985dm,Morris:1998da,Bervillier:2007tc,Braun:2010tt,Codello:2012ec,Demmel:2015oqa,Dietz:2015owa,Vacca:2015nta,Hellwig:2015woa,Borchardt:2015rxa,Ohta:2015efa,Litim:2016hlb}. In
the present work, we use the shooting method for an analysis of the
solution space of the \fgFPh\ potentials of the nonabelian Higgs
model.

In order to highlight the similarities and differences to conventional
models, let us start here with a short recap of the Wilson-Fisher
fixed-point potential of the Ising model below $d<4$. For the purpose of
illustration, we use the simple local-potential approximation (LPA),
ignoring anomalous dimensions, and use the piece-wise linear
regulator. The fixed-point equation for the potential $v(\varphi)$ for
the dimensionless $\mathbbm{Z}_2$ order parameter $\varphi$ then reads
\cite{Litim:2000ci}
\begin{equation}
0=-dv+\frac{d-2}{2} \varphi v'+\frac{4v_d}{d} \frac{1}{1+v''}.
\label{eq:vfp}
\end{equation}
It is illustrative to write this ordinary differential equation as
\begin{equation}
v''= -\frac{4 v_d}{d} \frac{e(v,v';\varphi)}{s(v,v';\varphi)}, \quad s(v,v';\varphi)=-dv+\frac{d-2}{2} \varphi v',
\label{eq:vfnd}
\end{equation}
and the numerator function being $e=1+ds/(4v_d)$. Being a second order
equation, the solution manifold is generally parametrized by two
initial conditions.  $\mathbbm{Z}_2$ symmetry, requiring
$v'(\varphi=0)=0$, reduces the solution manifold to only one
parameter, e.g., the choice of $\sigma:= v''(\varphi=0)$. Still, there
is no one-parameter family of fixed-point solutions, because of the
fact that the scaling terms $s(v,v';\varphi)$ in the denominator of
\Eqref{eq:vfnd} generically exhibits a zero at some finite field
amplitude $\varphi_s$ when integrating the differential equation from
$\varphi=0$ to larger field values for a generic $\sigma$. This zero
of $s(v,v';\varphi)$ indicates the presence of a \textit{movable}
singularity in the fixed point equation. This singularity can be
lifted, if the numerator also vanishes sufficiently fast at the same
field value, $e(v,v';\varphi_s)=0$. This later condition effectively
fixes (quantizes) the remaining parameter $\sigma$. For the Ising
model, only one choice of $\sigma\neq0$ yields a nontrivial fixed
point potential which is globally defined for all
$\varphi\in\mathbbm{R}$; this solution is stable, i.e., bounded from
below and also matches the large-field asymptotics
\cite{Hasenfratz:1985dm,Morris:1994ki}.

As a second illustrative step, we rephrase the same model in the
language of the $\mathbbm{Z}_2$ invariant $\trho=\varphi^2/2$. The
resulting flow equation for the potential $u(\trho)=v(\varphi)$ can be
obtained by rewriting \Eqref{eq:vfp} accordingly as well as from the
flow equation \eqref{floweq:potential} by dropping the gauge
contributions and setting $N=1/2$ (as $N$ counts complex scalars). In
the LPA, we can bring the fixed-point equation into the form similar
to \Eqref{eq:vfnd},
\begin{equation}
u''= -\frac{2 v_d}{d} \frac{e(u,u';\trho)}{\trho\, s(u,u';\trho)}, \quad s(u,u';\trho)=-du+(d-2) \trho u',
\label{eq:ufnd}
\end{equation}
and the numerator function $e=1+d(1+u')s/(4v_d)$. In addition to the
movable singularity arising from $s(u,u';\trho)=0$ for some
$\trho=\trho_s$, there is now also a \textit{fixed} singularity of the
fixed point equation at $\trho=0$. The latter even inhibits to naively
integrate from $\rho=0$ to larger field values. Also, the boundary
condition establishing $\mathbbm{Z}_2$ symmetry is seemingly lost, as
$u(\trho)$ is already manifestly $\mathbbm{Z}_2$ invariant. 

In fact, $\mathbbm{Z}_2$ invariance and the occurrence of the fixed
singularity are connected: e.g., if a linear $\mathbbm{Z}_2$-violating
term $v(\varphi)\sim \varphi$ occurred in the potential near zero
field, we would have $u(\trho)\sim \sqrt{\trho}$ and the derivatives
$u', u''$ exhibited singularities.

These issues of the formulation using the invariant field variable
$\trho$, can be resolved by constructing the fixed-point potential by
starting from a finite field value. The two-dimensional manifold of
solutions to \Eqref{eq:ufnd} can, for instance, be parametrized by
demanding for $u'(\trho=\kappa)=0$ at a fiducial value for the minimum
$\kappa$, and similarly $u''(\trho=\kappa)=\sigma$ for the curvature
at the minimum. (Here we anticipated that the Wilson-Fisher
fixed-point potential has a nonvanishing minimum. For other systems,
alternative initial conditions can be chosen in terms of two
parameters.) By starting from $\trho=\kappa$ and integrating
\Eqref{eq:ufnd} both to larger and smaller field values, i.e.,
``shooting'' outward and inward, we obtain a two-parameter
$(\kappa,\sigma)$ family of solutions of \Eqref{eq:ufnd}. This time,
the integrations yield global solutions, if both singularities are
lifted by suitable conditions. As before, the movable singularity at
$s(u,u';\trho_s)=0$ has to be lifted by the condition
$e(u,u';\trho_s)=0$. In addition, the fixed singularity at $\trho=0$
has to be lifted by the further condition $e(u,u';\trho=0)=0$. In
total, we again obtain two conditions for the two-parameter family,
which leads to a quantization of these parameters and a discrete set
of fixed-point solutions. For $d=3$, we have checked that this method
rediscovers the Wilson-Fisher fixed-point potential, finding accurate
agreement with the known literature values. E.g. the position of the
minimum $\kappa\simeq0.0306479$ agrees on this accuracy level with the
highest-precision data for the LPA obtained in \cite{Borchardt:2015rxa}.

\section{Asymptotic expansions for large field amplitudes}
\label{app:largefields}

One remarkable property of functional RG equations like~\Eqref{floweq:fpotential}
is that they apply to arbitrarily large constant field amplitudes.
The corresponding asymptotic properties of the potential often play a crucial role in
constraining and characterizing the set of physically acceptable solutions. 
For the Higgs model such an equation has not only a functional dependence on
the average field $x$ 
(as well as the location of the nontrivial minimum $x_0$,
which enters through the anomalous dimensions),
 but also a parametric dependence on the gauge coupling $g^2$,
and the interplay between these two quantities gives rise to a rich spectrum of possible behaviors.

In particular, we have been mainly concerned with the understanding of the set
of consistent boundary conditions for \fgFPh{}
potentials.
We have addressed the construction of these functional solutions, both in a weak coupling expansion
and beyond.
In the first case, the ODE defining the problem is provided by \Eqref{eq:fgFPcond}.
For the $P=1$ case, we had to solve it numerically, and then 
the knowledge of the right large-$x$ behavior serves to validate/construct
the corresponding solution. 
These \fgFPh{} potentials must have an asymptotic behavior for large $x$ of the kind
\begin{equation}\label{eq:weak_g2_asymptotics_1}
f(x)\widesim[1.8]{x\to \infty}f_{\infty}(x) = \xi_\infty\  x^{N_\infty}+\sum_{n=0}^{n_S}\xi_{-n}x^{-n}
\end{equation}
where $\xi_\infty$ is an arbitrary integration constant,
while all the rest is a function of $g^2$ and $x_0$.
We adopted an expansion up to $n_S=2$, which is enough for the level of accuracy we are interested in.
Retaining the full anomalous dimensions but neglecting terms of order $O(g^4)$, and denoting $D_0=(1+x_0/2)^{-1}$, one finds 
\begin{align}\label{eq:weak_g2_asymptotics_2}
N_{\infty }&=2-\frac{g^2}{64\pi^2}\left(\frac{88}{3}-18 D_0^4+9D_0^2+19D_0\right)\\
\xi_{-0}&=\frac{1}{128\pi^2}+\frac{21 g^2}{(128\pi^2)^2}\left(2D_0^4-D_0^2+D_0\right)\nonumber\\
\xi_{-1}&=\frac{3}{32 \pi^2}+\frac{9g^2}{2(32 \pi^2)^2}\left(1+\frac{x_0}{4}\right) \left(1+\frac{x_0^2}{4}\right) D_0^4\nonumber\\
\xi_{-2}&=\frac{-9}{64\pi^2}+\frac{g^2}{(32\pi^2)^2}\left(\frac{88+3 D_0+81 D_0^2-162 D_0^4}{16}\right)\nonumber\, .
\end{align}
Since $N_\infty$ is positive and close to $2$, 
the scalar potential is stable along such trajectories 
(the case $\xi_{\infty}=0$ is also stable but not bounded, just like the $P=1,\xi=0$ fixed point).
Thus, it is not possible to split such a solution in a fixed-point potential plus a small fluctuation,
since the latter would grow like $x^2$ for large field values.

As \Eqref{eq:fgFPcond} arises by an expansion in $g^2$,
with respect to the full gauge dependence, this large-field
asymptotics is valid in an intermediate field-amplitude regime, where
$g^2x$ is still sufficiently small. In other words, it applies in the double limit
$x\to\infty$ and $g^2\to0$, provided the latter is stronger than the former.

Clearly, to go beyond the weak gauge coupling approximation,
we need a different asymptotic expansion.
This can be obtained by directly analyzing the complete $\beta$ functional of~\Eqref{floweq:fpotential}, for any $P$ and $g^2$.
Furthermore, in computing this asymptotic expansion, it is convenient to keep
$\eta_W$ and $\eta_\phi$ implicit.  Hence, the result remains valid
also upon retaining the full expressions (or even better approximations) 
for the anomalous dimensions.
The first five terms of such an expansion turn out to be sufficient
for our purposes,
\begin{eqnarray}\label{eq:general_asymptotics_large_x}
f(x)\widesim[1.8]{x\to \infty}f_{\infty}(x) &=& \xi_\infty\  x^{N_\infty}+\xi_{1-N_\infty}x^{1-N_\infty}+\xi_{-1}x^{-1}\nonumber\\
&&\!\! +\ \xi_{2\left(1-N_\infty\right)}x^{2\left(1-N_\infty\right)}+\xi_{-2}x^{-2}.
\end{eqnarray}
Again $\xi_\infty$ stays free, while the other
parameters can be written as functions of $P$, $g^2$, $\eta_\phi$, $\eta_W$
as follows:
\begin{align}
N_{\infty}&=4/d_x\ ,\quad d_x=2+\eta_\phi-P\eta_W\label{eq:asexp1bis}\\
\xi_{1-N_{\infty}}&=\frac{d_x \left(12-d_x\right)\left(6-\eta_\phi \right)}{384\pi^2 \xi_\infty g^{2P}\left(8-d_x\right)^2}\nonumber\\
\xi_{-1}&=\frac{3 g^{2(P-1)}\left(6-\eta_W \right)}{32\pi^2 \left(4+d_x\right)}\nonumber\\
\xi_{2\left(1-N_{\infty}\right)}&=-\frac{d_x^2 \left(48-d_x\left(12-d_x\right)\right)\left(6-\eta_\phi \right)}{1536\pi^2 \xi_\infty^2 g^{4P}\left(8-d_x\right)^2\left(6-d_x\right)}\nonumber\\
\xi_{-2}&=-\frac{3 g^{4(P-1)}\left(6-\eta_W \right)}{32\pi^2 \left(2+d_x\right)}\, .\nonumber
\end{align}
At this point one could ask whether the weak coupling approximation
of this expression agrees with Eqs.~(\ref{eq:weak_g2_asymptotics_1})
and~(\ref{eq:weak_g2_asymptotics_2}) in the $P=1$ case.
The agreement does occur at leading order, 
but it is restricted to the
integer expansion terms in Eqs.~(\ref{eq:general_asymptotics_large_x})
and~(\ref{eq:asexp1bis}). The reason for the discrepancy of the remaining
terms lies in the fact the the two limits 
$x\to\infty$ and $g^2\to0$ do not commute, such that the two expansions
refer to different asymptotic regions.

\bibliography{bibliography}

\begin{thebibliography}{102}%
\makeatletter
\providecommand \@ifxundefined [1]{%
 \@ifx{#1\undefined}
}%
\providecommand \@ifnum [1]{%
 \ifnum #1\expandafter \@firstoftwo
 \else \expandafter \@secondoftwo
 \fi
}%
\providecommand \@ifx [1]{%
 \ifx #1\expandafter \@firstoftwo
 \else \expandafter \@secondoftwo
 \fi
}%
\providecommand \natexlab [1]{#1}%
\providecommand \enquote  [1]{``#1''}%
\providecommand \bibnamefont  [1]{#1}%
\providecommand \bibfnamefont [1]{#1}%
\providecommand \citenamefont [1]{#1}%
\providecommand \href@noop [0]{\@secondoftwo}%
\providecommand \href [0]{\begingroup \@sanitize@url \@href}%
\providecommand \@href[1]{\@@startlink{#1}\@@href}%
\providecommand \@@href[1]{\endgroup#1\@@endlink}%
\providecommand \@sanitize@url [0]{\catcode `\\12\catcode `\$12\catcode
  `\&12\catcode `\#12\catcode `\^12\catcode `\_12\catcode `\%12\relax}%
\providecommand \@@startlink[1]{}%
\providecommand \@@endlink[0]{}%
\providecommand \url  [0]{\begingroup\@sanitize@url \@url }%
\providecommand \@url [1]{\endgroup\@href {#1}{\urlprefix }}%
\providecommand \urlprefix  [0]{URL }%
\providecommand \Eprint [0]{\href }%
\providecommand \doibase [0]{http://dx.doi.org/}%
\providecommand \selectlanguage [0]{\@gobble}%
\providecommand \bibinfo  [0]{\@secondoftwo}%
\providecommand \bibfield  [0]{\@secondoftwo}%
\providecommand \translation [1]{[#1]}%
\providecommand \BibitemOpen [0]{}%
\providecommand \bibitemStop [0]{}%
\providecommand \bibitemNoStop [0]{.\EOS\space}%
\providecommand \EOS [0]{\spacefactor3000\relax}%
\providecommand \BibitemShut  [1]{\csname bibitem#1\endcsname}%
\let\auto@bib@innerbib\@empty
\bibitem [{\citenamefont {Gross}\ and\ \citenamefont
  {Wilczek}(1973{\natexlab{a}})}]{Gross:1973id}%
  \BibitemOpen
  \bibfield  {author} {\bibinfo {author} {\bibfnamefont {D.~J.}\ \bibnamefont
  {Gross}}\ and\ \bibinfo {author} {\bibfnamefont {F.}~\bibnamefont
  {Wilczek}},\ }\href {\doibase 10.1103/PhysRevLett.30.1343} {\bibfield
  {journal} {\bibinfo  {journal} {Phys. Rev. Lett.}\ }\textbf {\bibinfo
  {volume} {30}},\ \bibinfo {pages} {1343} (\bibinfo {year}
  {1973}{\natexlab{a}})}\BibitemShut {NoStop}%
\bibitem [{\citenamefont {Politzer}(1973)}]{Politzer:1973fx}%
  \BibitemOpen
  \bibfield  {author} {\bibinfo {author} {\bibfnamefont {H.~D.}\ \bibnamefont
  {Politzer}},\ }\href {\doibase 10.1103/PhysRevLett.30.1346} {\bibfield
  {journal} {\bibinfo  {journal} {Phys. Rev. Lett.}\ }\textbf {\bibinfo
  {volume} {30}},\ \bibinfo {pages} {1346} (\bibinfo {year}
  {1973})}\BibitemShut {NoStop}%
\bibitem [{\citenamefont {Gross}\ and\ \citenamefont
  {Wilczek}(1973{\natexlab{b}})}]{Gross:1973ju}%
  \BibitemOpen
  \bibfield  {author} {\bibinfo {author} {\bibfnamefont {D.~J.}\ \bibnamefont
  {Gross}}\ and\ \bibinfo {author} {\bibfnamefont {F.}~\bibnamefont
  {Wilczek}},\ }\href {\doibase 10.1103/PhysRevD.8.3633} {\bibfield  {journal}
  {\bibinfo  {journal} {Phys. Rev.}\ }\textbf {\bibinfo {volume} {D8}},\
  \bibinfo {pages} {3633} (\bibinfo {year} {1973}{\natexlab{b}})}\BibitemShut
  {NoStop}%
\bibitem [{\citenamefont {Gross}\ and\ \citenamefont
  {Wilczek}(1974)}]{Gross:1974cs}%
  \BibitemOpen
  \bibfield  {author} {\bibinfo {author} {\bibfnamefont {D.~J.}\ \bibnamefont
  {Gross}}\ and\ \bibinfo {author} {\bibfnamefont {F.}~\bibnamefont
  {Wilczek}},\ }\href {\doibase 10.1103/PhysRevD.9.980} {\bibfield  {journal}
  {\bibinfo  {journal} {Phys. Rev.}\ }\textbf {\bibinfo {volume} {D9}},\
  \bibinfo {pages} {980} (\bibinfo {year} {1974})}\BibitemShut {NoStop}%
\bibitem [{\citenamefont {Politzer}(1974)}]{Politzer:1974fr}%
  \BibitemOpen
  \bibfield  {author} {\bibinfo {author} {\bibfnamefont {H.~D.}\ \bibnamefont
  {Politzer}},\ }\href {\doibase 10.1016/0370-1573(74)90014-3} {\bibfield
  {journal} {\bibinfo  {journal} {Phys. Rept.}\ }\textbf {\bibinfo {volume}
  {14}},\ \bibinfo {pages} {129} (\bibinfo {year} {1974})}\BibitemShut
  {NoStop}%
\bibitem [{\citenamefont {Cheng}\ \emph {et~al.}(1974)\citenamefont {Cheng},
  \citenamefont {Eichten},\ and\ \citenamefont {Li}}]{Cheng:1973nv}%
  \BibitemOpen
  \bibfield  {author} {\bibinfo {author} {\bibfnamefont {T.~P.}\ \bibnamefont
  {Cheng}}, \bibinfo {author} {\bibfnamefont {E.}~\bibnamefont {Eichten}}, \
  and\ \bibinfo {author} {\bibfnamefont {L.-F.}\ \bibnamefont {Li}},\ }\href
  {\doibase 10.1103/PhysRevD.9.2259} {\bibfield  {journal} {\bibinfo  {journal}
  {Phys. Rev.}\ }\textbf {\bibinfo {volume} {D9}},\ \bibinfo {pages} {2259}
  (\bibinfo {year} {1974})}\BibitemShut {NoStop}%
\bibitem [{\citenamefont {Bais}\ and\ \citenamefont
  {Weldon}(1978)}]{Bais:1978fv}%
  \BibitemOpen
  \bibfield  {author} {\bibinfo {author} {\bibfnamefont {F.~A.}\ \bibnamefont
  {Bais}}\ and\ \bibinfo {author} {\bibfnamefont {H.~A.}\ \bibnamefont
  {Weldon}},\ }\href {\doibase 10.1103/PhysRevD.18.1199} {\bibfield  {journal}
  {\bibinfo  {journal} {Phys. Rev.}\ }\textbf {\bibinfo {volume} {D18}},\
  \bibinfo {pages} {1199} (\bibinfo {year} {1978})}\BibitemShut {NoStop}%
\bibitem [{\citenamefont {Callaway}(1988)}]{Callaway:1988ya}%
  \BibitemOpen
  \bibfield  {author} {\bibinfo {author} {\bibfnamefont {D.~J.~E.}\
  \bibnamefont {Callaway}},\ }\href {\doibase 10.1016/0370-1573(88)90008-7}
  {\bibfield  {journal} {\bibinfo  {journal} {Phys. Rept.}\ }\textbf {\bibinfo
  {volume} {167}},\ \bibinfo {pages} {241} (\bibinfo {year}
  {1988})}\BibitemShut {NoStop}%
\bibitem [{\citenamefont {Wilson}\ and\ \citenamefont
  {Kogut}(1974)}]{Wilson:1973jj}%
  \BibitemOpen
  \bibfield  {author} {\bibinfo {author} {\bibfnamefont {K.~G.}\ \bibnamefont
  {Wilson}}\ and\ \bibinfo {author} {\bibfnamefont {J.~B.}\ \bibnamefont
  {Kogut}},\ }\href {\doibase 10.1016/0370-1573(74)90023-4} {\bibfield
  {journal} {\bibinfo  {journal} {Phys. Rept.}\ }\textbf {\bibinfo {volume}
  {12}},\ \bibinfo {pages} {75} (\bibinfo {year} {1974})}\BibitemShut {NoStop}%
\bibitem [{\citenamefont {Frohlich}(1982)}]{Frohlich:1982tw}%
  \BibitemOpen
  \bibfield  {author} {\bibinfo {author} {\bibfnamefont {J.}~\bibnamefont
  {Frohlich}},\ }\href {\doibase 10.1016/0550-3213(82)90088-8} {\bibfield
  {journal} {\bibinfo  {journal} {Nucl. Phys.}\ }\textbf {\bibinfo {volume}
  {B200}},\ \bibinfo {pages} {281} (\bibinfo {year} {1982})}\BibitemShut
  {NoStop}%
\bibitem [{\citenamefont {Luscher}\ and\ \citenamefont
  {Weisz}(1987)}]{Luscher:1987ay}%
  \BibitemOpen
  \bibfield  {author} {\bibinfo {author} {\bibfnamefont {M.}~\bibnamefont
  {Luscher}}\ and\ \bibinfo {author} {\bibfnamefont {P.}~\bibnamefont
  {Weisz}},\ }\href {\doibase 10.1016/0550-3213(87)90177-5} {\bibfield
  {journal} {\bibinfo  {journal} {Nucl. Phys.}\ }\textbf {\bibinfo {volume}
  {B290}},\ \bibinfo {pages} {25} (\bibinfo {year} {1987})}\BibitemShut
  {NoStop}%
\bibitem [{\citenamefont {Luscher}\ and\ \citenamefont
  {Weisz}(1988)}]{Luscher:1987ek}%
  \BibitemOpen
  \bibfield  {author} {\bibinfo {author} {\bibfnamefont {M.}~\bibnamefont
  {Luscher}}\ and\ \bibinfo {author} {\bibfnamefont {P.}~\bibnamefont
  {Weisz}},\ }\href {\doibase 10.1016/0550-3213(88)90228-3} {\bibfield
  {journal} {\bibinfo  {journal} {Nucl. Phys.}\ }\textbf {\bibinfo {volume}
  {B295}},\ \bibinfo {pages} {65} (\bibinfo {year} {1988})}\BibitemShut
  {NoStop}%
\bibitem [{\citenamefont {Luscher}\ and\ \citenamefont
  {Weisz}(1989)}]{Luscher:1988uq}%
  \BibitemOpen
  \bibfield  {author} {\bibinfo {author} {\bibfnamefont {M.}~\bibnamefont
  {Luscher}}\ and\ \bibinfo {author} {\bibfnamefont {P.}~\bibnamefont
  {Weisz}},\ }\href {\doibase 10.1016/0550-3213(89)90637-8} {\bibfield
  {journal} {\bibinfo  {journal} {Nucl. Phys.}\ }\textbf {\bibinfo {volume}
  {B318}},\ \bibinfo {pages} {705} (\bibinfo {year} {1989})}\BibitemShut
  {NoStop}%
\bibitem [{\citenamefont {Hasenfratz}\ \emph {et~al.}(1987)\citenamefont
  {Hasenfratz}, \citenamefont {Jansen}, \citenamefont {Lang}, \citenamefont
  {Neuhaus},\ and\ \citenamefont {Yoneyama}}]{Hasenfratz:1987eh}%
  \BibitemOpen
  \bibfield  {author} {\bibinfo {author} {\bibfnamefont {A.}~\bibnamefont
  {Hasenfratz}}, \bibinfo {author} {\bibfnamefont {K.}~\bibnamefont {Jansen}},
  \bibinfo {author} {\bibfnamefont {C.~B.}\ \bibnamefont {Lang}}, \bibinfo
  {author} {\bibfnamefont {T.}~\bibnamefont {Neuhaus}}, \ and\ \bibinfo
  {author} {\bibfnamefont {H.}~\bibnamefont {Yoneyama}},\ }\href {\doibase
  10.1016/0370-2693(87)91622-4} {\bibfield  {journal} {\bibinfo  {journal}
  {Phys. Lett.}\ }\textbf {\bibinfo {volume} {B199}},\ \bibinfo {pages} {531}
  (\bibinfo {year} {1987})}\BibitemShut {NoStop}%
\bibitem [{\citenamefont {Heller}\ \emph {et~al.}(1993)\citenamefont {Heller},
  \citenamefont {Neuberger},\ and\ \citenamefont {Vranas}}]{Heller:1992js}%
  \BibitemOpen
  \bibfield  {author} {\bibinfo {author} {\bibfnamefont {U.~M.}\ \bibnamefont
  {Heller}}, \bibinfo {author} {\bibfnamefont {H.}~\bibnamefont {Neuberger}}, \
  and\ \bibinfo {author} {\bibfnamefont {P.~M.}\ \bibnamefont {Vranas}},\
  }\href {\doibase 10.1016/0550-3213(93)90499-F} {\bibfield  {journal}
  {\bibinfo  {journal} {Nucl. Phys.}\ }\textbf {\bibinfo {volume} {B399}},\
  \bibinfo {pages} {271} (\bibinfo {year} {1993})},\ \Eprint
  {http://arxiv.org/abs/hep-lat/9207024} {arXiv:hep-lat/9207024 [hep-lat]}
  \BibitemShut {NoStop}%
\bibitem [{\citenamefont {Wolff}(2009)}]{Wolff:2009ke}%
  \BibitemOpen
  \bibfield  {author} {\bibinfo {author} {\bibfnamefont {U.}~\bibnamefont
  {Wolff}},\ }\href {\doibase 10.1103/PhysRevD.79.105002} {\bibfield  {journal}
  {\bibinfo  {journal} {Phys. Rev.}\ }\textbf {\bibinfo {volume} {D79}},\
  \bibinfo {pages} {105002} (\bibinfo {year} {2009})},\ \Eprint
  {http://arxiv.org/abs/0902.3100} {arXiv:0902.3100 [hep-lat]} \BibitemShut
  {NoStop}%
\bibitem [{\citenamefont {Buividovich}(2011)}]{Buividovich:2011zy}%
  \BibitemOpen
  \bibfield  {author} {\bibinfo {author} {\bibfnamefont {P.~V.}\ \bibnamefont
  {Buividovich}},\ }\href {\doibase 10.1016/j.nuclphysb.2011.08.010} {\bibfield
   {journal} {\bibinfo  {journal} {Nucl. Phys.}\ }\textbf {\bibinfo {volume}
  {B853}},\ \bibinfo {pages} {688} (\bibinfo {year} {2011})},\ \Eprint
  {http://arxiv.org/abs/1104.3459} {arXiv:1104.3459 [hep-lat]} \BibitemShut
  {NoStop}%
\bibitem [{\citenamefont {Rosten}(2009)}]{Rosten:2008ts}%
  \BibitemOpen
  \bibfield  {author} {\bibinfo {author} {\bibfnamefont {O.~J.}\ \bibnamefont
  {Rosten}},\ }\href {\doibase 10.1088/1126-6708/2009/07/019} {\bibfield
  {journal} {\bibinfo  {journal} {JHEP}\ }\textbf {\bibinfo {volume} {07}},\
  \bibinfo {pages} {019} (\bibinfo {year} {2009})},\ \Eprint
  {http://arxiv.org/abs/0808.0082} {arXiv:0808.0082 [hep-th]} \BibitemShut
  {NoStop}%
\bibitem [{\citenamefont {Gockeler}\ \emph {et~al.}(1998)\citenamefont
  {Gockeler}, \citenamefont {Horsley}, \citenamefont {Linke}, \citenamefont
  {Rakow}, \citenamefont {Schierholz},\ and\ \citenamefont
  {Stuben}}]{Gockeler:1997dn}%
  \BibitemOpen
  \bibfield  {author} {\bibinfo {author} {\bibfnamefont {M.}~\bibnamefont
  {Gockeler}}, \bibinfo {author} {\bibfnamefont {R.}~\bibnamefont {Horsley}},
  \bibinfo {author} {\bibfnamefont {V.}~\bibnamefont {Linke}}, \bibinfo
  {author} {\bibfnamefont {P.~E.~L.}\ \bibnamefont {Rakow}}, \bibinfo {author}
  {\bibfnamefont {G.}~\bibnamefont {Schierholz}}, \ and\ \bibinfo {author}
  {\bibfnamefont {H.}~\bibnamefont {Stuben}},\ }\href {\doibase
  10.1103/PhysRevLett.80.4119} {\bibfield  {journal} {\bibinfo  {journal}
  {Phys. Rev. Lett.}\ }\textbf {\bibinfo {volume} {80}},\ \bibinfo {pages}
  {4119} (\bibinfo {year} {1998})},\ \Eprint
  {http://arxiv.org/abs/hep-th/9712244} {arXiv:hep-th/9712244 [hep-th]}
  \BibitemShut {NoStop}%
\bibitem [{\citenamefont {Gies}\ and\ \citenamefont
  {Jaeckel}(2004)}]{Gies:2004hy}%
  \BibitemOpen
  \bibfield  {author} {\bibinfo {author} {\bibfnamefont {H.}~\bibnamefont
  {Gies}}\ and\ \bibinfo {author} {\bibfnamefont {J.}~\bibnamefont {Jaeckel}},\
  }\href {\doibase 10.1103/PhysRevLett.93.110405} {\bibfield  {journal}
  {\bibinfo  {journal} {Phys. Rev. Lett.}\ }\textbf {\bibinfo {volume} {93}},\
  \bibinfo {pages} {110405} (\bibinfo {year} {2004})},\ \Eprint
  {http://arxiv.org/abs/hep-ph/0405183} {arXiv:hep-ph/0405183 [hep-ph]}
  \BibitemShut {NoStop}%
\bibitem [{\citenamefont {Landau}()}]{Landau:1955}%
  \BibitemOpen
  \bibfield  {author} {\bibinfo {author} {\bibfnamefont {L.~D.}\ \bibnamefont
  {Landau}},\ }in\ \href@noop {} {\emph {\bibinfo {booktitle} {{Niels Bohr and
  the Development of Physics, ed.~Wolfgang Pauli, London: Pergamon
  Press}}}}\BibitemShut {NoStop}%
\bibitem [{\citenamefont {Gell-Mann}\ and\ \citenamefont
  {Low}(1954)}]{GellMann:1954fq}%
  \BibitemOpen
  \bibfield  {author} {\bibinfo {author} {\bibfnamefont {M.}~\bibnamefont
  {Gell-Mann}}\ and\ \bibinfo {author} {\bibfnamefont {F.~E.}\ \bibnamefont
  {Low}},\ }\href {\doibase 10.1103/PhysRev.95.1300} {\bibfield  {journal}
  {\bibinfo  {journal} {Phys. Rev.}\ }\textbf {\bibinfo {volume} {95}},\
  \bibinfo {pages} {1300} (\bibinfo {year} {1954})}\BibitemShut {NoStop}%
\bibitem [{\citenamefont {Chang}(1974)}]{Chang:1974bv}%
  \BibitemOpen
  \bibfield  {author} {\bibinfo {author} {\bibfnamefont {N.-P.}\ \bibnamefont
  {Chang}},\ }\href {\doibase 10.1103/PhysRevD.10.2706} {\bibfield  {journal}
  {\bibinfo  {journal} {Phys. Rev.}\ }\textbf {\bibinfo {volume} {D10}},\
  \bibinfo {pages} {2706} (\bibinfo {year} {1974})}\BibitemShut {NoStop}%
\bibitem [{\citenamefont {Chang}\ and\ \citenamefont
  {Perez-Mercader}(1978)}]{Chang:1978nu}%
  \BibitemOpen
  \bibfield  {author} {\bibinfo {author} {\bibfnamefont {N.-P.}\ \bibnamefont
  {Chang}}\ and\ \bibinfo {author} {\bibfnamefont {J.}~\bibnamefont
  {Perez-Mercader}},\ }\href {\doibase 10.1103/PhysRevD.18.4721,
  10.1103/PhysRevD.19.2515} {\bibfield  {journal} {\bibinfo  {journal} {Phys.
  Rev.}\ }\textbf {\bibinfo {volume} {D18}},\ \bibinfo {pages} {4721} (\bibinfo
  {year} {1978})},\ \bibinfo {note} {[Erratum: Phys.
  Rev.D19,2515(1979)]}\BibitemShut {NoStop}%
\bibitem [{\citenamefont {Fradkin}\ and\ \citenamefont
  {Kalashnikov}(1975)}]{Fradkin:1975yt}%
  \BibitemOpen
  \bibfield  {author} {\bibinfo {author} {\bibfnamefont {E.~S.}\ \bibnamefont
  {Fradkin}}\ and\ \bibinfo {author} {\bibfnamefont {O.~K.}\ \bibnamefont
  {Kalashnikov}},\ }\href {\doibase 10.1088/0305-4470/8/11/017} {\bibfield
  {journal} {\bibinfo  {journal} {J. Phys.}\ }\textbf {\bibinfo {volume}
  {A8}},\ \bibinfo {pages} {1814} (\bibinfo {year} {1975})}\BibitemShut
  {NoStop}%
\bibitem [{\citenamefont {Salam}\ and\ \citenamefont
  {Strathdee}(1978)}]{Salam:1978dk}%
  \BibitemOpen
  \bibfield  {author} {\bibinfo {author} {\bibfnamefont {A.}~\bibnamefont
  {Salam}}\ and\ \bibinfo {author} {\bibfnamefont {J.~A.}\ \bibnamefont
  {Strathdee}},\ }\href {\doibase 10.1103/PhysRevD.18.4713} {\bibfield
  {journal} {\bibinfo  {journal} {Phys. Rev.}\ }\textbf {\bibinfo {volume}
  {D18}},\ \bibinfo {pages} {4713} (\bibinfo {year} {1978})}\BibitemShut
  {NoStop}%
\bibitem [{\citenamefont {Salam}\ and\ \citenamefont
  {Elias}(1980)}]{Salam:1980ss}%
  \BibitemOpen
  \bibfield  {author} {\bibinfo {author} {\bibfnamefont {A.}~\bibnamefont
  {Salam}}\ and\ \bibinfo {author} {\bibfnamefont {V.}~\bibnamefont {Elias}},\
  }\href {\doibase 10.1103/PhysRevD.22.1469} {\bibfield  {journal} {\bibinfo
  {journal} {Phys. Rev.}\ }\textbf {\bibinfo {volume} {D22}},\ \bibinfo {pages}
  {1469} (\bibinfo {year} {1980})}\BibitemShut {NoStop}%
\bibitem [{\citenamefont {Giudice}\ \emph {et~al.}(2015)\citenamefont
  {Giudice}, \citenamefont {Isidori}, \citenamefont {Salvio},\ and\
  \citenamefont {Strumia}}]{Giudice:2014tma}%
  \BibitemOpen
  \bibfield  {author} {\bibinfo {author} {\bibfnamefont {G.~F.}\ \bibnamefont
  {Giudice}}, \bibinfo {author} {\bibfnamefont {G.}~\bibnamefont {Isidori}},
  \bibinfo {author} {\bibfnamefont {A.}~\bibnamefont {Salvio}}, \ and\ \bibinfo
  {author} {\bibfnamefont {A.}~\bibnamefont {Strumia}},\ }\href {\doibase
  10.1007/JHEP02(2015)137} {\bibfield  {journal} {\bibinfo  {journal} {JHEP}\
  }\textbf {\bibinfo {volume} {02}},\ \bibinfo {pages} {137} (\bibinfo {year}
  {2015})},\ \Eprint {http://arxiv.org/abs/1412.2769} {arXiv:1412.2769
  [hep-ph]} \BibitemShut {NoStop}%
\bibitem [{\citenamefont {Holdom}\ \emph {et~al.}(2015)\citenamefont {Holdom},
  \citenamefont {Ren},\ and\ \citenamefont {Zhang}}]{Holdom:2014hla}%
  \BibitemOpen
  \bibfield  {author} {\bibinfo {author} {\bibfnamefont {B.}~\bibnamefont
  {Holdom}}, \bibinfo {author} {\bibfnamefont {J.}~\bibnamefont {Ren}}, \ and\
  \bibinfo {author} {\bibfnamefont {C.}~\bibnamefont {Zhang}},\ }\href
  {\doibase 10.1007/JHEP03(2015)028} {\bibfield  {journal} {\bibinfo  {journal}
  {JHEP}\ }\textbf {\bibinfo {volume} {03}},\ \bibinfo {pages} {028} (\bibinfo
  {year} {2015})},\ \Eprint {http://arxiv.org/abs/1412.5540} {arXiv:1412.5540
  [hep-ph]} \BibitemShut {NoStop}%
\bibitem [{\citenamefont {Zimmermann}(1985)}]{Zimmermann:1984sx}%
  \BibitemOpen
  \bibfield  {author} {\bibinfo {author} {\bibfnamefont {W.}~\bibnamefont
  {Zimmermann}},\ }\href {\doibase 10.1007/BF01206187} {\bibfield  {journal}
  {\bibinfo  {journal} {Commun. Math. Phys.}\ }\textbf {\bibinfo {volume}
  {97}},\ \bibinfo {pages} {211} (\bibinfo {year} {1985})}\BibitemShut
  {NoStop}%
\bibitem [{\citenamefont {Oehme}\ and\ \citenamefont
  {Zimmermann}(1985)}]{Oehme:1984yy}%
  \BibitemOpen
  \bibfield  {author} {\bibinfo {author} {\bibfnamefont {R.}~\bibnamefont
  {Oehme}}\ and\ \bibinfo {author} {\bibfnamefont {W.}~\bibnamefont
  {Zimmermann}},\ }\href {\doibase 10.1007/BF01221218} {\bibfield  {journal}
  {\bibinfo  {journal} {Commun. Math. Phys.}\ }\textbf {\bibinfo {volume}
  {97}},\ \bibinfo {pages} {569} (\bibinfo {year} {1985})}\BibitemShut
  {NoStop}%
\bibitem [{\citenamefont {Heinemeyer}\ \emph {et~al.}(2014)\citenamefont
  {Heinemeyer}, \citenamefont {Kubo}, \citenamefont {Mondragon}, \citenamefont
  {Piguet}, \citenamefont {Sibold}, \citenamefont {Zimmermann},\ and\
  \citenamefont {Zoupanos}}]{Heinemeyer:2014vxa}%
  \BibitemOpen
  \bibfield  {author} {\bibinfo {author} {\bibfnamefont {S.}~\bibnamefont
  {Heinemeyer}}, \bibinfo {author} {\bibfnamefont {J.}~\bibnamefont {Kubo}},
  \bibinfo {author} {\bibfnamefont {M.}~\bibnamefont {Mondragon}}, \bibinfo
  {author} {\bibfnamefont {O.}~\bibnamefont {Piguet}}, \bibinfo {author}
  {\bibfnamefont {K.}~\bibnamefont {Sibold}}, \bibinfo {author} {\bibfnamefont
  {W.}~\bibnamefont {Zimmermann}}, \ and\ \bibinfo {author} {\bibfnamefont
  {G.}~\bibnamefont {Zoupanos}},\ }\href@noop {} {\  (\bibinfo {year}
  {2014})},\ \Eprint {http://arxiv.org/abs/1411.7155} {arXiv:1411.7155
  [hep-ph]} \BibitemShut {NoStop}%
\bibitem [{\citenamefont {Hetzel}\ and\ \citenamefont
  {Stech}(2015)}]{Hetzel:2015bla}%
  \BibitemOpen
  \bibfield  {author} {\bibinfo {author} {\bibfnamefont {J.}~\bibnamefont
  {Hetzel}}\ and\ \bibinfo {author} {\bibfnamefont {B.}~\bibnamefont {Stech}},\
  }\href {\doibase 10.1103/PhysRevD.91.055026} {\bibfield  {journal} {\bibinfo
  {journal} {Phys. Rev.}\ }\textbf {\bibinfo {volume} {D91}},\ \bibinfo {pages}
  {055026} (\bibinfo {year} {2015})},\ \Eprint
  {http://arxiv.org/abs/1502.00919} {arXiv:1502.00919 [hep-ph]} \BibitemShut
  {NoStop}%
\bibitem [{\citenamefont {Pelaggi}\ \emph {et~al.}(2015)\citenamefont
  {Pelaggi}, \citenamefont {Strumia},\ and\ \citenamefont
  {Vignali}}]{Pelaggi:2015kna}%
  \BibitemOpen
  \bibfield  {author} {\bibinfo {author} {\bibfnamefont {G.~M.}\ \bibnamefont
  {Pelaggi}}, \bibinfo {author} {\bibfnamefont {A.}~\bibnamefont {Strumia}}, \
  and\ \bibinfo {author} {\bibfnamefont {S.}~\bibnamefont {Vignali}},\ }\href
  {\doibase 10.1007/JHEP08(2015)130} {\bibfield  {journal} {\bibinfo  {journal}
  {JHEP}\ }\textbf {\bibinfo {volume} {08}},\ \bibinfo {pages} {130} (\bibinfo
  {year} {2015})},\ \Eprint {http://arxiv.org/abs/1507.06848} {arXiv:1507.06848
  [hep-ph]} \BibitemShut {NoStop}%
\bibitem [{\citenamefont {Pica}\ \emph {et~al.}(2016)\citenamefont {Pica},
  \citenamefont {Ryttov},\ and\ \citenamefont {Sannino}}]{Pica:2016krb}%
  \BibitemOpen
  \bibfield  {author} {\bibinfo {author} {\bibfnamefont {C.}~\bibnamefont
  {Pica}}, \bibinfo {author} {\bibfnamefont {T.~A.}\ \bibnamefont {Ryttov}}, \
  and\ \bibinfo {author} {\bibfnamefont {F.}~\bibnamefont {Sannino}},\
  }\href@noop {} {\  (\bibinfo {year} {2016})},\ \Eprint
  {http://arxiv.org/abs/1605.04712} {arXiv:1605.04712 [hep-th]} \BibitemShut
  {NoStop}%
\bibitem [{\citenamefont {Molgaard}\ and\ \citenamefont
  {Sannino}(2016)}]{Molgaard:2016bqf}%
  \BibitemOpen
  \bibfield  {author} {\bibinfo {author} {\bibfnamefont {E.}~\bibnamefont
  {Molgaard}}\ and\ \bibinfo {author} {\bibfnamefont {F.}~\bibnamefont
  {Sannino}},\ }\href@noop {} {\  (\bibinfo {year} {2016})},\ \Eprint
  {http://arxiv.org/abs/1610.03130} {arXiv:1610.03130 [hep-ph]} \BibitemShut
  {NoStop}%
\bibitem [{\citenamefont {Gies}\ and\ \citenamefont
  {Zambelli}(2015)}]{Gies:2015lia}%
  \BibitemOpen
  \bibfield  {author} {\bibinfo {author} {\bibfnamefont {H.}~\bibnamefont
  {Gies}}\ and\ \bibinfo {author} {\bibfnamefont {L.}~\bibnamefont
  {Zambelli}},\ }\href {\doibase 10.1103/PhysRevD.92.025016} {\bibfield
  {journal} {\bibinfo  {journal} {Phys. Rev.}\ }\textbf {\bibinfo {volume}
  {D92}},\ \bibinfo {pages} {025016} (\bibinfo {year} {2015})},\ \Eprint
  {http://arxiv.org/abs/1502.05907} {arXiv:1502.05907 [hep-ph]} \BibitemShut
  {NoStop}%
\bibitem [{\citenamefont {Hasenfratz}\ and\ \citenamefont
  {Hasenfratz}(1986)}]{Hasenfratz:1985dm}%
  \BibitemOpen
  \bibfield  {author} {\bibinfo {author} {\bibfnamefont {A.}~\bibnamefont
  {Hasenfratz}}\ and\ \bibinfo {author} {\bibfnamefont {P.}~\bibnamefont
  {Hasenfratz}},\ }\href {\doibase 10.1016/0550-3213(86)90573-0} {\bibfield
  {journal} {\bibinfo  {journal} {Nucl. Phys.}\ }\textbf {\bibinfo {volume}
  {B270}},\ \bibinfo {pages} {687} (\bibinfo {year} {1986})}\BibitemShut
  {NoStop}%
\bibitem [{\citenamefont {Morris}(1998)}]{Morris:1998da}%
  \BibitemOpen
  \bibfield  {author} {\bibinfo {author} {\bibfnamefont {T.~R.}\ \bibnamefont
  {Morris}},\ }\bibfield  {booktitle} {\emph {\bibinfo {booktitle}
  {{Nonperturbative QCD: Structure of the QCD vacuum}}},\ }\href {\doibase
  10.1143/PTPS.131.395} {\bibfield  {journal} {\bibinfo  {journal} {Prog.
  Theor. Phys. Suppl.}\ }\textbf {\bibinfo {volume} {131}},\ \bibinfo {pages}
  {395} (\bibinfo {year} {1998})},\ \Eprint
  {http://arxiv.org/abs/hep-th/9802039} {arXiv:hep-th/9802039 [hep-th]}
  \BibitemShut {NoStop}%
\bibitem [{\citenamefont {Morris}(1997)}]{Morris:1996xq}%
  \BibitemOpen
  \bibfield  {author} {\bibinfo {author} {\bibfnamefont {T.~R.}\ \bibnamefont
  {Morris}},\ }\href {\doibase 10.1016/S0550-3213(97)00233-2} {\bibfield
  {journal} {\bibinfo  {journal} {Nucl. Phys.}\ }\textbf {\bibinfo {volume}
  {B495}},\ \bibinfo {pages} {477} (\bibinfo {year} {1997})},\ \Eprint
  {http://arxiv.org/abs/hep-th/9612117} {arXiv:hep-th/9612117 [hep-th]}
  \BibitemShut {NoStop}%
\bibitem [{\citenamefont {Morris}\ and\ \citenamefont
  {Turner}(1998)}]{Morris:1997xj}%
  \BibitemOpen
  \bibfield  {author} {\bibinfo {author} {\bibfnamefont {T.~R.}\ \bibnamefont
  {Morris}}\ and\ \bibinfo {author} {\bibfnamefont {M.~D.}\ \bibnamefont
  {Turner}},\ }\href {\doibase 10.1016/S0550-3213(97)00640-8} {\bibfield
  {journal} {\bibinfo  {journal} {Nucl. Phys.}\ }\textbf {\bibinfo {volume}
  {B509}},\ \bibinfo {pages} {637} (\bibinfo {year} {1998})},\ \Eprint
  {http://arxiv.org/abs/hep-th/9704202} {arXiv:hep-th/9704202 [hep-th]}
  \BibitemShut {NoStop}%
\bibitem [{\citenamefont {Dietz}\ and\ \citenamefont
  {Morris}(2013)}]{Dietz:2012ic}%
  \BibitemOpen
  \bibfield  {author} {\bibinfo {author} {\bibfnamefont {J.~A.}\ \bibnamefont
  {Dietz}}\ and\ \bibinfo {author} {\bibfnamefont {T.~R.}\ \bibnamefont
  {Morris}},\ }\href {\doibase 10.1007/JHEP01(2013)108} {\bibfield  {journal}
  {\bibinfo  {journal} {JHEP}\ }\textbf {\bibinfo {volume} {01}},\ \bibinfo
  {pages} {108} (\bibinfo {year} {2013})},\ \Eprint
  {http://arxiv.org/abs/1211.0955} {arXiv:1211.0955 [hep-th]} \BibitemShut
  {NoStop}%
\bibitem [{\citenamefont {Demmel}\ \emph {et~al.}(2015)\citenamefont {Demmel},
  \citenamefont {Saueressig},\ and\ \citenamefont {Zanusso}}]{Demmel:2015oqa}%
  \BibitemOpen
  \bibfield  {author} {\bibinfo {author} {\bibfnamefont {M.}~\bibnamefont
  {Demmel}}, \bibinfo {author} {\bibfnamefont {F.}~\bibnamefont {Saueressig}},
  \ and\ \bibinfo {author} {\bibfnamefont {O.}~\bibnamefont {Zanusso}},\ }\href
  {\doibase 10.1007/JHEP08(2015)113} {\bibfield  {journal} {\bibinfo  {journal}
  {JHEP}\ }\textbf {\bibinfo {volume} {08}},\ \bibinfo {pages} {113} (\bibinfo
  {year} {2015})},\ \Eprint {http://arxiv.org/abs/1504.07656} {arXiv:1504.07656
  [hep-th]} \BibitemShut {NoStop}%
\bibitem [{\citenamefont {Gies}\ and\ \citenamefont
  {Scherer}(2010)}]{Gies:2009hq}%
  \BibitemOpen
  \bibfield  {author} {\bibinfo {author} {\bibfnamefont {H.}~\bibnamefont
  {Gies}}\ and\ \bibinfo {author} {\bibfnamefont {M.~M.}\ \bibnamefont
  {Scherer}},\ }\href {\doibase 10.1140/epjc/s10052-010-1256-z} {\bibfield
  {journal} {\bibinfo  {journal} {Eur. Phys. J.}\ }\textbf {\bibinfo {volume}
  {C66}},\ \bibinfo {pages} {387} (\bibinfo {year} {2010})},\ \Eprint
  {http://arxiv.org/abs/0901.2459} {arXiv:0901.2459 [hep-th]} \BibitemShut
  {NoStop}%
\bibitem [{\citenamefont {Gies}\ \emph {et~al.}(2010)\citenamefont {Gies},
  \citenamefont {Rechenberger},\ and\ \citenamefont {Scherer}}]{Gies:2009sv}%
  \BibitemOpen
  \bibfield  {author} {\bibinfo {author} {\bibfnamefont {H.}~\bibnamefont
  {Gies}}, \bibinfo {author} {\bibfnamefont {S.}~\bibnamefont {Rechenberger}},
  \ and\ \bibinfo {author} {\bibfnamefont {M.~M.}\ \bibnamefont {Scherer}},\
  }\href {\doibase 10.1140/epjc/s10052-010-1257-y} {\bibfield  {journal}
  {\bibinfo  {journal} {Eur. Phys. J.}\ }\textbf {\bibinfo {volume} {C66}},\
  \bibinfo {pages} {403} (\bibinfo {year} {2010})},\ \Eprint
  {http://arxiv.org/abs/0907.0327} {arXiv:0907.0327 [hep-th]} \BibitemShut
  {NoStop}%
\bibitem [{\citenamefont {Gies}\ \emph {et~al.}(2013)\citenamefont {Gies},
  \citenamefont {Rechenberger}, \citenamefont {Scherer},\ and\ \citenamefont
  {Zambelli}}]{Gies:2013pma}%
  \BibitemOpen
  \bibfield  {author} {\bibinfo {author} {\bibfnamefont {H.}~\bibnamefont
  {Gies}}, \bibinfo {author} {\bibfnamefont {S.}~\bibnamefont {Rechenberger}},
  \bibinfo {author} {\bibfnamefont {M.~M.}\ \bibnamefont {Scherer}}, \ and\
  \bibinfo {author} {\bibfnamefont {L.}~\bibnamefont {Zambelli}},\ }\href
  {\doibase 10.1140/epjc/s10052-013-2652-y} {\bibfield  {journal} {\bibinfo
  {journal} {Eur. Phys. J.}\ }\textbf {\bibinfo {volume} {C73}},\ \bibinfo
  {pages} {2652} (\bibinfo {year} {2013})},\ \Eprint
  {http://arxiv.org/abs/1306.6508} {arXiv:1306.6508 [hep-th]} \BibitemShut
  {NoStop}%
\bibitem [{\citenamefont {Browne}\ \emph {et~al.}(1975)\citenamefont {Browne},
  \citenamefont {O'Raifeartaigh},\ and\ \citenamefont
  {Sherry}}]{Browne:1975js}%
  \BibitemOpen
  \bibfield  {author} {\bibinfo {author} {\bibfnamefont {S.}~\bibnamefont
  {Browne}}, \bibinfo {author} {\bibfnamefont {L.}~\bibnamefont
  {O'Raifeartaigh}}, \ and\ \bibinfo {author} {\bibfnamefont {T.}~\bibnamefont
  {Sherry}},\ }\href {\doibase 10.1016/0550-3213(75)90060-7} {\bibfield
  {journal} {\bibinfo  {journal} {Nucl. Phys.}\ }\textbf {\bibinfo {volume}
  {B99}},\ \bibinfo {pages} {150} (\bibinfo {year} {1975})}\BibitemShut
  {NoStop}%
\bibitem [{\citenamefont {Kaplunovsky}(1983)}]{Kaplunovsky:1982mf}%
  \BibitemOpen
  \bibfield  {author} {\bibinfo {author} {\bibfnamefont {V.}~\bibnamefont
  {Kaplunovsky}},\ }\href {\doibase 10.1016/0550-3213(83)90410-8} {\bibfield
  {journal} {\bibinfo  {journal} {Nucl. Phys.}\ }\textbf {\bibinfo {volume}
  {B211}},\ \bibinfo {pages} {297} (\bibinfo {year} {1983})}\BibitemShut
  {NoStop}%
\bibitem [{\citenamefont {Lee}\ and\ \citenamefont
  {Weisberger}(1974)}]{Lee:1974gua}%
  \BibitemOpen
  \bibfield  {author} {\bibinfo {author} {\bibfnamefont {B.~W.}\ \bibnamefont
  {Lee}}\ and\ \bibinfo {author} {\bibfnamefont {W.~I.}\ \bibnamefont
  {Weisberger}},\ }\href {\doibase 10.1103/PhysRevD.10.2530} {\bibfield
  {journal} {\bibinfo  {journal} {Phys. Rev.}\ }\textbf {\bibinfo {volume}
  {D10}},\ \bibinfo {pages} {2530} (\bibinfo {year} {1974})}\BibitemShut
  {NoStop}%
\bibitem [{\citenamefont {Coleman}\ and\ \citenamefont
  {Weinberg}(1973)}]{Coleman:1973jx}%
  \BibitemOpen
  \bibfield  {author} {\bibinfo {author} {\bibfnamefont {S.~R.}\ \bibnamefont
  {Coleman}}\ and\ \bibinfo {author} {\bibfnamefont {E.~J.}\ \bibnamefont
  {Weinberg}},\ }\href {\doibase 10.1103/PhysRevD.7.1888} {\bibfield  {journal}
  {\bibinfo  {journal} {Phys. Rev.}\ }\textbf {\bibinfo {volume} {D7}},\
  \bibinfo {pages} {1888} (\bibinfo {year} {1973})}\BibitemShut {NoStop}%
\bibitem [{\citenamefont {Fei}\ \emph {et~al.}(2014)\citenamefont {Fei},
  \citenamefont {Giombi},\ and\ \citenamefont {Klebanov}}]{Fei:2014yja}%
  \BibitemOpen
  \bibfield  {author} {\bibinfo {author} {\bibfnamefont {L.}~\bibnamefont
  {Fei}}, \bibinfo {author} {\bibfnamefont {S.}~\bibnamefont {Giombi}}, \ and\
  \bibinfo {author} {\bibfnamefont {I.~R.}\ \bibnamefont {Klebanov}},\ }\href
  {\doibase 10.1103/PhysRevD.90.025018} {\bibfield  {journal} {\bibinfo
  {journal} {Phys. Rev.}\ }\textbf {\bibinfo {volume} {D90}},\ \bibinfo {pages}
  {025018} (\bibinfo {year} {2014})},\ \Eprint {http://arxiv.org/abs/1404.1094}
  {arXiv:1404.1094 [hep-th]} \BibitemShut {NoStop}%
\bibitem [{\citenamefont {Fei}\ \emph {et~al.}(2015)\citenamefont {Fei},
  \citenamefont {Giombi}, \citenamefont {Klebanov},\ and\ \citenamefont
  {Tarnopolsky}}]{Fei:2014xta}%
  \BibitemOpen
  \bibfield  {author} {\bibinfo {author} {\bibfnamefont {L.}~\bibnamefont
  {Fei}}, \bibinfo {author} {\bibfnamefont {S.}~\bibnamefont {Giombi}},
  \bibinfo {author} {\bibfnamefont {I.~R.}\ \bibnamefont {Klebanov}}, \ and\
  \bibinfo {author} {\bibfnamefont {G.}~\bibnamefont {Tarnopolsky}},\ }\href
  {\doibase 10.1103/PhysRevD.91.045011} {\bibfield  {journal} {\bibinfo
  {journal} {Phys. Rev.}\ }\textbf {\bibinfo {volume} {D91}},\ \bibinfo {pages}
  {045011} (\bibinfo {year} {2015})},\ \Eprint {http://arxiv.org/abs/1411.1099}
  {arXiv:1411.1099 [hep-th]} \BibitemShut {NoStop}%
\bibitem [{\citenamefont {Percacci}\ and\ \citenamefont
  {Vacca}(2014)}]{Percacci:2014tfa}%
  \BibitemOpen
  \bibfield  {author} {\bibinfo {author} {\bibfnamefont {R.}~\bibnamefont
  {Percacci}}\ and\ \bibinfo {author} {\bibfnamefont {G.~P.}\ \bibnamefont
  {Vacca}},\ }\href {\doibase 10.1103/PhysRevD.90.107702} {\bibfield  {journal}
  {\bibinfo  {journal} {Phys. Rev.}\ }\textbf {\bibinfo {volume} {D90}},\
  \bibinfo {pages} {107702} (\bibinfo {year} {2014})},\ \Eprint
  {http://arxiv.org/abs/1405.6622} {arXiv:1405.6622 [hep-th]} \BibitemShut
  {NoStop}%
\bibitem [{\citenamefont {Herbut}\ and\ \citenamefont
  {Janssen}(2016)}]{Herbut:2015zqa}%
  \BibitemOpen
  \bibfield  {author} {\bibinfo {author} {\bibfnamefont {I.~F.}\ \bibnamefont
  {Herbut}}\ and\ \bibinfo {author} {\bibfnamefont {L.}~\bibnamefont
  {Janssen}},\ }\href {\doibase 10.1103/PhysRevD.93.085005} {\bibfield
  {journal} {\bibinfo  {journal} {Phys. Rev.}\ }\textbf {\bibinfo {volume}
  {D93}},\ \bibinfo {pages} {085005} (\bibinfo {year} {2016})},\ \Eprint
  {http://arxiv.org/abs/1510.05691} {arXiv:1510.05691 [hep-th]} \BibitemShut
  {NoStop}%
\bibitem [{\citenamefont {Mati}(2016)}]{Mati:2016wjn}%
  \BibitemOpen
  \bibfield  {author} {\bibinfo {author} {\bibfnamefont {P.}~\bibnamefont
  {Mati}},\ }\href@noop {} {\  (\bibinfo {year} {2016})},\ \Eprint
  {http://arxiv.org/abs/1601.00450} {arXiv:1601.00450 [hep-th]} \BibitemShut
  {NoStop}%
\bibitem [{\citenamefont {Eichhorn}\ \emph {et~al.}(2016)\citenamefont
  {Eichhorn}, \citenamefont {Janssen},\ and\ \citenamefont
  {Scherer}}]{Eichhorn:2016hdi}%
  \BibitemOpen
  \bibfield  {author} {\bibinfo {author} {\bibfnamefont {A.}~\bibnamefont
  {Eichhorn}}, \bibinfo {author} {\bibfnamefont {L.}~\bibnamefont {Janssen}}, \
  and\ \bibinfo {author} {\bibfnamefont {M.~M.}\ \bibnamefont {Scherer}},\
  }\href@noop {} {\  (\bibinfo {year} {2016})},\ \Eprint
  {http://arxiv.org/abs/1604.03561} {arXiv:1604.03561 [hep-th]} \BibitemShut
  {NoStop}%
\bibitem [{\citenamefont {Bridle}\ and\ \citenamefont
  {Morris}(2016)}]{Bridle:2016nsu}%
  \BibitemOpen
  \bibfield  {author} {\bibinfo {author} {\bibfnamefont {I.~H.}\ \bibnamefont
  {Bridle}}\ and\ \bibinfo {author} {\bibfnamefont {T.~R.}\ \bibnamefont
  {Morris}},\ }\href@noop {} {\  (\bibinfo {year} {2016})},\ \Eprint
  {http://arxiv.org/abs/1605.06075} {arXiv:1605.06075 [hep-th]} \BibitemShut
  {NoStop}%
\bibitem [{\citenamefont {Rosten}(2012)}]{Rosten:2010vm}%
  \BibitemOpen
  \bibfield  {author} {\bibinfo {author} {\bibfnamefont {O.~J.}\ \bibnamefont
  {Rosten}},\ }\href {\doibase 10.1016/j.physrep.2011.12.003} {\bibfield
  {journal} {\bibinfo  {journal} {Phys. Rept.}\ }\textbf {\bibinfo {volume}
  {511}},\ \bibinfo {pages} {177} (\bibinfo {year} {2012})},\ \Eprint
  {http://arxiv.org/abs/1003.1366} {arXiv:1003.1366 [hep-th]} \BibitemShut
  {NoStop}%
\bibitem [{\citenamefont {Halpern}\ and\ \citenamefont
  {Huang}(1995)}]{Halpern:1994vw}%
  \BibitemOpen
  \bibfield  {author} {\bibinfo {author} {\bibfnamefont {K.}~\bibnamefont
  {Halpern}}\ and\ \bibinfo {author} {\bibfnamefont {K.}~\bibnamefont
  {Huang}},\ }\href {\doibase 10.1103/PhysRevLett.74.3526} {\bibfield
  {journal} {\bibinfo  {journal} {Phys. Rev. Lett.}\ }\textbf {\bibinfo
  {volume} {74}},\ \bibinfo {pages} {3526} (\bibinfo {year} {1995})},\ \Eprint
  {http://arxiv.org/abs/hep-th/9406199} {arXiv:hep-th/9406199 [hep-th]}
  \BibitemShut {NoStop}%
\bibitem [{\citenamefont {Halpern}\ and\ \citenamefont
  {Huang}(1996{\natexlab{a}})}]{Halpern:1995vf}%
  \BibitemOpen
  \bibfield  {author} {\bibinfo {author} {\bibfnamefont {K.}~\bibnamefont
  {Halpern}}\ and\ \bibinfo {author} {\bibfnamefont {K.}~\bibnamefont
  {Huang}},\ }\href {\doibase 10.1103/PhysRevD.53.3252} {\bibfield  {journal}
  {\bibinfo  {journal} {Phys. Rev.}\ }\textbf {\bibinfo {volume} {D53}},\
  \bibinfo {pages} {3252} (\bibinfo {year} {1996}{\natexlab{a}})},\ \Eprint
  {http://arxiv.org/abs/hep-th/9510240} {arXiv:hep-th/9510240 [hep-th]}
  \BibitemShut {NoStop}%
\bibitem [{\citenamefont {Periwal}(1996)}]{Periwal:1995hw}%
  \BibitemOpen
  \bibfield  {author} {\bibinfo {author} {\bibfnamefont {V.}~\bibnamefont
  {Periwal}},\ }\href {\doibase 10.1142/S0217732396002885} {\bibfield
  {journal} {\bibinfo  {journal} {Mod. Phys. Lett.}\ }\textbf {\bibinfo
  {volume} {A11}},\ \bibinfo {pages} {2915} (\bibinfo {year} {1996})},\ \Eprint
  {http://arxiv.org/abs/hep-th/9512108} {arXiv:hep-th/9512108 [hep-th]}
  \BibitemShut {NoStop}%
\bibitem [{\citenamefont {Bonanno}(2000)}]{Bonanno:2000sy}%
  \BibitemOpen
  \bibfield  {author} {\bibinfo {author} {\bibfnamefont {A.}~\bibnamefont
  {Bonanno}},\ }\href {\doibase 10.1103/PhysRevD.62.027701} {\bibfield
  {journal} {\bibinfo  {journal} {Phys. Rev.}\ }\textbf {\bibinfo {volume}
  {D62}},\ \bibinfo {pages} {027701} (\bibinfo {year} {2000})},\ \Eprint
  {http://arxiv.org/abs/hep-th/0001060} {arXiv:hep-th/0001060 [hep-th]}
  \BibitemShut {NoStop}%
\bibitem [{\citenamefont {Gies}(2001)}]{Gies:2000xr}%
  \BibitemOpen
  \bibfield  {author} {\bibinfo {author} {\bibfnamefont {H.}~\bibnamefont
  {Gies}},\ }\href {\doibase 10.1103/PhysRevD.63.065011} {\bibfield  {journal}
  {\bibinfo  {journal} {Phys. Rev.}\ }\textbf {\bibinfo {volume} {D63}},\
  \bibinfo {pages} {065011} (\bibinfo {year} {2001})},\ \Eprint
  {http://arxiv.org/abs/hep-th/0009041} {arXiv:hep-th/0009041 [hep-th]}
  \BibitemShut {NoStop}%
\bibitem [{\citenamefont {Morris}(1996)}]{Morris:1996nx}%
  \BibitemOpen
  \bibfield  {author} {\bibinfo {author} {\bibfnamefont {T.~R.}\ \bibnamefont
  {Morris}},\ }\href {\doibase 10.1103/PhysRevLett.77.1658} {\bibfield
  {journal} {\bibinfo  {journal} {Phys. Rev. Lett.}\ }\textbf {\bibinfo
  {volume} {77}},\ \bibinfo {pages} {1658} (\bibinfo {year} {1996})},\ \Eprint
  {http://arxiv.org/abs/hep-th/9601128} {arXiv:hep-th/9601128 [hep-th]}
  \BibitemShut {NoStop}%
\bibitem [{\citenamefont {Halpern}\ and\ \citenamefont
  {Huang}(1996{\natexlab{b}})}]{Halpern:1996dh}%
  \BibitemOpen
  \bibfield  {author} {\bibinfo {author} {\bibfnamefont {K.}~\bibnamefont
  {Halpern}}\ and\ \bibinfo {author} {\bibfnamefont {K.}~\bibnamefont
  {Huang}},\ }\href {\doibase 10.1103/PhysRevLett.77.1659} {\bibfield
  {journal} {\bibinfo  {journal} {Phys. Rev. Lett.}\ }\textbf {\bibinfo
  {volume} {77}},\ \bibinfo {pages} {1659} (\bibinfo {year}
  {1996}{\natexlab{b}})}\BibitemShut {NoStop}%
\bibitem [{\citenamefont {Wetterich}(1993)}]{Wetterich:1992yh}%
  \BibitemOpen
  \bibfield  {author} {\bibinfo {author} {\bibfnamefont {C.}~\bibnamefont
  {Wetterich}},\ }\href {\doibase 10.1016/0370-2693(93)90726-X} {\bibfield
  {journal} {\bibinfo  {journal} {Phys. Lett.}\ }\textbf {\bibinfo {volume}
  {B301}},\ \bibinfo {pages} {90} (\bibinfo {year} {1993})}\BibitemShut
  {NoStop}%
\bibitem [{\citenamefont {Aoki}(2000)}]{Aoki:2000wm}%
  \BibitemOpen
  \bibfield  {author} {\bibinfo {author} {\bibfnamefont {K.}~\bibnamefont
  {Aoki}},\ }\bibfield  {booktitle} {\emph {\bibinfo {booktitle} {{Methods of
  renormalization group. Proceedings, Summer School on mathematical physics,
  Tokyo, Japan, September 23-26, 1999}}},\ }\href {\doibase
  10.1016/S0217-9792(00)00092-3} {\bibfield  {journal} {\bibinfo  {journal}
  {Int. J. Mod. Phys.}\ }\textbf {\bibinfo {volume} {B14}},\ \bibinfo {pages}
  {1249} (\bibinfo {year} {2000})}\BibitemShut {NoStop}%
\bibitem [{\citenamefont {Berges}\ \emph {et~al.}(2002)\citenamefont {Berges},
  \citenamefont {Tetradis},\ and\ \citenamefont {Wetterich}}]{Berges:2000ew}%
  \BibitemOpen
  \bibfield  {author} {\bibinfo {author} {\bibfnamefont {J.}~\bibnamefont
  {Berges}}, \bibinfo {author} {\bibfnamefont {N.}~\bibnamefont {Tetradis}}, \
  and\ \bibinfo {author} {\bibfnamefont {C.}~\bibnamefont {Wetterich}},\ }\href
  {\doibase 10.1016/S0370-1573(01)00098-9} {\bibfield  {journal} {\bibinfo
  {journal} {Phys. Rept.}\ }\textbf {\bibinfo {volume} {363}},\ \bibinfo
  {pages} {223} (\bibinfo {year} {2002})},\ \Eprint
  {http://arxiv.org/abs/hep-ph/0005122} {arXiv:hep-ph/0005122 [hep-ph]}
  \BibitemShut {NoStop}%
\bibitem [{\citenamefont {Pawlowski}(2007)}]{Pawlowski:2005xe}%
  \BibitemOpen
  \bibfield  {author} {\bibinfo {author} {\bibfnamefont {J.~M.}\ \bibnamefont
  {Pawlowski}},\ }\href {\doibase 10.1016/j.aop.2007.01.007} {\bibfield
  {journal} {\bibinfo  {journal} {Annals Phys.}\ }\textbf {\bibinfo {volume}
  {322}},\ \bibinfo {pages} {2831} (\bibinfo {year} {2007})},\ \Eprint
  {http://arxiv.org/abs/hep-th/0512261} {arXiv:hep-th/0512261 [hep-th]}
  \BibitemShut {NoStop}%
\bibitem [{\citenamefont {Gies}(2012)}]{Gies:2006wv}%
  \BibitemOpen
  \bibfield  {author} {\bibinfo {author} {\bibfnamefont {H.}~\bibnamefont
  {Gies}},\ }\bibfield  {booktitle} {\emph {\bibinfo {booktitle} {{ECT* School
  on Renormalization Group and Effective Field Theory Approaches to Many-Body
  Systems Trento, Italy, February 27-March 10, 2006}}},\ }\href {\doibase
  10.1007/978-3-642-27320-9_6} {\bibfield  {journal} {\bibinfo  {journal}
  {Lect. Notes Phys.}\ }\textbf {\bibinfo {volume} {852}},\ \bibinfo {pages}
  {287} (\bibinfo {year} {2012})},\ \Eprint
  {http://arxiv.org/abs/hep-ph/0611146} {arXiv:hep-ph/0611146 [hep-ph]}
  \BibitemShut {NoStop}%
\bibitem [{\citenamefont {Delamotte}(2012)}]{Delamotte:2007pf}%
  \BibitemOpen
  \bibfield  {author} {\bibinfo {author} {\bibfnamefont {B.}~\bibnamefont
  {Delamotte}},\ }\href {\doibase 10.1007/978-3-642-27320-9_2} {\bibfield
  {journal} {\bibinfo  {journal} {Lect. Notes Phys.}\ }\textbf {\bibinfo
  {volume} {852}},\ \bibinfo {pages} {49} (\bibinfo {year} {2012})},\ \Eprint
  {http://arxiv.org/abs/cond-mat/0702365} {arXiv:cond-mat/0702365
  [cond-mat.stat-mech]} \BibitemShut {NoStop}%
\bibitem [{\citenamefont {Braun}(2012)}]{Braun:2011pp}%
  \BibitemOpen
  \bibfield  {author} {\bibinfo {author} {\bibfnamefont {J.}~\bibnamefont
  {Braun}},\ }\href {\doibase 10.1088/0954-3899/39/3/033001} {\bibfield
  {journal} {\bibinfo  {journal} {J. Phys.}\ }\textbf {\bibinfo {volume}
  {G39}},\ \bibinfo {pages} {033001} (\bibinfo {year} {2012})},\ \Eprint
  {http://arxiv.org/abs/1108.4449} {arXiv:1108.4449 [hep-ph]} \BibitemShut
  {NoStop}%
\bibitem [{\citenamefont {Ellwanger}\ \emph {et~al.}(1996)\citenamefont
  {Ellwanger}, \citenamefont {Hirsch},\ and\ \citenamefont
  {Weber}}]{Ellwanger:1995qf}%
  \BibitemOpen
  \bibfield  {author} {\bibinfo {author} {\bibfnamefont {U.}~\bibnamefont
  {Ellwanger}}, \bibinfo {author} {\bibfnamefont {M.}~\bibnamefont {Hirsch}}, \
  and\ \bibinfo {author} {\bibfnamefont {A.}~\bibnamefont {Weber}},\ }\href
  {\doibase 10.1007/s002880050073} {\bibfield  {journal} {\bibinfo  {journal}
  {Z. Phys.}\ }\textbf {\bibinfo {volume} {C69}},\ \bibinfo {pages} {687}
  (\bibinfo {year} {1996})},\ \Eprint {http://arxiv.org/abs/hep-th/9506019}
  {arXiv:hep-th/9506019 [hep-th]} \BibitemShut {NoStop}%
\bibitem [{\citenamefont {Litim}\ and\ \citenamefont
  {Pawlowski}(1998)}]{Litim:1998qi}%
  \BibitemOpen
  \bibfield  {author} {\bibinfo {author} {\bibfnamefont {D.~F.}\ \bibnamefont
  {Litim}}\ and\ \bibinfo {author} {\bibfnamefont {J.~M.}\ \bibnamefont
  {Pawlowski}},\ }\href {\doibase 10.1016/S0370-2693(98)00761-8} {\bibfield
  {journal} {\bibinfo  {journal} {Phys.Lett.}\ }\textbf {\bibinfo {volume}
  {B435}},\ \bibinfo {pages} {181} (\bibinfo {year} {1998})},\ \Eprint
  {http://arxiv.org/abs/hep-th/9802064} {arXiv:hep-th/9802064 [hep-th]}
  \BibitemShut {NoStop}%
\bibitem [{\citenamefont {Abbott}(1981)}]{Abbott:1980hw}%
  \BibitemOpen
  \bibfield  {author} {\bibinfo {author} {\bibfnamefont {L.~F.}\ \bibnamefont
  {Abbott}},\ }\href {\doibase 10.1016/0550-3213(81)90371-0} {\bibfield
  {journal} {\bibinfo  {journal} {Nucl. Phys.}\ }\textbf {\bibinfo {volume}
  {B185}},\ \bibinfo {pages} {189} (\bibinfo {year} {1981})}\BibitemShut
  {NoStop}%
\bibitem [{\citenamefont {Litim}(2000)}]{Litim:2000ci}%
  \BibitemOpen
  \bibfield  {author} {\bibinfo {author} {\bibfnamefont {D.~F.}\ \bibnamefont
  {Litim}},\ }\href {\doibase 10.1016/S0370-2693(00)00748-6} {\bibfield
  {journal} {\bibinfo  {journal} {Phys. Lett.}\ }\textbf {\bibinfo {volume}
  {B486}},\ \bibinfo {pages} {92} (\bibinfo {year} {2000})},\ \Eprint
  {http://arxiv.org/abs/hep-th/0005245} {arXiv:hep-th/0005245 [hep-th]}
  \BibitemShut {NoStop}%
\bibitem [{\citenamefont {Litim}(2001)}]{Litim:2001up}%
  \BibitemOpen
  \bibfield  {author} {\bibinfo {author} {\bibfnamefont {D.~F.}\ \bibnamefont
  {Litim}},\ }\href {\doibase 10.1103/PhysRevD.64.105007} {\bibfield  {journal}
  {\bibinfo  {journal} {Phys. Rev.}\ }\textbf {\bibinfo {volume} {D64}},\
  \bibinfo {pages} {105007} (\bibinfo {year} {2001})},\ \Eprint
  {http://arxiv.org/abs/hep-th/0103195} {arXiv:hep-th/0103195 [hep-th]}
  \BibitemShut {NoStop}%
\bibitem [{\citenamefont {Bervillier}\ \emph {et~al.}(2008)\citenamefont
  {Bervillier}, \citenamefont {Boisseau},\ and\ \citenamefont
  {Giacomini}}]{Bervillier:2007tc}%
  \BibitemOpen
  \bibfield  {author} {\bibinfo {author} {\bibfnamefont {C.}~\bibnamefont
  {Bervillier}}, \bibinfo {author} {\bibfnamefont {B.}~\bibnamefont
  {Boisseau}}, \ and\ \bibinfo {author} {\bibfnamefont {H.}~\bibnamefont
  {Giacomini}},\ }\href {\doibase 10.1016/j.nuclphysb.2007.07.005} {\bibfield
  {journal} {\bibinfo  {journal} {Nucl. Phys.}\ }\textbf {\bibinfo {volume}
  {B789}},\ \bibinfo {pages} {525} (\bibinfo {year} {2008})},\ \Eprint
  {http://arxiv.org/abs/0706.0990} {arXiv:0706.0990 [hep-th]} \BibitemShut
  {NoStop}%
\bibitem [{\citenamefont {Codello}\ and\ \citenamefont
  {D'Odorico}(2013)}]{Codello:2012ec}%
  \BibitemOpen
  \bibfield  {author} {\bibinfo {author} {\bibfnamefont {A.}~\bibnamefont
  {Codello}}\ and\ \bibinfo {author} {\bibfnamefont {G.}~\bibnamefont
  {D'Odorico}},\ }\href {\doibase 10.1103/PhysRevLett.110.141601} {\bibfield
  {journal} {\bibinfo  {journal} {Phys. Rev. Lett.}\ }\textbf {\bibinfo
  {volume} {110}},\ \bibinfo {pages} {141601} (\bibinfo {year} {2013})},\
  \Eprint {http://arxiv.org/abs/1210.4037} {arXiv:1210.4037 [hep-th]}
  \BibitemShut {NoStop}%
\bibitem [{\citenamefont {Dietz}\ and\ \citenamefont
  {Morris}(2015)}]{Dietz:2015owa}%
  \BibitemOpen
  \bibfield  {author} {\bibinfo {author} {\bibfnamefont {J.~A.}\ \bibnamefont
  {Dietz}}\ and\ \bibinfo {author} {\bibfnamefont {T.~R.}\ \bibnamefont
  {Morris}},\ }\href {\doibase 10.1007/JHEP04(2015)118} {\bibfield  {journal}
  {\bibinfo  {journal} {JHEP}\ }\textbf {\bibinfo {volume} {04}},\ \bibinfo
  {pages} {118} (\bibinfo {year} {2015})},\ \Eprint
  {http://arxiv.org/abs/1502.07396} {arXiv:1502.07396 [hep-th]} \BibitemShut
  {NoStop}%
\bibitem [{\citenamefont {Vacca}\ and\ \citenamefont
  {Zambelli}(2015)}]{Vacca:2015nta}%
  \BibitemOpen
  \bibfield  {author} {\bibinfo {author} {\bibfnamefont {G.~P.}\ \bibnamefont
  {Vacca}}\ and\ \bibinfo {author} {\bibfnamefont {L.}~\bibnamefont
  {Zambelli}},\ }\href {\doibase 10.1103/PhysRevD.91.125003} {\bibfield
  {journal} {\bibinfo  {journal} {Phys. Rev.}\ }\textbf {\bibinfo {volume}
  {D91}},\ \bibinfo {pages} {125003} (\bibinfo {year} {2015})},\ \Eprint
  {http://arxiv.org/abs/1503.09136} {arXiv:1503.09136 [hep-th]} \BibitemShut
  {NoStop}%
\bibitem [{\citenamefont {Hellwig}\ \emph {et~al.}(2015)\citenamefont
  {Hellwig}, \citenamefont {Wipf},\ and\ \citenamefont
  {Zanusso}}]{Hellwig:2015woa}%
  \BibitemOpen
  \bibfield  {author} {\bibinfo {author} {\bibfnamefont {T.}~\bibnamefont
  {Hellwig}}, \bibinfo {author} {\bibfnamefont {A.}~\bibnamefont {Wipf}}, \
  and\ \bibinfo {author} {\bibfnamefont {O.}~\bibnamefont {Zanusso}},\
  }\href@noop {} {\  (\bibinfo {year} {2015})},\ \Eprint
  {http://arxiv.org/abs/1508.02547} {arXiv:1508.02547 [hep-th]} \BibitemShut
  {NoStop}%
\bibitem [{\citenamefont {Borchardt}\ and\ \citenamefont
  {Knorr}(2015)}]{Borchardt:2015rxa}%
  \BibitemOpen
  \bibfield  {author} {\bibinfo {author} {\bibfnamefont {J.}~\bibnamefont
  {Borchardt}}\ and\ \bibinfo {author} {\bibfnamefont {B.}~\bibnamefont
  {Knorr}},\ }\href {\doibase 10.1103/PhysRevD.91.105011} {\bibfield  {journal}
  {\bibinfo  {journal} {Phys. Rev.}\ }\textbf {\bibinfo {volume} {D91}},\
  \bibinfo {pages} {105011} (\bibinfo {year} {2015})},\ \Eprint
  {http://arxiv.org/abs/1502.07511} {arXiv:1502.07511 [hep-th]} \BibitemShut
  {NoStop}%
\bibitem [{\citenamefont {Litim}\ and\ \citenamefont
  {Marchais}(2016)}]{Litim:2016hlb}%
  \BibitemOpen
  \bibfield  {author} {\bibinfo {author} {\bibfnamefont {D.~F.}\ \bibnamefont
  {Litim}}\ and\ \bibinfo {author} {\bibfnamefont {E.}~\bibnamefont
  {Marchais}},\ }\href@noop {} {\  (\bibinfo {year} {2016})},\ \Eprint
  {http://arxiv.org/abs/1607.02030} {arXiv:1607.02030 [hep-th]} \BibitemShut
  {NoStop}%
\bibitem [{\citenamefont {Morris}(1994)}]{Morris:1994ki}%
  \BibitemOpen
  \bibfield  {author} {\bibinfo {author} {\bibfnamefont {T.~R.}\ \bibnamefont
  {Morris}},\ }\href {\doibase 10.1016/0370-2693(94)90700-5} {\bibfield
  {journal} {\bibinfo  {journal} {Phys. Lett.}\ }\textbf {\bibinfo {volume}
  {B334}},\ \bibinfo {pages} {355} (\bibinfo {year} {1994})},\ \Eprint
  {http://arxiv.org/abs/hep-th/9405190} {arXiv:hep-th/9405190 [hep-th]}
  \BibitemShut {NoStop}%
\bibitem [{\citenamefont {Gies}\ \emph {et~al.}(2014)\citenamefont {Gies},
  \citenamefont {Gneiting},\ and\ \citenamefont {Sondenheimer}}]{Gies:2013fua}%
  \BibitemOpen
  \bibfield  {author} {\bibinfo {author} {\bibfnamefont {H.}~\bibnamefont
  {Gies}}, \bibinfo {author} {\bibfnamefont {C.}~\bibnamefont {Gneiting}}, \
  and\ \bibinfo {author} {\bibfnamefont {R.}~\bibnamefont {Sondenheimer}},\
  }\href {\doibase 10.1103/PhysRevD.89.045012} {\bibfield  {journal} {\bibinfo
  {journal} {Phys. Rev.}\ }\textbf {\bibinfo {volume} {D89}},\ \bibinfo {pages}
  {045012} (\bibinfo {year} {2014})},\ \Eprint {http://arxiv.org/abs/1308.5075}
  {arXiv:1308.5075 [hep-ph]} \BibitemShut {NoStop}%
\bibitem [{\citenamefont {Gies}\ and\ \citenamefont
  {Sondenheimer}(2015)}]{Gies:2014xha}%
  \BibitemOpen
  \bibfield  {author} {\bibinfo {author} {\bibfnamefont {H.}~\bibnamefont
  {Gies}}\ and\ \bibinfo {author} {\bibfnamefont {R.}~\bibnamefont
  {Sondenheimer}},\ }\href {\doibase 10.1140/epjc/s10052-015-3284-1} {\bibfield
   {journal} {\bibinfo  {journal} {Eur. Phys. J.}\ }\textbf {\bibinfo {volume}
  {C75}},\ \bibinfo {pages} {68} (\bibinfo {year} {2015})},\ \Eprint
  {http://arxiv.org/abs/1407.8124} {arXiv:1407.8124 [hep-ph]} \BibitemShut
  {NoStop}%
\bibitem [{\citenamefont {Eichhorn}\ \emph {et~al.}(2015)\citenamefont
  {Eichhorn}, \citenamefont {Gies}, \citenamefont {Jaeckel}, \citenamefont
  {Plehn}, \citenamefont {Scherer},\ and\ \citenamefont
  {Sondenheimer}}]{Eichhorn:2015kea}%
  \BibitemOpen
  \bibfield  {author} {\bibinfo {author} {\bibfnamefont {A.}~\bibnamefont
  {Eichhorn}}, \bibinfo {author} {\bibfnamefont {H.}~\bibnamefont {Gies}},
  \bibinfo {author} {\bibfnamefont {J.}~\bibnamefont {Jaeckel}}, \bibinfo
  {author} {\bibfnamefont {T.}~\bibnamefont {Plehn}}, \bibinfo {author}
  {\bibfnamefont {M.~M.}\ \bibnamefont {Scherer}}, \ and\ \bibinfo {author}
  {\bibfnamefont {R.}~\bibnamefont {Sondenheimer}},\ }\href {\doibase
  10.1007/JHEP04(2015)022} {\bibfield  {journal} {\bibinfo  {journal} {JHEP}\
  }\textbf {\bibinfo {volume} {04}},\ \bibinfo {pages} {022} (\bibinfo {year}
  {2015})},\ \Eprint {http://arxiv.org/abs/1501.02812} {arXiv:1501.02812
  [hep-ph]} \BibitemShut {NoStop}%
\bibitem [{\citenamefont {Borchardt}\ \emph {et~al.}(2016)\citenamefont
  {Borchardt}, \citenamefont {Gies},\ and\ \citenamefont
  {Sondenheimer}}]{Borchardt:2016xju}%
  \BibitemOpen
  \bibfield  {author} {\bibinfo {author} {\bibfnamefont {J.}~\bibnamefont
  {Borchardt}}, \bibinfo {author} {\bibfnamefont {H.}~\bibnamefont {Gies}}, \
  and\ \bibinfo {author} {\bibfnamefont {R.}~\bibnamefont {Sondenheimer}},\
  }\href@noop {} {\  (\bibinfo {year} {2016})},\ \Eprint
  {http://arxiv.org/abs/1603.05861} {arXiv:1603.05861 [hep-ph]} \BibitemShut
  {NoStop}%
\bibitem [{\citenamefont {Jakovac}\ \emph {et~al.}(2015)\citenamefont
  {Jakovac}, \citenamefont {Kaposvari},\ and\ \citenamefont
  {Patkos}}]{Jakovac:2015kka}%
  \BibitemOpen
  \bibfield  {author} {\bibinfo {author} {\bibfnamefont {A.}~\bibnamefont
  {Jakovac}}, \bibinfo {author} {\bibfnamefont {I.}~\bibnamefont {Kaposvari}},
  \ and\ \bibinfo {author} {\bibfnamefont {A.}~\bibnamefont {Patkos}},\
  }\href@noop {} {\  (\bibinfo {year} {2015})},\ \Eprint
  {http://arxiv.org/abs/1508.06774} {arXiv:1508.06774 [hep-th]} \BibitemShut
  {NoStop}%
\bibitem [{\citenamefont {Borchardt}\ and\ \citenamefont
  {Knorr}(2016)}]{Borchardt:2016pif}%
  \BibitemOpen
  \bibfield  {author} {\bibinfo {author} {\bibfnamefont {J.}~\bibnamefont
  {Borchardt}}\ and\ \bibinfo {author} {\bibfnamefont {B.}~\bibnamefont
  {Knorr}},\ }\href {\doibase 10.1103/PhysRevD.94.025027} {\bibfield  {journal}
  {\bibinfo  {journal} {Phys. Rev.}\ }\textbf {\bibinfo {volume} {D94}},\
  \bibinfo {pages} {025027} (\bibinfo {year} {2016})},\ \Eprint
  {http://arxiv.org/abs/1603.06726} {arXiv:1603.06726 [hep-th]} \BibitemShut
  {NoStop}%
\bibitem [{\citenamefont {Maas}(2013)}]{Maas:2012tj}%
  \BibitemOpen
  \bibfield  {author} {\bibinfo {author} {\bibfnamefont {A.}~\bibnamefont
  {Maas}},\ }\href {\doibase 10.1142/S0217732313501034} {\bibfield  {journal}
  {\bibinfo  {journal} {Mod. Phys. Lett.}\ }\textbf {\bibinfo {volume} {A28}},\
  \bibinfo {pages} {1350103} (\bibinfo {year} {2013})},\ \Eprint
  {http://arxiv.org/abs/1205.6625} {arXiv:1205.6625 [hep-lat]} \BibitemShut
  {NoStop}%
\bibitem [{\citenamefont {Maas}\ and\ \citenamefont
  {Mufti}(2014)}]{Maas:2013aia}%
  \BibitemOpen
  \bibfield  {author} {\bibinfo {author} {\bibfnamefont {A.}~\bibnamefont
  {Maas}}\ and\ \bibinfo {author} {\bibfnamefont {T.}~\bibnamefont {Mufti}},\
  }\href {\doibase 10.1007/JHEP04(2014)006} {\bibfield  {journal} {\bibinfo
  {journal} {JHEP}\ }\textbf {\bibinfo {volume} {04}},\ \bibinfo {pages} {006}
  (\bibinfo {year} {2014})},\ \Eprint {http://arxiv.org/abs/1312.4873}
  {arXiv:1312.4873 [hep-lat]} \BibitemShut {NoStop}%
\bibitem [{\citenamefont {Maas}\ and\ \citenamefont
  {Mufti}(2015)}]{Maas:2014pba}%
  \BibitemOpen
  \bibfield  {author} {\bibinfo {author} {\bibfnamefont {A.}~\bibnamefont
  {Maas}}\ and\ \bibinfo {author} {\bibfnamefont {T.}~\bibnamefont {Mufti}},\
  }\href {\doibase 10.1103/PhysRevD.91.113011} {\bibfield  {journal} {\bibinfo
  {journal} {Phys. Rev.}\ }\textbf {\bibinfo {volume} {D91}},\ \bibinfo {pages}
  {113011} (\bibinfo {year} {2015})},\ \Eprint {http://arxiv.org/abs/1412.6440}
  {arXiv:1412.6440 [hep-lat]} \BibitemShut {NoStop}%
\bibitem [{\citenamefont {Maas}(2015)}]{Maas:2015gma}%
  \BibitemOpen
  \bibfield  {author} {\bibinfo {author} {\bibfnamefont {A.}~\bibnamefont
  {Maas}},\ }\href {\doibase 10.1142/S0217732315501357} {\bibfield  {journal}
  {\bibinfo  {journal} {Mod. Phys. Lett.}\ }\textbf {\bibinfo {volume} {A30}},\
  \bibinfo {pages} {1550135} (\bibinfo {year} {2015})},\ \Eprint
  {http://arxiv.org/abs/1502.02421} {arXiv:1502.02421 [hep-ph]} \BibitemShut
  {NoStop}%
\bibitem [{\citenamefont {Maas}\ and\ \citenamefont
  {T{\"o}rek}(2016)}]{Maas:2016ngo}%
  \BibitemOpen
  \bibfield  {author} {\bibinfo {author} {\bibfnamefont {A.}~\bibnamefont
  {Maas}}\ and\ \bibinfo {author} {\bibfnamefont {P.}~\bibnamefont
  {T{\"o}rek}},\ }\href@noop {} {\  (\bibinfo {year} {2016})},\ \Eprint
  {http://arxiv.org/abs/1607.05860} {arXiv:1607.05860 [hep-lat]} \BibitemShut
  {NoStop}%
\bibitem [{\citenamefont {Litim}\ and\ \citenamefont
  {Sannino}(2014)}]{Litim:2014uca}%
  \BibitemOpen
  \bibfield  {author} {\bibinfo {author} {\bibfnamefont {D.~F.}\ \bibnamefont
  {Litim}}\ and\ \bibinfo {author} {\bibfnamefont {F.}~\bibnamefont
  {Sannino}},\ }\href {\doibase 10.1007/JHEP12(2014)178} {\bibfield  {journal}
  {\bibinfo  {journal} {JHEP}\ }\textbf {\bibinfo {volume} {12}},\ \bibinfo
  {pages} {178} (\bibinfo {year} {2014})},\ \Eprint
  {http://arxiv.org/abs/1406.2337} {arXiv:1406.2337 [hep-th]} \BibitemShut
  {NoStop}%
\bibitem [{\citenamefont {Bond}\ and\ \citenamefont
  {Litim}(2016)}]{Bond:2016dvk}%
  \BibitemOpen
  \bibfield  {author} {\bibinfo {author} {\bibfnamefont {A.~D.}\ \bibnamefont
  {Bond}}\ and\ \bibinfo {author} {\bibfnamefont {D.~F.}\ \bibnamefont
  {Litim}},\ }\href@noop {} {\  (\bibinfo {year} {2016})},\ \Eprint
  {http://arxiv.org/abs/1608.00519} {arXiv:1608.00519 [hep-th]} \BibitemShut
  {NoStop}%
\bibitem [{\citenamefont {Codello}\ \emph {et~al.}(2016)\citenamefont
  {Codello}, \citenamefont {Langæble}, \citenamefont {Litim},\ and\
  \citenamefont {Sannino}}]{Codello:2016muj}%
  \BibitemOpen
  \bibfield  {author} {\bibinfo {author} {\bibfnamefont {A.}~\bibnamefont
  {Codello}}, \bibinfo {author} {\bibfnamefont {K.}~\bibnamefont {Langæble}},
  \bibinfo {author} {\bibfnamefont {D.~F.}\ \bibnamefont {Litim}}, \ and\
  \bibinfo {author} {\bibfnamefont {F.}~\bibnamefont {Sannino}},\ }\href
  {\doibase 10.1007/JHEP07(2016)118} {\bibfield  {journal} {\bibinfo  {journal}
  {JHEP}\ }\textbf {\bibinfo {volume} {07}},\ \bibinfo {pages} {118} (\bibinfo
  {year} {2016})},\ \Eprint {http://arxiv.org/abs/1603.03462} {arXiv:1603.03462
  [hep-th]} \BibitemShut {NoStop}%
\bibitem [{\citenamefont {Bajc}\ and\ \citenamefont
  {Sannino}(2016)}]{Bajc:2016efj}%
  \BibitemOpen
  \bibfield  {author} {\bibinfo {author} {\bibfnamefont {B.}~\bibnamefont
  {Bajc}}\ and\ \bibinfo {author} {\bibfnamefont {F.}~\bibnamefont {Sannino}},\
  }\href@noop {} {\  (\bibinfo {year} {2016})},\ \Eprint
  {http://arxiv.org/abs/1610.09681} {arXiv:1610.09681 [hep-th]} \BibitemShut
  {NoStop}%
\bibitem [{\citenamefont {Braun}\ \emph {et~al.}(2011)\citenamefont {Braun},
  \citenamefont {Gies},\ and\ \citenamefont {Scherer}}]{Braun:2010tt}%
  \BibitemOpen
  \bibfield  {author} {\bibinfo {author} {\bibfnamefont {J.}~\bibnamefont
  {Braun}}, \bibinfo {author} {\bibfnamefont {H.}~\bibnamefont {Gies}}, \ and\
  \bibinfo {author} {\bibfnamefont {D.~D.}\ \bibnamefont {Scherer}},\ }\href
  {\doibase 10.1103/PhysRevD.83.085012} {\bibfield  {journal} {\bibinfo
  {journal} {Phys. Rev.}\ }\textbf {\bibinfo {volume} {D83}},\ \bibinfo {pages}
  {085012} (\bibinfo {year} {2011})},\ \Eprint {http://arxiv.org/abs/1011.1456}
  {arXiv:1011.1456 [hep-th]} \BibitemShut {NoStop}%
\bibitem [{\citenamefont {Ohta}\ \emph {et~al.}(2015)\citenamefont {Ohta},
  \citenamefont {Percacci},\ and\ \citenamefont {Vacca}}]{Ohta:2015efa}%
  \BibitemOpen
  \bibfield  {author} {\bibinfo {author} {\bibfnamefont {N.}~\bibnamefont
  {Ohta}}, \bibinfo {author} {\bibfnamefont {R.}~\bibnamefont {Percacci}}, \
  and\ \bibinfo {author} {\bibfnamefont {G.~P.}\ \bibnamefont {Vacca}},\ }\href
  {\doibase 10.1103/PhysRevD.92.061501} {\bibfield  {journal} {\bibinfo
  {journal} {Phys. Rev.}\ }\textbf {\bibinfo {volume} {D92}},\ \bibinfo {pages}
  {061501} (\bibinfo {year} {2015})},\ \Eprint
  {http://arxiv.org/abs/1507.00968} {arXiv:1507.00968 [hep-th]} \BibitemShut
  {NoStop}%
\end{thebibliography}%

\end{document}